\documentclass[12pt]{article}
\usepackage[utf8]{inputenc}
\usepackage[T1]{fontenc}

\usepackage{graphicx}
\graphicspath{ {./plot/} }

\usepackage{authblk}

\usepackage{caption}

\usepackage{float}
\usepackage{subcaption}
\usepackage{amsmath, amssymb, amsthm}
\usepackage{mathtools}
\usepackage{bbm}
\usepackage{bm}
\usepackage{mathrsfs}
\usepackage{tikz}
\usepackage{pgfplots}
\usepgfplotslibrary{dateplot}

\usepackage{xr-hyper}

\makeatletter
\newcommand*{\addFileDependency}[1]{
  \typeout{(#1)}
  \@addtofilelist{#1}
  \IfFileExists{#1}{}{\typeout{No file #1.}}
}
\makeatother

\usepackage{afterpage}
\usepackage{hyperref}
\hypersetup{hidelinks}
\hypersetup{
colorlinks=false,
}
\usepackage{xr}

\usepackage{multirow}

\usepackage{enumerate, enumitem}
\usepackage{fancyhdr, graphicx, proof, comment, multicol}
\usepackage[none]{hyphenat} 

\usepackage{microtype} 
\usepackage{mdframed} 

\usepackage[round]{natbib}
\setcitestyle{authoryear,open={(},close={)}}

\newtheorem{thm}{Theorem}
\newtheorem{lemma}{Lemma}
\newtheorem{proposition}{Proposition}
\newtheorem{corollary}{Corollary}

\begin{document}
\def\spacingset#1{\renewcommand{\baselinestretch}%
{#1}\small\normalsize} \spacingset{1}






\title{\bf Structured Mixture of Continuation-ratio Logits Models for Ordinal Regression}
\author{Jizhou Kang and Athanasios Kottas\thanks{Jizhou Kang (jkang37@ucsc.edu)
is Ph.D. student, and Athanasios Kottas (thanos@soe.ucsc.edu) is Professor, Department 
of Statistics, University of California, Santa Cruz.
The research was supported in part by the National Science Foundation under award SES 1950902.
The authors wish to thank an Associate Editor and three reviewers 
for several useful comments.
} \\
Department of Statistics, University of California, Santa Cruz\\
}
%
\maketitle

\bigskip
\begin{abstract}
We develop a nonparametric Bayesian modeling approach to ordinal regression based on 
priors placed directly on the discrete distribution of the ordinal responses. The prior 
probability models are built from a structured mixture of multinomial distributions. We 
leverage the continuation-ratio logits representation to formulate the mixture kernel, with 
mixture weights defined through the logit stick-breaking process that incorporates the 
covariates through a linear function. The implied regression functions for the response 
probabilities can be expressed as weighted sums of parametric regression functions, with 
covariate-dependent weights. Thus, the modeling approach achieves flexible ordinal regression 
relationships, avoiding linearity or additivity assumptions in the covariate effects. 
Model flexibility is formally explored through the Kullback-Leibler support 
of the prior probability model.
A key model feature is that the parameters for both the mixture 
kernel and the mixture weights can be associated with the continuation-ratio logits regression
structure. Hence, an efficient and relatively easy to implement posterior simulation method
can be designed, using Pólya-Gamma data augmentation. Moreover, the model is built from a 
conditional independence structure for category-specific parameters, which results in 
additional computational efficiency gains through partial parallel sampling. In addition to 
the general mixture structure, we study simplified model versions that incorporate 
covariate dependence only in the mixture kernel parameters or only in the mixture weights.
For all proposed models, we discuss approaches to prior specification and develop Markov 
chain Monte Carlo methods for posterior simulation. The methodology is illustrated with 
several synthetic and real data examples. 
\end{abstract}

\noindent%

\newpage
\spacingset{1.75}

\section{Introduction}
\label{sec:intro}

Ordinal responses are widely encountered in many fields, including econometrics and the 
biomedical and social sciences, typically accompanied by covariate information. Hence, 
estimation and prediction of ordinal regression relationships remains a methodologically 
and practically relevant problem. The typical ordinal regression setting consists 
of a univariate ordinal response $Y$ with $C$ categories, and a covariate vector $\mathbf{x}$. 
The modeling challenge for the ordinal regression problem involves capturing 
general regression relationships in the response probabilities (especially for moderate
to large $C$), while at the same time appropriately accounting for the ordinal nature 
of the response.

A commonly used approach involves cumulative link models \citep[e.g.,][]{Agresti2012}, under which 
the ordinal responses can be viewed as a discretized version of latent continuous responses, typically 
assumed normally distributed resulting in popular cumulative probit models.
For Bayesian inference, such data augmentation facilitates posterior simulation 
\citep{AlbertChib1993}. However, probit models preclude a flexible analysis of 
probability response curves, since covariate effects enter linearly and additively, 
and the normality assumption implies restrictions on the marginal response probabilities 
\citep[e.g.,][]{Boes2006}. In general, parametric ordinal regression models sacrifice 
flexibility in the response distribution and/or the regression functions for the 
response probabilities.


To overcome such limitations, early work in the Bayesian nonparametrics literature has 
explored semiparametric models, focusing mostly on binary regression.
Such methods relax parametric assumptions for the distribution of the latent variables 
\citep[e.g.,][]{BasuChib2003} or for the regression function \citep[e.g.,][]{Choudhuri2007}.
As a further extension, \citet*{ChibGreenberg2010} modeled covariate effects additively by 
cubic splines, combined with a scale normal mixture for the latent responses, using the 
Dirichlet process (DP) prior \citep{Ferguson1973} for the mixing distribution.
More general DP mixture priors for the distribution of the latent continuous responses 
have been considered in \cite{BaoHanson} and \cite{DeYoreoKottas2018}. The latter involves 
a fully nonparametric Bayesian method under the density regression framework, modeling  
the joint distribution of covariates and latent responses with a DP mixture of multivariate 
normals. \cite{MDAKbook} provide a review of the joint response-covariate modeling 
approach with categorical variables. 
The density regression modeling framework is appealing with regard to the scope of ordinal 
regression inferences. However, it involves computationally intensive posterior simulation 
which does not scale with the number of covariates, and it is not suitable for applications 
where the assumption of random covariates is not relevant.
%

%
%
The ``logits regression family'' \citep{Agresti2012} offers an alternative approach to ordinal 
regression, based on direct modeling of the response distribution. Of particular interest 
to our methodology are continuation-ratio logits models. 
The continuation-ratio logits parameterization of the multinomial
distribution implies a sequential mechanism, such that the ordinal response is determined through 
a sequence of binary outcomes. Starting from the lowest category, each binary outcome indicates 
whether the ordinal response belongs to that category or to one of the higher categories. 
The continuation-ratio logit for response category $j$ is the logit of the conditional 
probability of response $j$, given that the response is $j$ or higher. A key consequence is 
that, in a multinomial continuation-ratio logits regression model, the response distribution 
can be factorized into complete conditionals defined by Binomial logistic regression models. 
%

To our knowledge, continuation-ratio logits have not been explored for general Bayesian 
nonparametric methods for ordinal regression. 
For nominal regression, \citet{Linderman2015} discussed a semiparametric model that, under the 
multinomial response distribution, replaces the linear covariate effects within the 
continuation-ratio logits by Gaussian process priors. 
More relevant to our methodology is the dependent DP mixture model in \cite{KF2013}, based on a 
trinomial kernel that builds from the continuation-ratio logits formulation. This modeling 
approach was developed specifically in the context of developmental toxicity studies, rather
than for general ordinal regression problems. 

The continuation-ratio logits structure is 
attractive as a building block for general nonparametric Bayesian ordinal regression modeling, 
and this is the primary motivation for our methodology. We build the response distribution 
from a nonparametric mixture of multinomial distributions, mixing on the regression coefficients 
under the continuation-ratio logits formulation for the mixture kernel. Model flexibility is enhanced 
through covariate dependent mixture weights, 
assigned a logit stick-breaking prior \citep{RigonDurante2021}. The stick-breaking structure, along 
with the logistic form for the underlying covariate dependent variables, yields a continuation-ratio 
logits regression representation also for the mixture weights. The similarity in the structure of 
the mixture kernel and the mixture weights is a distinguishing feature of the methodology, in terms 
of model properties and model implementation. 
We take advantage of this structure, as well as a latent variable model formulation, 
to explore the Kullback-Leibler support of the prior probability model.

Regarding model implementation, using the Pólya-Gamma data augmentation approach for logistic 
regression \citep{Polson2013}, we design an efficient Gibbs sampling algorithm for posterior inference. 
The posterior simulation method is ready to implement, in particular, it does not require specialized 
techniques or tuning of Metropolis-Hastings steps. Moreover, the product of Binomials formulation of 
the multinomial kernel yields a Gibbs sampler which, given all other model parameters, allows for 
separate updates for each set of mixture kernel parameters associated with each response category. 
Hence, the complexity of the inference procedure is not unduly increasing with the number of response 
categories.

The model yields flexible probability response curves expressed as weighted sums of parametric 
regression functions with local, covariate-dependent weights. As simplified versions of the 
general model structure, we explore mixture models that incorporate the covariates only in the 
kernel parameters or only in the weights. We study model properties and use synthetic and real 
data examples to compare the different model formulations.

Our objective is to develop a general toolbox for ordinal regression that allows flexibility in 
both the response distribution and the ordinal regression relationships. The toolbox comprises
models of different complexity, all of which can be implemented with relatively straightforward
posterior simulation methods. It also includes prior specification methods that range from a
fairly non-informative choice to more informative options that enable incorporation of 
monotonicity trends for the probability response functions.

The article is organized as follows. In Section \ref{sec:orimodel}, we formulate the general modeling
approach, and discuss prior specification, posterior inference, 
and model properties, including Kullback-Leibler support (with technical details given 
in the Supplementary Material). Section \ref{sec:specorimodel} presents the two simplified mixture 
models. The methodology is illustrated in Section \ref{sec:oriexample} with synthetic and real 
data examples. Section \ref{sec:summary} concludes with discussion.

\section{General methodology}
\label{sec:orimodel}

\subsection{From building blocks to general model}
\label{subsec:standardmodel}

Consider an ordinal response $Y$ with $C$ categories, recorded along with a 
covariate vector $\mathbf{x}\in\mathbb{R}^p$. We can equivalently encode the response 
as a vector of binary variables $\mathbf{Y}=$ $(Y_{1},\ldots,Y_{C})$, such that $Y=j$ 
is equivalent to $Y_{j}=1$ and $Y_{k}=0$ for any $k\neq j$. 

The continuation-ratio logits regression model builds from the factorization of the 
multinomial distribution in terms of Binomial distributions, 
%
\begin{equation}
Mult(\mathbf{Y} \mid 1,\pi_1,\ldots,\pi_C) \, = \,
Bin(Y_1 \mid m_1,\varphi(\theta_{1})) \ldots
Bin(Y_{C-1} \mid m_{C-1},\varphi(\theta_{C-1})),
\label{eq:paramcontilogits}
\end{equation}
where $m_{1}=1$, and $m_{j}=1-\sum_{k=1}^{j-1}Y_{k}$, for $j=2,\ldots,C-1$, 
$\theta_{j} \equiv$ $\theta_{j}(\mathbf{x}) =$
$\mathbf{x}^T\boldsymbol{\beta}_j$, and $\varphi(\theta) = $ $\exp(\theta)/\{ 1 + \exp(\theta) \}$ 
denotes the standard logistic function. 
The two parameterizations are linked through $\pi_1=\varphi(\theta_1)$, $\pi_j=$
$\varphi(\theta_j) \prod_{k=1}^{j-1} \{ 1-\varphi(\theta_k) \}$, for $j=2,\ldots,C-1$, 
and $\pi_C=$ $\prod_{k=1}^{C-1} \{ 1-\varphi(\theta_k) \}$.
For notation simplicity, we use $K(\mathbf{Y} \mid \boldsymbol{\theta})$, where
$\boldsymbol{\theta} =$ $(\theta_{1},\ldots,\theta_{C-1})$, for the continuation-ratio 
logits representation of the multinomial distribution.

The parametric model is limited in the response distribution and the form of covariate effects. 
A strategy that surpasses these limitations and achieves flexible inference is to 
generalize the model via Bayesian nonparametric mixing. Using the kernel function in 
(\ref{eq:paramcontilogits}) in conjunction with a nonparametric prior for the covariate-dependent 
mixing distribution, we achieve the general nonparametric extension of the continuation-ratio 
logits model, 
\begin{equation}
\mathbf{Y} \mid G_{\mathbf{x}} \, \sim \, 
\int K(\mathbf{Y} \mid \boldsymbol{\theta}) \, dG_{\mathbf{x}}(\boldsymbol{\theta})
\, = \, \sum_{\ell=1}^{\infty} \omega_{\ell}(\mathbf{x}) \, 
K(\mathbf{Y} \mid \boldsymbol{\theta}_{\ell}(\mathbf{x})).
\label{eq:npbext}
\end{equation}
Here, the countable mixture form emerges under the nonparametric prior formulation for the 
mixing distribution that represents it as a discrete distribution, $G_{\mathbf{x}}=$
$\sum_{\ell=1}^{\infty} \omega_{\ell}(\mathbf{x}) \, \delta_{\boldsymbol{\theta}_{\ell}(\mathbf{x})}$,
with covariate-dependent atoms, $\boldsymbol{\theta}_{\ell}(\mathbf{x})$, and weights, 
$\omega_{\ell}(\mathbf{x})$.

%
%

The prior formulation for $G_{\mathbf{x}}$ in (\ref{eq:npbext}) is generic. There are several 
options for building the model for the atoms and weights, a stick-breaking formulation for the 
latter being the more common strategy. The dependent DP (DDP) prior and related 
models \citep{MacEachern2000,Quintana2022} has been explored in different applications, including 
simplified ``common-weights'' or ``common-atoms'' versions under which only the atoms or the 
weights, respectively, depend on the covariates. Other options include the kernel stick-breaking 
process \citep{Dunson2008}, the probit stick-breaking process \citep{DunsonAbel2011}, and the 
logit stick-breaking process \citep{RigonDurante2021}.

As discussed below, for the ordinal regression problem with mixture kernel 
$K(\mathbf{Y} \mid \boldsymbol{\theta})$, the logit stick-breaking process (LSBP) 
prior offers a key advantage in model structure and in posterior simulation.  
Therefore, for the general model in (\ref{eq:npbext}), we assume the 
following LSBP prior for the covariate-dependent weights:
\begin{equation}
%
\omega_{1}(\mathbf{x}) \, = \, \varphi(\mathbf{x}^T\boldsymbol{\gamma}_1), \,\,\,
\omega_{\ell}(\mathbf{x}) \, = \,
\varphi(\mathbf{x}^T\boldsymbol{\gamma}_{\ell}) 
\prod_{h=1}^{\ell-1} (1-\varphi(\mathbf{x}^T\boldsymbol{\gamma}_h)), \,\,\, \ell \geq 2; 
\,\,\,\,\,
\boldsymbol{\gamma}_{\ell} \, \stackrel{i.i.d.}{\sim} \, N(\boldsymbol{\gamma}_0,\Gamma_0)
\label{eq:generalmodori}
\end{equation}
%
In addition, the atoms, 
$\boldsymbol{\theta}_{\ell}(\mathbf{x}) = 
(\theta_{1\ell}(\mathbf{x}),\ldots,\theta_{C-1,\ell}(\mathbf{x}))$, 
are built through a linear regression structure, 
\begin{equation}
\theta_{j\ell}(\mathbf{x}) \, = \, \mathbf{x}^T\boldsymbol{\beta}_{j\ell}
\mid \boldsymbol{\mu}_j, \Sigma_j \, \stackrel{ind.}{\sim} \, 
N(\mathbf{x}^T\boldsymbol{\mu}_j,\mathbf{x}^T\Sigma_j\mathbf{x}), \,\,\,\,\,
j=1,\ldots,C-1, \,\,\, \ell \geq 1,
\label{general_model_atoms}
\end{equation}
with the random variables that define the atoms assumed a priori independent of those 
that define the weights. 
%
The model is completed with the conjugate prior for the collection of hyperparameters 
$\boldsymbol{\psi} =$ $\{\boldsymbol{\mu}_j,\Sigma_j\}_{j=1}^{C-1}$, that is, 
\begin{equation}
\Sigma_j \stackrel{ind.}{\sim} IW(\nu_{0j},\Lambda_{0j}^{-1}), \,\,\,\,\,\,\,
\boldsymbol{\mu}_j \mid \Sigma_j \stackrel{ind.}{\sim} 
N(\boldsymbol{\mu}_{0j},\Sigma_j/\kappa_{0j}), \,\,\,\,\, j=1,\ldots, C-1.
\label{general_model_hyperparameters}
\end{equation}
In Section \ref{subsec:prispecori}, we discuss prior specification for 
$\{ \nu_{0j},\Lambda_{0j},\boldsymbol{\mu}_{0j},\kappa_{0j} \}_{j=1}^{C-1}$, and for 
$\boldsymbol{\gamma}_0,\Gamma_0$.

\begin{figure}[t!]
\centering
\begin{tikzpicture}[thick,scale=0.88, every node/.style={scale=0.88}]

\node {Generate $\mathcal{H}_{1}\sim Bern(\Delta_1), \Delta_1=\varphi(\mathbf{x}^T\boldsymbol{\beta}_1),\boldsymbol{\beta}_1\sim N(\boldsymbol{\mu}_1,\Sigma_1)$}[sibling distance = 6.4cm]
    child {node {$Y=1$} edge from parent node [left] {$\mathcal{H}_{1}=1$}}
    child {node {Generate $\mathcal{H}_{2}\sim Bern(\Delta_2),\Delta_2=\varphi(\mathbf{x}^T\boldsymbol{\beta}_2),\boldsymbol{\beta}_2\sim N(\boldsymbol{\mu}_2,\Sigma_2)$}[sibling distance = 6.4cm] 
    child {node {$Y=2$} edge from parent node [left] {$\mathcal{H}_{2}=1$}}
    child {node {Generate $\mathcal{H}_{3}\sim Bern(\Delta_3),\Delta_3=\varphi(\mathbf{x}^T\boldsymbol{\beta}_3),\boldsymbol{\beta}_3\sim N(\boldsymbol{\mu}_3,\Sigma_3)$}[sibling distance = 6.4cm] 
    child {node {$Y=3$} edge from parent node [left] {$\mathcal{H}_{3}=1$}}
    child {node {$\cdots$} edge from parent node [right] {$\mathcal{H}_{3}=0$}}
    edge from parent node [right] {$\mathcal{H}_{2}=0$}}
    edge from parent [black] node [right] {$\mathcal{H}_{1}=0$}};
\end{tikzpicture}
\caption{
{\small Illustration of the continuation-ratio logits structure.} 
}
\label{fig:illustratestructure}
\end{figure}
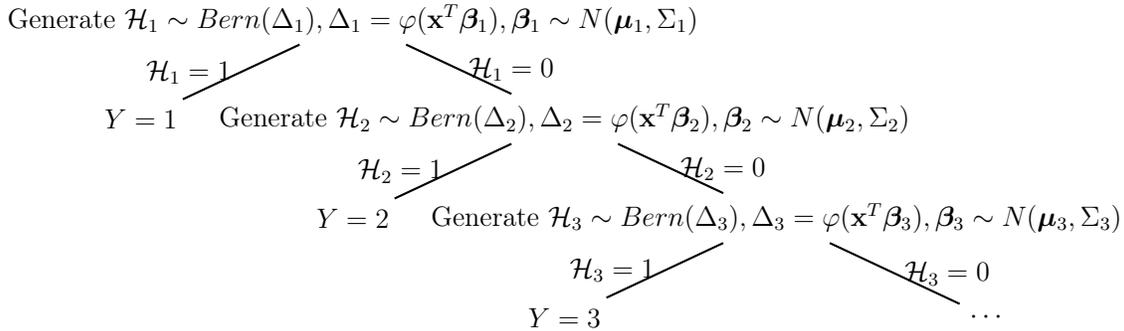

To point to the benefit of working with the LSBP prior, we examine the 
continuation-ratio logits structure in (\ref{eq:paramcontilogits}). As illustrated 
in Figure \ref{fig:illustratestructure}, such structure implies a sequential mechanism 
in determining the ordinal response ${Y}$. At a generic step $j$, a Bernoulli variable 
$\mathcal{H}_{j}\sim Bern(\Delta_{j})$ is generated to either set ${Y}=j$ if $\mathcal{H}_{j}=1$, 
or to allocate ${Y}$ to $\{k: k>j\}$ when $\mathcal{H}_{j}=0$. The $j$-th step can only be 
reached if ${Y}$ has not been assigned to $1,\ldots,j-1$. To bring in the covariate effects, 
we place a logit-normal prior on $\Delta_{j}$, that is, $\Delta_{j}=$
$\varphi(\mathbf{x}^T \boldsymbol{\beta}_j)$ and 
$\boldsymbol{\beta}_j \sim N(\boldsymbol{\mu}_j,\Sigma_j)$. This procedure provides 
a natural way of defining a stick-breaking process, engendering 
the LSBP as mentioned in \citet{RigonDurante2021}. Consider a configuration variable 
$\mathcal{L}$, corresponding to $\mathbf{Y}$, that indicates the mixture component 
in (\ref{eq:npbext}) from which $\mathbf{Y}$ is generated. The same sequential 
generative process applies to $\mathcal{L}$. At generic step $\ell$, a Bernoulli variable 
$\mathcal{H}^{*}_{\ell}\sim Bern(\eta_{\ell})$ is generated, serving the same role 
as $\mathcal{H}_j$ in determining whether $\mathcal{L}$ locates at the current stage, 
or moves to later stages. Treating $\eta_{\ell}$ as the stick-breaking proportion, 
the covariate effects are incorporated through $\eta_{\ell}(\mathbf{x})=\varphi(\mathbf{x}^T\boldsymbol{\gamma}_{\ell})$. The resulting 
nonparametric model admits the countable mixture representation in (\ref{eq:npbext}), 
with weights and atoms depending on covariates in a similar fashion. We highlight this 
correspondence because it paves the way in developing tractable posterior inference 
strategies, which will be discussed in Section \ref{subsec:postinfori}.


In this section, we consider properties under the general model formulation in (\ref{eq:npbext}) 
comprising the covariate-dependent weights and atoms in (\ref{eq:generalmodori}) and 
(\ref{general_model_atoms}), respectively. In Section \ref{sec:specorimodel}, we discuss 
the simpler common-weights and common-atoms models as a means to address the trade-off
between the flexibility of model (\ref{eq:npbext}) and its potential computational cost.
Our study of model properties and data illustrations explore such trade-off and suggest 
scenarios for which the simpler models may be suitable. 
%
%

\subsection{Model properties}
\label{subsec:modpropori}

The covariate-response relationship can be studied through the marginal 
probability response curves $\text{Pr}(\mathbf{Y}=j \mid G_{\mathbf{x}})$, for $j=1,\ldots,C$. 
Given the ordinal nature of the response, also of interest are the conditional 
probability response curves, $\text{Pr}(\mathbf{Y}=j \mid \mathbf{Y}\geq j,G_{\mathbf{x}})$. 
Here, we slightly abuse notation by writing $\mathbf{Y} = j$, while it is actually 
$\mathbf{Y}=$ $\mathbf{1}_j$, the unit vector in $\mathbb{R}^C$ with the $j$th element equal to 1.

Based on the particular mixture of multinomial distributions for the general model 
in (\ref{eq:npbext}), the marginal probability response curve for $j=1,\ldots,C$
can be expressed as 
\begin{equation}
\text{Pr}(\mathbf{Y}=j \mid G_{\mathbf{x}}) \, = \, \sum_{\ell=1}^{\infty} 
\omega_{\ell}(\mathbf{x}) \, \left\{ \varphi(\theta_{j\ell}(\mathbf{x})) \prod\nolimits_{k=1}^{j-1}[1-\varphi(\theta_{k\ell}(\mathbf{x}))]
\right\},
\label{eq:margcurvegen}
\end{equation}
where the weights, $\omega_{\ell}(\mathbf{x})$, and atoms, $\theta_{j\ell}(\mathbf{x})$,
are defined in (\ref{eq:generalmodori}) and (\ref{general_model_atoms}), respectively, 
and we set $\varphi(\theta_{C\ell}(\mathbf{x}))\equiv1$. Moreover, the conditional probability 
response curves are given by
\begin{equation}
\text{Pr}(\mathbf{Y}=j \mid \mathbf{Y}\geq j,G_{\mathbf{x}}) \, = \,
\sum_{\ell=1}^{\infty} w_{j\ell}(\mathbf{x}) \, \varphi(\theta_{j\ell}(\mathbf{x}));
\quad  w_{j\ell}(\mathbf{x}) = \frac{\omega_{\ell}(\mathbf{x})
\prod_{k=1}^{j-1}[1-\varphi(\theta_{k\ell}(\mathbf{x}))]}
{\sum_{\ell=1}^{\infty}\omega_{\ell}(\mathbf{x})
\prod_{k=1}^{j-1}[1-\varphi(\theta_{k\ell}(\mathbf{x}))]}
\label{eq:condcurvegen}
\end{equation}
Both the marginal and conditional probability response curves admit a weighted sum 
representation with component regression functions that correspond to the parametric 
continuation-ratio logits model. The covariate-dependent weights in (\ref{eq:margcurvegen}) 
and (\ref{eq:condcurvegen}) allow for local adjustment over the covariate space, thus 
enabling non-standard regression relationships and relaxing the restrictions on 
the covariate effects under the parametric model.
%

A useful observation is that the continuation-ratio logits model plays the role of a 
parametric backbone for the nonparametric model, in the sense of prior expectation. 
More specifically, using (\ref{eq:margcurvegen}), and the assumptions of the prior 
model in (\ref{eq:npbext}), (\ref{eq:generalmodori}) and (\ref{general_model_atoms}), 
\begin{equation}
\begin{split}
\text{E}(\text{Pr}(\mathbf{Y}=j \mid G_{\mathbf{x}})) & =
\sum_{\ell=1}^{\infty} \text{E}(\omega_{\ell}(\mathbf{x})) \,
\text{E}\left\{
\varphi(\theta_{j\ell}(\mathbf{x}))
\prod\nolimits_{k=1}^{j-1}[1-\varphi(\theta_{k\ell}(\mathbf{x}))]
\right\} \\
& = \text{E}\left\{
\varphi(\mathbf{x}^T\boldsymbol{\beta}_j)\prod\nolimits_{k=1}^{j-1}
[1-\varphi(\mathbf{x}^T\boldsymbol{\beta}_k)] \right\},
\end{split}    
\label{eq:expprobrespcurvegen}
\end{equation}
where the last expectation is taken with respect to 
$\boldsymbol{\beta}_j \stackrel{ind.}{\sim} N(\boldsymbol{\mu}_j,\Sigma_j)$, 
$j=1,\ldots,C-1$. Hence, the prior expectation for the marginal probability response 
curves under the nonparametric model reduces to the prior expectation under the 
parametric model. This property facilitates prior specification, as discussed in 
Section \ref{subsec:prispecori}.

The general model can capture a spectrum of inferences, with the parameters 
$\boldsymbol{\gamma}_{\ell}$ controling the number of effective mixture components. 
Suppose the covariates take values in a bounded region. If $\boldsymbol{\gamma}_1$ 
results in $\varphi(\mathbf{x}^T\boldsymbol{\gamma}_1)$ effectively equal to one, then 
the nonparametric model collapses to its parametric backbone. On the other hand, if the 
first several $\boldsymbol{\gamma}_{\ell}$ are such that $\varphi(\mathbf{x}^T\boldsymbol{\gamma}_{\ell})$ are relatively small, a larger number
of effective components is favored, in the extreme utilizing a distinct multinomial 
component for each ordinal response. In practice, the strength of the nonparametric 
model lies between these two extremes.

\subsection{Prior specification}
\label{subsec:prispecori}

%

To implement the general model in (\ref{eq:generalmodori}), (\ref{general_model_atoms}) 
and (\ref{general_model_hyperparameters}), we need to specify the parameters of the hyperpriors,
that is, $(\boldsymbol{\gamma}_0,\Gamma_0)$ and 
$\{ \nu_{0j},\Lambda_{0j},\boldsymbol{\mu}_{0j},\kappa_{0j} \}_{j=1}^{C-1}$.

We set $\kappa_{0j}=\nu_{0j}=p+2$ for all $j$, where $p$ is the dimension of the covariate 
vector $\mathbf{x}$ (including the intercept). For the other prior hyperparameters, the proposed 
strategy is developed by first considering the prior expected probability response curves to 
specify $\{\boldsymbol{\mu}_{0j},\Lambda_{0j}\}_{j=1}^{C-1}$, and then using the prior expected 
weight placed on each mixing component to determine $\boldsymbol{\gamma}_0$ and $\Gamma_0$.

The weights and atoms of the mixture model have the same structure. Specifically, the weights 
are generated from a stick-breaking process with breaking proportion $\eta_{\ell}(\mathbf{x})=$
$\varphi(\mathbf{x}^T\boldsymbol{\gamma}_{\ell})$, while the atoms can also be viewed as 
possessing a stick-breaking form with breaking proportion $\Delta_{j\ell}(\mathbf{x})=$ $\varphi(\mathbf{x}^T\boldsymbol{\beta}_{j\ell})$. Taking the prior into consideration, we have $\eta_{\ell}(\mathbf{x}) \sim LN(\mathbf{x}^T\boldsymbol{\gamma}_0,\mathbf{x}^T\Gamma_0\mathbf{x})$ 
and $\Delta_{j\ell}(\mathbf{x})\sim LN(\mathbf{x}^T\boldsymbol{\mu}_{0j},(\kappa_{0j}+1)/(\kappa_{0j}(\nu_{0j}-p-1))\mathbf{x}^T\Lambda_{0j}\mathbf{x})$, where $LN(\cdot,\cdot)$ denotes the logit-normal distribution. Therefore, 
a key quantity in prior specification is the expectation of a logit-normal distributed random 
variable, which does not have analytical form in general.

Nonetheless, if $Z\sim N(0,\sigma^2)$, then $\text{E}(\varphi(Z)) = 0.5$, for any value of 
the variance $\sigma^{2}$ \citep{Pirjol2013}.
%
%
%
%
This result motivates the default choice of hyperparameters we use in practice, that is, 
$\boldsymbol{\mu}_{0j} =$ $\boldsymbol{\gamma}_0=\mathbf{0}_p$, and $\Lambda_{0j} =$
$\Gamma_0 = 10^2 \, \mathbf{I}_p$. We refer to this specification as the ``baseline'' prior,
which yields $\text{E}(\text{Pr}(\mathbf{Y}=j\mid G_{\mathbf{x}})) = 2^{-j}$, for $j=1,\ldots,C-1$, 
and $\text{E}(\text{Pr}(\mathbf{Y}=C\mid G_{\mathbf{x}})) = 2^{-(C-1)}$, for all $\mathbf{x}$. 
The prior expectation of the weight associated with the $\ell$th mixing component is given by 
$2^{-\ell}$, for any $\ell$.

In general, both the shape of the prior expected probability response curves and the 
prior expected weight placed on each mixing component depend on the expectation of 
the logit-normal distribution. Even though that expectation does not have analytical 
form, it can be readily obtained by simulation. Therefore, we can tune the prior 
hyperparameters and evaluate the prior expectation of $\eta_{\ell}(\mathbf{x})$ and $\Delta_{j\ell}(\mathbf{x})$. For instance, we can favor prior expected probability 
response curves possessing some specific pattern (such as monotonicity) and/or a 
certain number of mixture components. The following proposition, which can be obtained 
using results from \citet{Pirjol2013}, facilitates the tuning of prior hyperparameters. 

\begin{proposition}
\label{prop:logitnormalexpbound}
If $Z\sim N(\mu,\sigma^2)$, then $\varphi(\mu-\sigma^2/2)\leq \text{E}(\varphi(Z))\leq\varphi(\mu+\sigma^2/2)$.
\end{proposition}

\begin{figure}[t!]
\centering
\includegraphics[width=15.5cm,height=3.5cm]{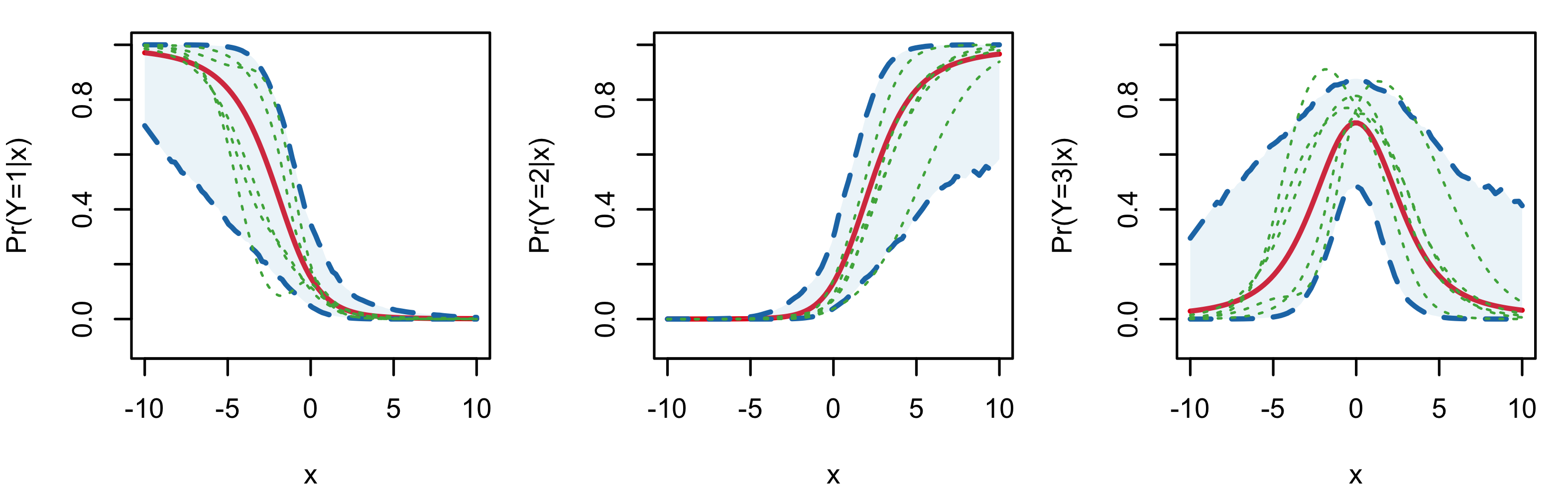}
\caption{
{\small Illustration of the prior specification strategy. 
In each panel, the red solid line is the prior expected probability 
response curve, the blue dashed lines and shaded region indicate the prior $95\%$ interval 
estimate, and the green dotted lines show 5 prior realizations.}
}
\label{fig:priorpiest}
\end{figure}

As an illustration, consider an ordinal response with $C=3$ categories, and a single covariate 
taking values in $(-10,10)$. Suppose the prior information is that the first marginal 
probability response function is decreasing, whereas the second is increasing. Using a 
particular prior choice, Figure \ref{fig:priorpiest} shows point and interval estimates that 
reflect such prior information, with a fair amount of variability. The details for determining 
the hyperparameters, using Proposition 1, are presented in the Supplementary Material.
   
\subsection{Posterior inference}
\label{subsec:postinfori}

%
%

For Markov chain Monte Carlo (MCMC) posterior simulation, we work with a truncation 
approximation of the mixing distribution in the spirit of blocked Gibbs sampling for 
stick-breaking priors \citep{IshwaranJames2001}.
We favor the blocked Gibbs sampler as it results in practical model implementation 
and it allows for full posterior inference for general regression functionals. 
Hence, for posterior simulation, the mixing distribution $G_{\mathbf{x}}$ in (\ref{eq:npbext}) 
is replaced by $G_{\mathbf{x}}^L=$ 
$\sum_{\ell=1}^L p_{\ell}(\mathbf{x}) \, \delta_{\boldsymbol{\theta}_{\ell}(\mathbf{x})}$, 
with $\boldsymbol{\theta}_{\ell}(\mathbf{x})$ 
defined as before, and $p_{\ell}(\mathbf{x}) =$ $\omega_{\ell}(\mathbf{x})$, for 
$\ell = 1,...,L-1$, whereas $p_{L}(\mathbf{x})=$ $1-\sum^{L-1}_{\ell=1}p_{\ell}(\mathbf{x})$.

The truncation level $L$ can be chosen to achieve any desired level of accuracy. For normal 
mixtures with LSBP weights, \cite{RigonDurante2021} show that, for fixed sample size and 
covariates, the $L^{1}$ distance between the prior predictive distribution of the sample under 
$G_{\mathbf{x}}$ and $G_{\mathbf{x}}^L$ decreases exponentially in $L$. The proof for this result (Theorem 1 in \cite{RigonDurante2021}) applies to essentially any 
mixture kernel, and it thus also holds for the multinomial LSBP mixture model defined in 
(\ref{eq:npbext}), (\ref{eq:generalmodori}) and (\ref{general_model_atoms}).


In practice, we can 
specify $L$ using the prior expectation for the partial sum of weights. Under the prior 
in (\ref{eq:generalmodori}), $\text{E}(\sum_{\ell=1}^L\omega_{\ell}(\mathbf{x}))=$
$1 - \{ 1 - \text{E}(\varphi(\mathbf{x}^T \boldsymbol{\gamma})) \}^{L}$, where the expectation on 
the right-hand-side is with respect to $\boldsymbol{\gamma} \sim N(\boldsymbol{\gamma}_0,\Gamma_0)$. 
Hence, $L$ can be selected by computing the expectation at a few representative values 
in the covariate space. Note that, when $\boldsymbol{\gamma}_0=\mathbf{0}_{p}$, 
$\text{E}(\varphi(\mathbf{x}^T \boldsymbol{\gamma})) = 0.5$, for any $\mathbf{x}$. 
We also recommend monitoring the posterior samples for $p_{L}(\mathbf{x})$ for different 
values $\mathbf{x}$ in the covariate space. Using a combination of such strategies, 
we worked with the (conservative) truncation level of $L=50$ for the data examples of 
Section \ref{sec:oriexample}.

Denote by $\mathbf{Y}_{i} =$ $(Y_{i1},\ldots,Y_{iC})$, where $Y_{ij} \in \{ 0,1 \}$ 
with $\sum_{j=1}^{C} Y_{ij} = 1$, the $i$th observed response, and by $\mathbf{x}_{i}$ the 
corresponding covariate vector, for $i=1,\ldots,n$. We introduce latent configuration 
variables, $\{ \mathcal{L}_i \}$, such that $\mathcal{L}_i=\ell$ if and only if 
$\mathbf{Y}_{i}$ is assigned to the $\ell$th mixture component. Then, the hierarchical 
model for the data can be expressed as
\begin{equation}
\begin{array}{rcl}
\mathbf{Y}_i \mid \{ \boldsymbol{\beta}_{j \ell} \},\mathcal{L}_i & \stackrel{ind.}{\sim} & 
K(\mathbf{Y}_i \mid \boldsymbol{\theta}_{\mathcal{L}_i}) = \prod\limits_{j=1}^{C-1} 
Bin(Y_{ij} \mid m_{ij},\varphi(\mathbf{x}_i^T\boldsymbol{\beta}_{j\mathcal{L}_i})), 
\,\,\, i=1,\ldots,n
\\ 
\mathcal{L}_i \mid \{ \boldsymbol{\gamma}_{\ell} \} & \stackrel{ind.}{\sim} &  
\sum\limits_{\ell=1}^L p_{i\ell} \, \delta_{\ell}(\mathcal{L}_i), \,\,\, i=1,\ldots,n
\\
\boldsymbol{\beta}_{j\ell} \mid (\boldsymbol{\mu}_j,\Sigma_j) & \stackrel{ind.}{\sim} & N(\boldsymbol{\mu}_j,\Sigma_j), \,\,\, j=1,\ldots,C-1, \,\,\, \ell=1,\ldots,L 
\\
\boldsymbol{\gamma}_{\ell} & \stackrel{i.i.d.}{\sim} & N(\boldsymbol{\gamma}_0,\Gamma_0),
\,\,\, \ell = 1,\ldots,L-1
\\
(\boldsymbol{\mu}_j,\Sigma_j) & \stackrel{ind.}{\sim} & 
N(\boldsymbol{\mu}_j \mid \boldsymbol{\mu}_{0j},\Sigma_j/\kappa_{0j}) \,
IW(\Sigma_j \mid \nu_{0j},\Lambda_{0j}^{-1}), \,\,\, j = 1,\ldots,C-1
\label{eq:hiergenmod}
\end{array}
\end{equation}
where $m_{i1}=1$, $m_{ij}=$ $1 - \sum_{k=1}^{j-1} Y_{ik}$, for $j=2,\ldots,C-1$,
$p_{i\ell}=$ $\varphi(\mathbf{x}_i^T \boldsymbol{\gamma}_{\ell}) 
\prod_{h=1}^{\ell-1} (1-\varphi(\mathbf{x}_i^T\boldsymbol{\gamma}_{h}))$, 
for $\ell = 1,\ldots,L-1$, and $p_{iL}=$
$\prod_{\ell=1}^{L-1} (1-\varphi(\mathbf{x}_i^T\boldsymbol{\gamma}_{\ell}))$.

Akin to the ordinal response $\mathbf{Y}_i$ and its original form $Y_i$, we can view the 
latent configuration variable $\mathcal{L}_i$ as the allocation of its multivariate form 
$\boldsymbol{\mathcal{L}}_i = (\mathcal{L}_{i1},\ldots,\mathcal{L}_{iL})
\in \mathbb{R}^L$, with the connection defined as $\mathcal{L}_i=\ell\Longleftrightarrow\boldsymbol{\mathcal{L}}_i=\mathbf{1}_{\ell}$, 
the unit vector in $\mathbb{R}^L$ with the $\ell$th element equal to $1$. 
An important observation is that the prior model for the $\mathcal{L}_i$ in (\ref{eq:hiergenmod}) 
can be equivalently defined through a continuation-ratio logits regression model for their 
multivariate images $\boldsymbol{\mathcal{L}}_i$. More specifically, 
\begin{equation*}
\boldsymbol{\mathcal{L}}_{i} \mid \{ \boldsymbol{\gamma}_{\ell} \} \stackrel{ind.}{\sim} 
Bin(\mathcal{L}_{i1} \mid 1,\eta_{1}(\mathbf{x}_i)) \,
Bin(\mathcal{L}_{i2} \mid 1 - \mathcal{L}_{i1},\eta_{2}(\mathbf{x}_i)) \ldots 
Bin\left( \mathcal{L}_{i,L-1}\mid 1-\sum_{k=1}^{L-2}\mathcal{L}_{i k},\eta_{L-1}(\mathbf{x}_i) \right)
\end{equation*}
where $\eta_{\ell}(\mathbf{x}_i)=$ $\varphi(\mathbf{x}_i^T \boldsymbol{\gamma}_{\ell})$,
for $\ell=1,\ldots,L-1$.

The form of the hierarchical model for the data, along with the observation above, 
elucidate the key model property discussed in Section \ref{subsec:standardmodel}. 
Under the (truncated) LSBP prior for the covariate-dependent 
weights, we achieve effectively the same structure for the weights and atoms of the general 
mixture model. In turn, this allows us to use the Pólya-Gamma data augmentation approach
\citep{Polson2013} to update both the atoms parameters as well as the ones for the weights. 
In particular, for each response $\mathbf{Y}_i$, we introduce two sets of Pólya-Gamma latent 
variables, such that conditionally conjugate updates emerge for the parameters defining 
both the weights and the atoms. Therefore, all model parameters can be updated via Gibbs 
sampling. Moreover, taking advantage of the continuation-ratio logits model structure for 
the mixture kernel, parallel computing for the different mixing components can be adopted, 
facilitating implementation in applications where the number of response categories is 
moderate to large. Details of the posterior simulation method are presented in the 
Supplementary Material.

Using the posterior samples for model parameters, we can obtain full inference for any 
regression functional of interest. 
%
The MCMC posterior samples can also be used to estimate the posterior predictive 
distribution for new response $\mathbf{Y}_*$ given new covariate vector $\mathbf{x}_*$.
Using superscript $(t)$ to indicate the $t$th posterior sample for the model 
parameters, the $t$th posterior predictive sample is obtained by first sampling 
the corresponding configuration variable $\mathcal{L}^{(t)}_*$, such that 
$\mathcal{L}^{(t)}_* = \ell$ with probability 
$\varphi(\mathbf{x}_{*}^T \boldsymbol{\gamma}_{\ell}^{(t)}) 
\prod_{h=1}^{\ell-1} (1-\varphi(\mathbf{x}_{*}^T\boldsymbol{\gamma}_{h}^{(t)}))$, 
for $\ell=1,\ldots,L-1$ (and $\mathcal{L}^{(t)}_* = L$ with the remaining probability), 
and then sampling $\mathbf{Y}^{(t)}_*$ from $K(\cdot \mid \boldsymbol{\theta}_*^{(t)})$, 
with the $j$th element of $\boldsymbol{\theta}_*^{(t)}$ given by
$\varphi(\mathbf{x}^{T}_*\boldsymbol{\beta}_{j\mathcal{L}^{(t)}_*}^{(t)})$, 
for $j=1,\ldots,C-1$.

\subsection{Assessing model flexibility}
\label{subsec:equreps}

In Section \ref{subsubsec:klsupport}, we study the Kullback-Leibler (KL) support of the proposed 
prior model, using results from \cite{Barrientos2012} for nonparametric mixtures for continuous 
responses. This study yields two results of independent interest: a formulation of the ordinal LSBP 
mixture model in terms of latent continuous responses (Section \ref{latent_variable_representation});
and, a connection between the KL support of a prior for continuous responses and the induced prior 
for categorical outcomes arising from discretizing the continuous responses. Moreover, in 
Section \ref{subsubsec:equivrep}, we contrast continuation-ratio logit and cumulative logit models. 
The purpose is to provide further motivation for the kernel choice of the LSBP mixture 
model. The proofs for all theoretical results are given in the Supplementary Material.

\subsubsection{Latent variable representation}
\label{latent_variable_representation}

Recall that the continuation-ratio logits structure implies a sequential mechanism involving 
binary steps to determine which of its $C$ levels the ordinal response admits. The mechanism 
can also be represented through latent continuous variables, 
$\boldsymbol{\mathcal{Z}} =$ $(\mathcal{Z}_{1},\ldots,\mathcal{Z}_{C-1})$, and a sequential, 
binary partition of $\mathbb{R}^{C-1}$, comprising sets
\begin{equation}
\mathcal{R}_1 = \mathbb{R}^{+} \times \mathbb{R}^{C-2};  \,\,\,\,
\mathcal{R}_j = (\mathbb{R}^{-})^{j-1} \times \mathbb{R}^{+} \times \mathbb{R}^{C-j-1}, \,\,
j=2,\ldots,C-1; \,\,\,\,
\mathcal{R}_C = (\mathbb{R}^{-})^{C-1}.
\label{eq:partition}
\end{equation}
%
%
Hence, $\boldsymbol{\mathcal{Z}}\in\mathcal{R}_j$ if its first $j-1$ elements take 
negative values, and the $j$ component is positive valued. Referring to the description of 
the continuation-ratio logits structure from Section \ref{subsec:standardmodel}, variable 
$\mathcal{Z}_j$ plays a similar role to Bernoulli variable $\mathcal{H}_{j}$, where now it is 
the sign of $\mathcal{Z}_j$ that specifies the ordinal response category, such that 
$\mathbf{Y}=j$ if-f $\mathcal{Z}_j > 0$, given that $\mathcal{Z}_k \leq 0$, for $k=1,\ldots,j-1$.
As stated in the following proposition, the multinomial model in (\ref{eq:paramcontilogits}) 
emerges for independent logistic variables $\mathcal{Z}_j$.

\begin{proposition}
\label{prop:seqbreaklogit}
Consider ordinal response $\mathbf{Y}=$ $(Y_{1},\ldots,Y_{C})$, where $Y_{j} \in \{ 0,1 \}$ 
with $\sum_{j=1}^{C} Y_{j} = 1$, and continuous random vector 
$\boldsymbol{\mathcal{Z}} =$ $(\mathcal{Z}_{1},\ldots,\mathcal{Z}_{C-1}) \in \mathbb{R}^{C-1}$. 
Assume that: $\mathbf{Y} \mid \boldsymbol{\mathcal{Z}}\sim \boldsymbol{1}(\mathbf{Y}=j 
\Longleftrightarrow \boldsymbol{\mathcal{Z}}\in\mathcal{R}_j)$, for $j=1,\ldots,C$, with the 
$\mathcal{R}_j$ defined in (\ref{eq:partition}); and, 
$\boldsymbol{\mathcal{Z}}\mid \boldsymbol{\theta} \sim$ 
$\prod_{j=1}^{C-1} \mathfrak{L}(\mathcal{Z}_j \mid \theta_j)$, where 
$\boldsymbol{\theta} =$ $(\theta_{1},\ldots,\theta_{C-1})$, and $\mathfrak{L}(\cdot\mid\theta)$ 
denotes the logistic distribution with mean $\theta$ and scale parameter $1$. 
Then, marginalizing over $\boldsymbol{\mathcal{Z}}$, $\mathbf{Y} \mid \boldsymbol{\theta}$
follows the multinomial distribution with the continuation-ratio logits parameterization 
in (\ref{eq:paramcontilogits}).
\end{proposition}

Proposition \ref{prop:seqbreaklogit} formalizes the latent variable representation 
discussed in \cite{Tutz1991}, in particular, it provides the explicit connection between the 
values of $\mathbf{Y}$ and $\boldsymbol{\mathcal{Z}}$, and the complete distributional 
assumptions (including independence) for $\boldsymbol{\mathcal{Z}}$. 
This result further highlights the benefits of the binary choice, 
sequential structure. Because the order of the response variable is preserved in the sequential
mechanism, order restrictions for the latent $\mathcal{Z}_j$ are not required. This is in 
contrast with cumulative link models where the cut-off variables that discretize the single latent 
continuous response must be ordered. 
The proposition also suggests a direction for constructing more flexible models by relaxing the 
parametric assumption for the distribution of the $\mathcal{Z}_j$. Indeed, the next result 
shows that we can recover the ordinal regression model of Section \ref{subsec:standardmodel}
through a LSBP mixture model for $\boldsymbol{\mathcal{Z}}$ with a product logistic 
mixture kernel.

\begin{proposition}
\label{prop:lsbpeql}
Consider ordinal response $\mathbf{Y}=$ $(Y_{1},\ldots,Y_{C})$, where $Y_{j} \in \{ 0,1 \}$ 
with $\sum_{j=1}^{C} Y_{j} = 1$, and continuous random vector 
$\boldsymbol{\mathcal{Z}} =$ $(\mathcal{Z}_{1},\ldots,\mathcal{Z}_{C-1}) \in \mathbb{R}^{C-1}$. 
Assume that: $\mathbf{Y} \mid \boldsymbol{\mathcal{Z}} \sim \boldsymbol{1}(\mathbf{Y}=j 
\Longleftrightarrow \boldsymbol{\mathcal{Z}}\in\mathcal{R}_j)$, for $j=1,\ldots,C$, 
with the $\mathcal{R}_j$ defined in (\ref{eq:partition}); and, 
$\boldsymbol{\mathcal{Z}} \mid G_{\mathbf{x}} \sim$ 
$\sum_{\ell=1}^{\infty} \omega_{\ell}(\mathbf{x}) \left\{
\prod_{j=1}^{C-1} \mathfrak{L}(\mathcal{Z}_j\mid\theta_{j\ell}(\mathbf{x})) \right\}$, 
where the $\omega_{\ell}(\mathbf{x})$ and $\theta_{j\ell}(\mathbf{x})$ are defined 
in (\ref{eq:generalmodori}) and (\ref{general_model_atoms}), respectively. 
Then, marginalizing over $\boldsymbol{\mathcal{Z}}$, $\mathbf{Y} \mid G_{\mathbf{x}}$ follows 
the multinomial LSBP mixture model in (\ref{eq:npbext}). 
\end{proposition}

We note that Proposition \ref{prop:lsbpeql} does not simplify posterior simulation; Gibbs sampling 
for the model augmented with latent $\boldsymbol{\mathcal{Z}}_{i}$ for each observed response
$\mathbf{Y}_{i}$ would require imputing the $\boldsymbol{\mathcal{Z}}_{i}$, and it would still 
involve two sets of Pólya-Gamma latent variables. However, the latent variable formulation offers 
an alternative perspective to model structure, as well as a useful tool to study model properties, 
such as KL support discussed in the next section.

\subsubsection{Kullback-Leibler support of the LSBP mixture model}
\label{subsubsec:klsupport}

Consider a prior $\mathcal{F}$ on a space of densities $\mathfrak{F}$. Density 
$f^0\in\mathfrak{F}$ is in the KL 
support of $\mathcal{F}$ if $\mathcal{F}( N_{\epsilon}(f^0) ) > 0$, for any $\epsilon>0$, 
where $N_{\epsilon}(f^0) =$ $\{ f: \int f^0(\mathbf{z})\log(f^0(\mathbf{z})/
f(\mathbf{z})) \, \text{d}\mathbf{z} < \epsilon\}$ is the (size $\epsilon$) KL neighborhood of $f^0$. 
Keeping the focus on continuous distributions, the regression setting targets collections of 
densities $\{f^0_{\mathbf{x}}: \mathbf{x} \in \mathcal{X}\}$, indexed by values in the covariate 
space $\mathcal{X}$. We defer technical details to the Supplementary Material, but note that the 
extension of the KL support definition considers the KL divergence (in the standard definition above) 
at arbitrary, finite sets of values in $\mathcal{X}$ \citep[e.g.,][]{Barrientos2012}.

Theorem \ref{thm:KLgeneralmodel} establishes the KL support of the multinomial LSBP mixture prior 
model defined in (\ref{eq:npbext}), (\ref{eq:generalmodori}) and (\ref{general_model_atoms}).
The theorem builds from results in \cite{Barrientos2012} who examined the KL support of 
stick-breaking process mixture models for covariate-dependent densities. It can be shown that 
the LSBP mixture with the product logistic kernel, given in Proposition \ref{prop:lsbpeql} for 
the continuous random vector $\boldsymbol{\mathcal{Z}}$, satisfies the various conditions required 
for the KL results in \cite{Barrientos2012}. Thus, the latent variable representation of the 
ordinal regression LSBP mixture model yields the key step towards establishing its KL support. 
The other step is provided by Lemma \ref{lem:klsupport} which connects the KL support of priors
for continuous distributions with the KL support of priors for distributions of discrete variables 
induced by discretizing the corresponding continuous variables.

For our purposes, the connection is achieved by starting with a generic prior $\mathcal{F}_{\mathbf{x}}$ 
for the covariate-dependent distribution of continuous random vector $\mathbf{Z} \in \mathbb{R}^{C-1}$. 
Then, prior $\mathcal{P}_{\mathbf{x}}$ for the distribution of ordinal response $\mathbf{Y}=$ 
$(Y_{1},\ldots,Y_{C})$ is induced through 
$\boldsymbol{1}(\mathbf{Y}=j \Longleftrightarrow \mathbf{Z} \in \mathcal{R}_j)$, 
for $j=1,\ldots,C$, with the $\mathcal{R}_j$ defined in (\ref{eq:partition}). Hence, 
prior $\mathcal{F}_{\mathbf{x}}$ for densities $f_{\mathbf{x}}$ gives rise to prior 
$\mathcal{P}_{\mathbf{x}}$ for ordinal probabilities $p_{\mathbf{x}}$ via the mapping
\begin{equation}
f_{\mathbf{x}} \mapsto p_{\mathbf{x}}(y) =
\int_{\mathcal{R}_y} f_{\mathbf{x}}(\mathbf{z}) \, \text{d}\mathbf{z}, \,\,\,\,\,
\text{for} \,\,\, y=1,\ldots,C. 
\label{map_KL_support}
\end{equation}
The following lemma relates the KL support of priors $\mathcal{F}_{\mathbf{x}}$ and
$\mathcal{P}_{\mathbf{x}}$.

\begin{lemma}
\label{lem:klsupport} 
Consider prior $\mathcal{F}_{\mathbf{x}}$ for densities $f_{\mathbf{x}}$, and the prior 
$\mathcal{P}_{\mathbf{x}}$ for ordinal probabilities $p_{\mathbf{x}}$ induced from 
(\ref{map_KL_support}). Assume that densities $\{f^0_{\mathbf{x}}:\mathbf{x}\in\mathcal{X}\}$ 
are in the KL support of $\mathcal{F}_{\mathbf{x}}$, and consider probability mass functions  
$\{p^0_{\mathbf{x}}:\mathbf{x}\in\mathcal{X}\}$, where $p^0_{\mathbf{x}}$ is defined from 
$f^0_{\mathbf{x}}$ according to (\ref{map_KL_support}). Then, the probability mass functions
$\{p^0_{\mathbf{x}}:\mathbf{x}\in\mathcal{X}\}$ are in the KL support of $\mathcal{P}_{\mathbf{x}}$. 
\end{lemma}

Key to Lemma \ref{lem:klsupport} is an inequality that allows us to bound the 
sum in the KL divergence for probability mass functions by the integral in the KL divergence 
for densities, when the densities and mass functions are related as in (\ref{map_KL_support}).
The result is not restricted to the specific partition in (\ref{eq:partition}), and it thus 
offers general scope to study the KL support of priors for categorical distributions arising 
through discretization of latent continuous responses.

Finally, combining Lemma \ref{lem:klsupport}, Proposition \ref{prop:lsbpeql}, and results from 
\cite{Barrientos2012}, we can derive the KL property for our model.

\begin{thm}
\label{thm:KLgeneralmodel}
Denote by $\mathcal{P}_{\mathbf{x}}$ the LSBP mixture prior defined in 
(\ref{eq:npbext}), (\ref{eq:generalmodori}) and (\ref{general_model_atoms}), and consider 
$\{p^0_{\mathbf{x}}:\mathbf{x}\in\mathcal{X}\}$, a generic collection of covariate-dependent 
probabilities for an ordinal response with $C$ categories. Assume that the probability of each 
response category is strictly positive. Then, the mass functions
$\{p^0_{\mathbf{x}}:\mathbf{x}\in\mathcal{X}\}$ are in the KL support of 
$\mathcal{P}_{\mathbf{x}}$. 
\end{thm}

Full KL support is a key theoretical property of the prior model. For priors on spaces of 
continuous densities, it is typically the case that various regularity conditions are required
for a generic density to be in the KL support of the prior. In our context, the underlying 
regularity conditions in the results from \cite{Barrientos2012} reduce to the condition that 
response probabilities are strictly positive. Finally, as discussed in the Supplementary Material, 
KL support results can also be obtained for the simplified models of Section \ref{sec:specorimodel}.

\subsubsection{Continuation-ratio logits vs cumulative logits}
\label{subsubsec:equivrep}

As discussed in the Introduction, cumulative link models provide a common approach to ordinal 
regression, with the inverse link typically specified through the distribution function of a 
continuous variable $Z$. Then, the ordinal response $Y$ values can be developed through 
discretization of the latent continuous response $Z$, in particular, $Y=j$ if and only if 
$Z\in(\varkappa_{j-1},\varkappa_j]$, for $j=1,\ldots,C$. Here 
$-\infty = \varkappa_0 < \varkappa_1< \ldots <\varkappa_{C-1} <\varkappa_C = \infty$ are 
cut-off points, where, typically, $\varkappa_1=0$ for identifiability. Key examples are cumulative 
probit and cumulative logit models for which the continuous distribution is $N(Z \mid \vartheta,1)$ 
and $\mathfrak{L}(Z \mid \vartheta)$, respectively. In the absence of covariates, the parameters of 
cumulative logit and continuation-ratio logits models can be related as shown in the following result.

\begin{proposition}
\label{prop:equdiscont}
Consider the two distinct model formulations for an ordinal response with $C$ categories given by 
the cumulative logit model with parameters $(\vartheta,\varkappa_2,...,\varkappa_{C-1})$, and the 
continuation-ratio logits model in (\ref{eq:paramcontilogits}) with parameters 
$(\theta_1,...,\theta_{C-1})$. Then, the parameters of the two models are connected through 
$\vartheta=-\theta_1$ and the recursive expression $\varkappa_j=$ 
$\log(e^{\varkappa_{j-1}}+e^{\varkappa_{j-1}+\theta_j}+e^{\theta_j-\theta_1})$, for $j=2,\ldots,C-1$.
\end{proposition}

Despite the one-to-one correspondence between the parameters of the two models, there 
is a key difference in regression modeling. Under continuation-ratio logits,
covariate effects are modeled through $\theta_j=$ $\mathbf{x}^T\boldsymbol{\beta}_j$, 
for $j=1,...,C-1$. Here, the order for the response outcomes is induced by the binary, 
sequential mechanism, and thus the regression model specification is not constrained by 
restrictions on its parameters. In contrast, under cumulative link models, the order for the 
response values requires ordered cut-off points, which makes it challenging to model them as 
functions of the covariates. Indeed, cumulative link models typically incorporate covariates 
only through the location parameter of the latent continuous response, e.g, the proportional odds 
regression model arises under the $\mathfrak{L}(Z \mid \mathbf{x}^T\boldsymbol{\beta})$ distribution.

On the other hand, sequential models, such as continuation-ratio logits models, are not invariant 
under reversal of the order of the response categories, while cumulative link models are 
\citep{Peyhardi2015}. Nonetheless, for several applications, the order of the ordinal response 
categories is determined by the context of the problem and/or the relevant scientific questions. 
We provide an example in Section \ref{subsec:retinopathy}, where the order of the response 
(disease severeness) is encoded from the mildest to the most severe, because of primary interest 
is study of covariate effects on the progress from mild to severe levels. 

We note that existing nonparametric Bayesian methods for ordinal regression (reviewed in the 
Introduction) build from cumulative link models. In particular, the fully nonparametric models
in \cite{BaoHanson} and \cite{DeYoreoKottas2018} can be expressed as mixtures of cumulative 
probit regressions, the former with mixture weights that do not depend on the covariates, the 
latter with covariate-dependent mixture weights. The earlier discussion suggests that the 
continuation-ratio logits formulation offers wider scope as a building block in nonparametric 
mixture modeling for ordinal regression.

\section{Specific models for ordinal regression}
\label{sec:specorimodel}

%
%

Here, we study simplified model versions, which are naturally suggested given the two 
building blocks of the general model. In particular, we discuss ordinal regression models 
that arise by retaining covariate dependence only in the atoms 
(Section \ref{subsec:lddpmodel}) or only in the weights 
(Section \ref{subsec:comatomlsbpmodel}). The different model versions are empirically 
compared in Section \ref{sec:oriexample}. 
%

\subsection{The common-weights model}
\label{subsec:lddpmodel}

As a first simplification, we can remove the covariate dependence from the mixture weights.
That is, the ordinal regression mixture model is built from the common-weights mixing 
distribution $G_{\mathbf{x}}=$ 
$\sum_{\ell=1}^{\infty} \omega_{\ell} \, \delta_{\boldsymbol{\theta}_{\ell}(\mathbf{x})}$, 
such that 
$\mathbf{Y} \mid G_{\mathbf{x}} \sim$ $\sum_{\ell=1}^{\infty} \omega_{\ell} \, 
K(\mathbf{Y} \mid \boldsymbol{\theta}_{\ell}(\mathbf{x}))$, where the covariate-dependent 
atoms are defined as in the general model in (\ref{general_model_atoms}) 
and (\ref{general_model_hyperparameters}).

Regarding the prior model for the weights, one option would be to keep the LSBP structure,
that is, reduce $\mathbf{x}^T \boldsymbol{\gamma}_{\ell}$ in (\ref{eq:generalmodori}) to 
scalar parameter $\gamma_{\ell}$, with the $\gamma_{\ell}$ independent and identically 
normally distributed. We work instead with 
the DP prior for the weights: $\omega_{1} = V_{1}$, and $\omega_{\ell} =$ 
$V_{\ell} \prod_{h=1}^{\ell-1} (1 - V_{h})$, for $\ell \geq 2$, where 
$V_{\ell} \mid \alpha \stackrel{i.i.d.}{\sim} \, \text{Beta}(1,\alpha)$. 

%
%

Using the DP-induced prior for the weights allows connections with the well-established
literature on DDP mixtures, including the early work with common-weights DDP priors, e.g.,
the ANOVA DDP \citep{DeIorio2004} and the spatial DP \citep{GelfandKottas2005}.
In particular, the common-weights model can be equivalently written as a DP mixture model: 
\begin{equation*}
\mathbf{Y} \mid F \, \sim \,
\int K(\mathbf{Y} \mid  \mathbf{x}^{T} \boldsymbol{\beta}_{1},\ldots,
\mathbf{x}^{T} \boldsymbol{\beta}_{C-1} ) \, 
dF(\boldsymbol{\beta}_{1},\ldots,\boldsymbol{\beta}_{C-1})
\end{equation*}
where $F$ follows a DP prior with total mass parameter $\alpha$, and centering 
distribution defined through 
$\boldsymbol{\beta}_j \mid (\boldsymbol{\mu}_j,\Sigma_j) 
\stackrel{ind.}{\sim} N(\boldsymbol{\mu}_j,\Sigma_j)$, 
for $j=1,\ldots,C-1$. The model is completed with a $\text{Gamma}(a_{\alpha},b_{\alpha})$ 
hyperprior for $\alpha$, and the prior for the $(\boldsymbol{\mu}_j,\Sigma_j)$ in (\ref{general_model_hyperparameters}). For prior specification, we combine the approach 
for the atoms in the general model with techniques for specifying the prior for the total 
mass DP parameter. The posterior simulation method replaces the steps for updating the 
weights with the update for the DP weights under blocked Gibbs sampling. The details 
can be found in the Supplementary Material.

%
%

With the expression for the weights appropriately adjusted, the common-weights model 
inherits the properties of the general model, discussed in Section \ref{subsec:modpropori}.
The prior expectation in (\ref{eq:expprobrespcurvegen}) is not affected by the form 
of the weights. However, the probability response curves admit a potentially less flexible 
form than the one in (\ref{eq:margcurvegen}) under the general model. We still have a 
weighted combination of parametric regression functions, but now without the local adjustment 
afforded by covariate-dependent weights. The data analyses in Section \ref{sec:oriexample} 
demonstrate the practical utility of the general model, but also include examples 
where the common-weights model yields practical, sufficiently flexible inference.


\subsection{The common-atoms model}
\label{subsec:comatomlsbpmodel}

The alternative way to simplify the general model is to use mixing 
distribution $G_{\mathbf{x}}=$ $\sum_{\ell=1}^{\infty} \omega_{\ell}(\mathbf{x}) \, \delta_{\boldsymbol{\theta}_{\ell}}$, resulting in the common-atoms mixture model:
\begin{equation*}
\mathbf{Y} \mid G_{\mathbf{x}} \, \sim \, 
\sum_{\ell=1}^{\infty} \omega_{\ell}(\mathbf{x}) \, 
K(\mathbf{Y} \mid \boldsymbol{\theta}_{\ell})
\end{equation*}
where $\boldsymbol{\theta}_{\ell} =$ $(\theta_{1 \ell},\ldots,\theta_{C-1,\ell})$. 
The covariate-dependent weights are defined using the LSBP prior in
(\ref{eq:generalmodori}). The prior model for the atoms is built from 
$\theta_{j \ell} \mid \mu_{j},\sigma_{j}^{2} \stackrel{ind.}{\sim} N(\mu_j,\sigma_j^2)$, 
for $j=1,\ldots,C-1$, and $\ell \geq 1$. The model is completed with the conjugate prior 
for the hyperparameters: $\sigma_j^2 \stackrel{ind.}{\sim} IG(a_{0j},b_{0j})$, 
with $a_{0j} > 1$, and
$\mu_j \mid \sigma_j^2 \stackrel{ind.}{\sim} N(\mu_{0j},\sigma_j^2/\nu_{0j})$, for 
$j=1,\ldots\,C-1$, where $IG(\cdot,\cdot)$ denotes the inverse-gamma distribution. 

Model implementation builds from the general model, with appropriate adjustments 
for the atoms. Here, $\text{E}(\text{Pr}(\mathbf{Y}=j \mid G_{\mathbf{x}}))=$ 
$\text{E} \left\{ \varphi(\theta_j) \prod_{k=1}^{j-1}(1-\varphi(\theta_k)) \right\}$, for 
$j=1,\ldots,C$, where the expectation is taken with respect to 
$\theta_j \stackrel{ind.}{\sim} N(\mu_{0j},(\nu_{0j}+1)b_{0j}/\nu_{0j}(a_{0j}-1))$ 
(obtained by marginalizing over the prior for $(\mu_j,\sigma_j^2)$). Hence, the prior 
expected marginal probability response curves are constants over the covariate space. 
The prior specification strategy utilizes this property, by setting 
$\{\mu_{0j},\nu_{0j},a_{0j},b_{0j}\}_{j=1}^{C-1}$ such that these constants correspond
to prior information for the ordinal response probabilities. The key quantity is again 
the expectation of a logit-normal distributed random variable (discussed earlier in 
Section \ref{subsec:prispecori}). The posterior sampling scheme is adapted from the 
general model, with the normal-inverse-Wishart update for the atoms parameters
replaced by the univariate normal-inverse-Gamma analogue. Details are given in 
the Supplementary Material.

The common-atoms mixture structure offers a parsimonious model formulation, especially for 
problems with a moderate to large number of response categories. On the other hand, the 
simplified model form involves a potential limitation. The marginal and conditional probability 
response curves have the form in (\ref{eq:margcurvegen}) and (\ref{eq:condcurvegen}), 
respectively, with $\theta_{j\ell}(\mathbf{x})$ replaced by $\theta_{j\ell}$. Hence, the 
covariates inform the shape of the regression curves only through the mixture weights. 
As a practical consequence, the common-atoms model typically activates a larger number of 
effective mixture components to estimate the regression relationship, and it thus encounters 
a higher risk of overfitting for problems with a moderate to large number of 
covariates. This point is illustrated with the data examples of Section \ref{sec:oriexample}.

\section{Data illustrations}
\label{sec:oriexample}

\subsection{Synthetic data examples}
\label{subsec:simstudy}

We consider three simulation examples to demonstrate the modeling framework, including comparative 
study of the common-weights, common-atoms, and general models. 
The first example is designed to highlight the benefits of local, covariate-dependent 
weights in capturing non-standard shapes of probability response curves.
The objective of the second example is to study how the different models handle the challenge 
of recovering standard regression relationships for which the nonparametric mixture model structure 
is not necessary. 
The third example compares the effectiveness of the different models in 
capturing non-linear and non-additive covariate effects, including comparison with the 
density regression model from \citet{DeYoreoKottas2018}. Here, we introduce the first simulation 
example, while the other two examples are presented in the Supplementary Material, where we also 
report on MCMC diagnostics and computing times.

\begin{figure}[t!]
\centering
\begin{subfigure}{\textwidth}
  \centering 
  \includegraphics[width=15.4cm,height=2.85cm]{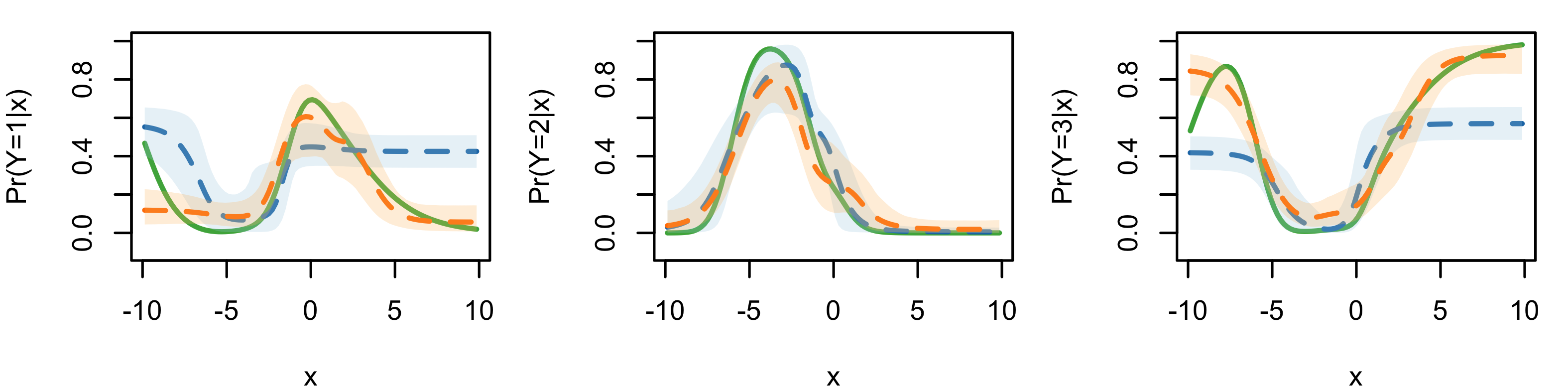}
  \caption{Common-weights and common-atoms models ($n=200$).}
  \label{subfig:sim2post200lsbp}
\end{subfigure}
\begin{subfigure}{\textwidth}
  \centering 
  \includegraphics[width=15.4cm,height=2.85cm]{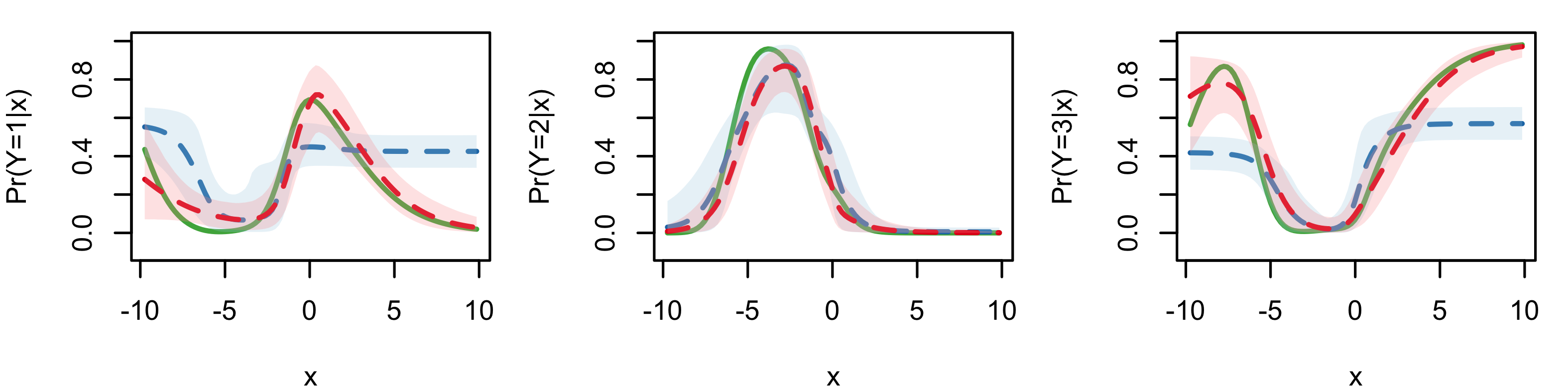}  
  \caption{Common-weights and general models ($n=200$).}
  \label{subfig:sim2post200general}
\end{subfigure}
\begin{subfigure}{\textwidth}
  \centering
  \includegraphics[width=15.4cm,height=2.85cm]{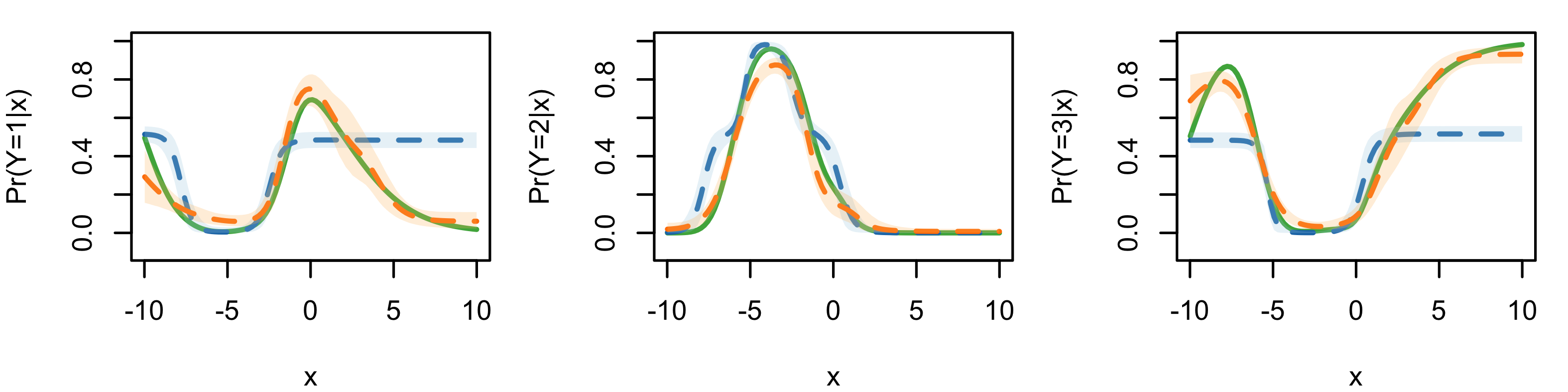} 
  \caption{Common-weights and common-atoms models ($n=800$).}
  \label{subfig:sim2post800lsbp}
\end{subfigure}
\begin{subfigure}{\textwidth}
  \centering
  \includegraphics[width=15.4cm,height=2.85cm]{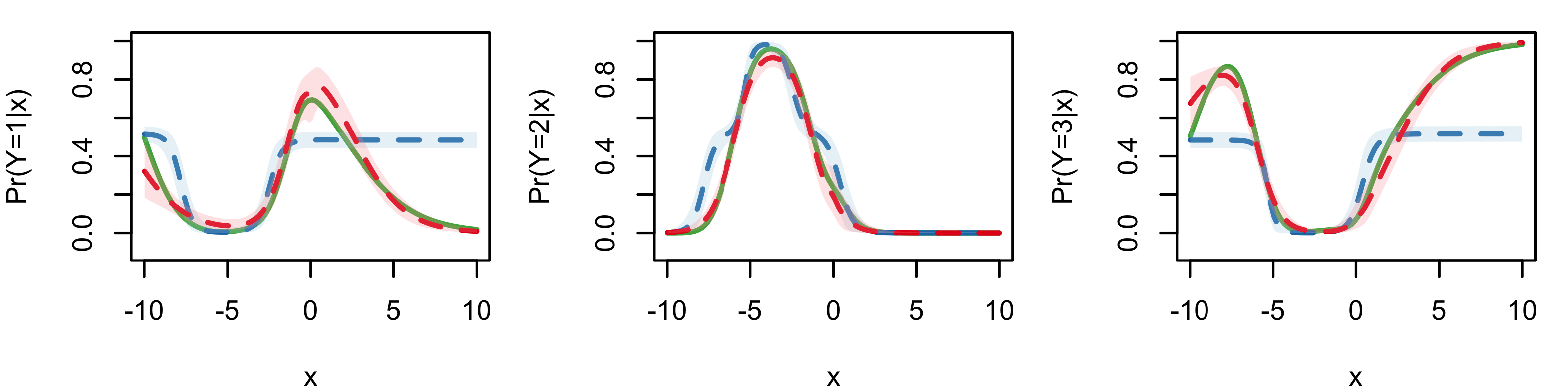}   
  \caption{Common-weights and general models ($n=800$).}
  \label{subfig:sim2post800general}
\end{subfigure}
\caption{
{\small Synthetic data example. Posterior mean and $95\%$ credible interval estimates 
for the marginal probability response curves under the common-weights (blue line and shaded 
region), common-atoms (orange line and shaded region), and general (red line and shaded 
region) models. In each panel, the green solid line is the true regression function.}
}
\label{fig:sim2postgeneral}
\end{figure}

For the first experiment, to facilitate graphical illustrations, we consider an ordinal 
response with $C=3$ categories, and one (continuous) covariate, where 
$x_i\stackrel{i.i.d.}{\sim} U(-10,10)$, such that with the intercept, the covariate 
vector is $\mathbf{x}_i=(1,x_i)^T$. 
%
We generate the responses from a three component mixture of multinomial distributions, expressed 
in their continuation-ratio logits form. That is, $\mathbf{Y} \sim$
$\sum_{k=1}^3 w_k(\mathbf{x}) \, K(\mathbf{Y} \mid \boldsymbol{\theta}_k(\mathbf{x}))$, 
where $\theta_{jk}(\mathbf{x})=$ $b_{jk0}+b_{jk1}x$, for $j=1,2$ and $k=1,2,3$. 
The covariate dependence is introduced in the weights by computing $p_{j\mathbf{x}}=$
$\Phi(a_{j0}+a_{j1}x)$, for $j=1,2$, where $\Phi$ is the N$(0,1)$ distribution 
function, and setting $(w_1(\mathbf{x}),w_2(\mathbf{x}),w_3(\mathbf{x})) =$
$(p_{1\mathbf{x}},(1-p_{1\mathbf{x}})p_{2\mathbf{x}},(1-p_{1\mathbf{x}})(1-p_{2\mathbf{x}}))$. 
The weights and atoms parameters are chosen such that the probability response 
curves have non-standard shapes (see Figure \ref{fig:sim2postgeneral}). We consider
two sample sizes, $n=200$ and $n=800$.

The prior hyperparameters for the atoms are set according to the baseline choice. 
For the common-atoms and general models, we specify the LSBP prior hyperparameters 
$(\boldsymbol{\gamma}_0,\Gamma_0)$ to favor a priori more mixture components in the interval 
of covariate values $(-10,0)$ where there is more variation in the regression functions. 
We note however that the prior specification is still fairly non-informative regarding the 
shape of the regression functions. In particular, under all three models, the prior mean 
estimates for the probability response curves are flat, and the prior $95\%$ interval 
estimates span a substantial portion of the unit interval (the prior estimates are
shown in the Supplementary Material). 

\begin{figure}[t!]
\centering
\centering
\begin{subfigure}{\textwidth}
  \centering 
  \includegraphics[width=15cm,height=4.5cm]{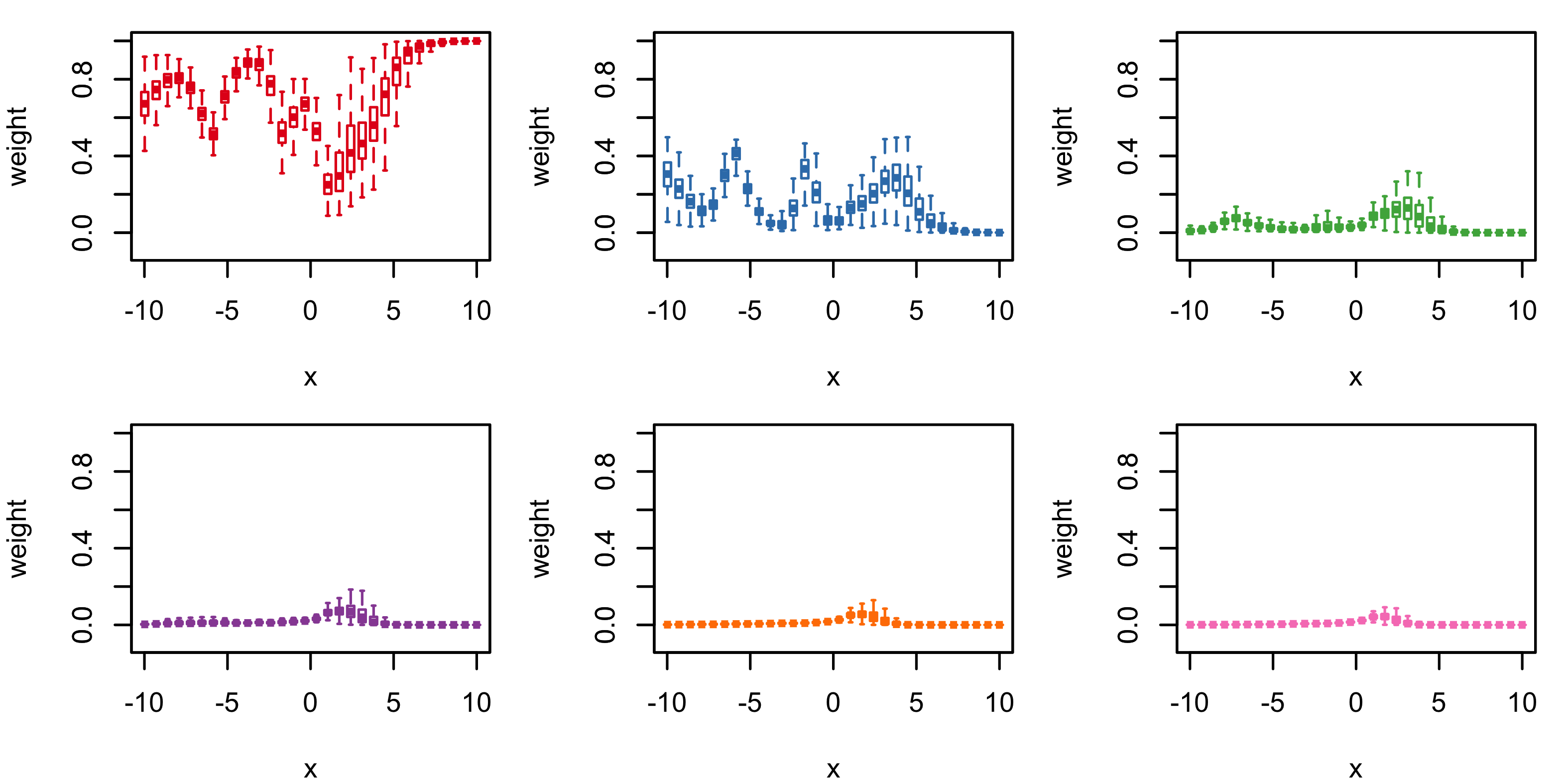}
  \caption{Common-atoms model.}
  \label{subfig:sim2lsbpweightbox}
\end{subfigure}
\begin{subfigure}{\textwidth}
  \centering
  \includegraphics[width=15cm,height=4.5cm]{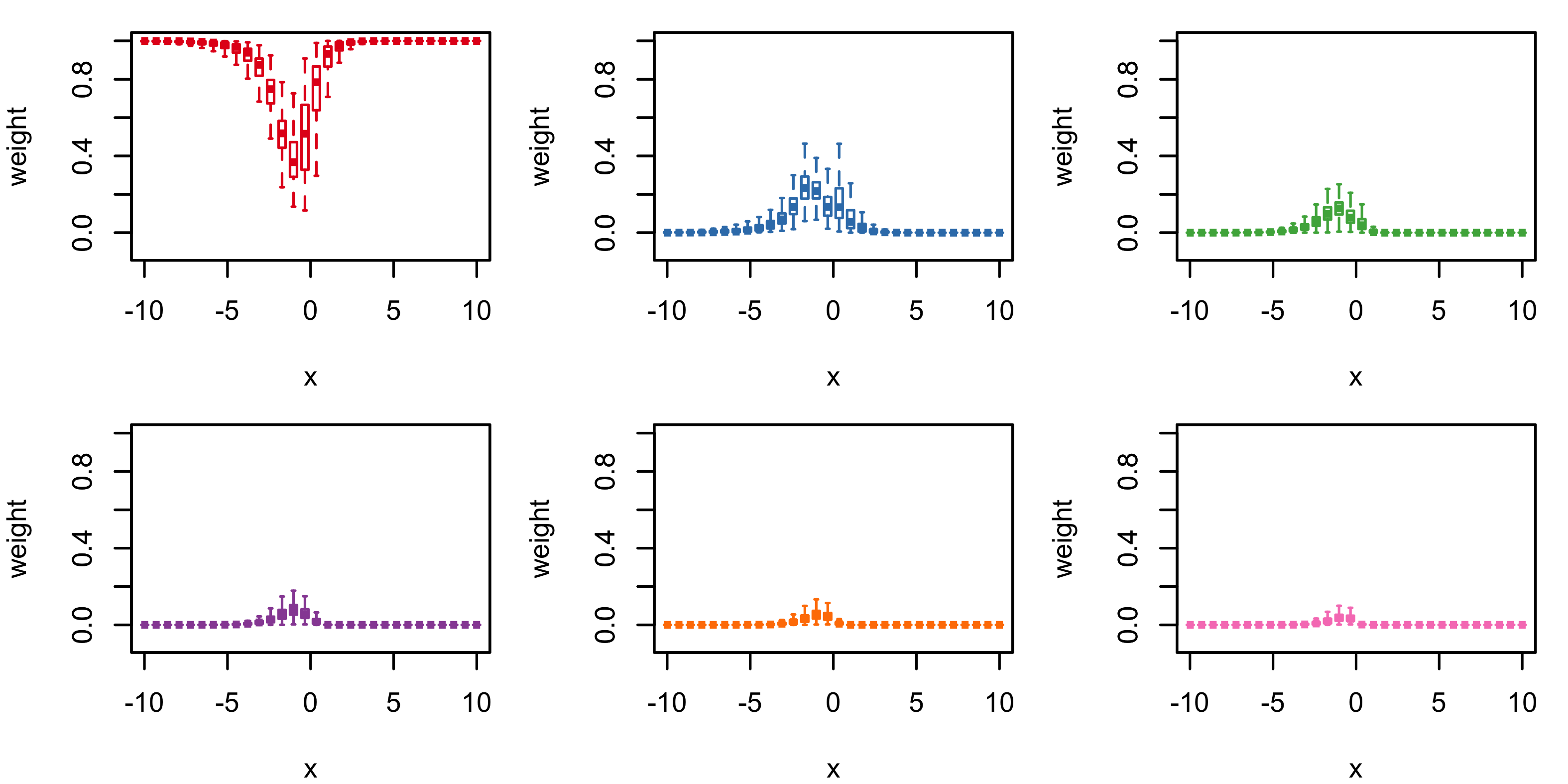} 
  \caption{General model.}
  \label{subfig:sim2generalweightbox}
\end{subfigure}
\caption{
{\small Synthetic data example ($n=800$). Box plots of the posterior samples 
for the six largest mixture weights, under the common-atoms and general models.}
}
\label{fig:weightsboxsim2}
\end{figure}

%
%

Inference results under the general and common-atoms models are contrasted with the 
common-weights model in Figure \ref{fig:sim2postgeneral}. As expected, the common-weights
model does not recover well the non-standard regression functions for the first and third 
response categories. The two models that use covariate-dependent mixture weights perform 
notably better, with the general model resulting overall in more accurate estimation. 
As shown in the Supplementary Material, this ranking in model performance is supported 
by the posterior predictive loss criterion from \cite{GelfandGhosh1998}. Increasing the sample 
size results in more precise point estimates and more narrow posterior uncertainty bands. 

Focusing on the models with covariate-dependent mixture weights (and the data set 
with $n=800$), Figure \ref{fig:weightsboxsim2} explores the posterior distribution 
of the six largest weights over the covariate space. 
For both models, it is essentially the first three largest weights that, given the data, 
define the probability vector of weights. However, we note the more local adjustment 
in the two largest weights under the common-atoms model, which becomes more pronounced 
in parts of the covariate space where the probability curves change more drastically. 
This is compatible with the common-atoms model's structure that seeks to fit the regression 
functions with atoms that do not change across the covariate space.  

\subsection{Credit ratings of U.S. firms}

We consider data on Standard and Poor's (S$\&$P) credit ratings for 921 U.S. firms in 
2005 \citep{Verbeek2008}. 
The ordinal response is the firm's credit rating, originally recorded on a scale with 
seven categories. Since there were only 17 firms with rating of 1 or 7, and to facilitate
illustration of inference results, we combine the responses in the first two and last two 
categories. We thus obtain an ordinal response scale ranging from 1 to 5, with higher 
ratings indicating higher creditworthiness. The data set includes five company characteristics 
that serve as covariates: book leverage (ratio of debt to assets), $x_1$; earnings before 
interest and taxes divided by total assets, $x_2$; standardized log-sales (proxy for firm size), 
$x_3$; retained earnings divided by total assets (proxy for historical profitability), $x_4$; 
and working capital divided by total assets (proxy for short-term liquidity), $x_5$.

%
%

The three nonparametric models were applied to the data, using the baseline choice for  
the atoms prior hyperparameters, and priors for the weights hyperparameters that 
favor a moderate to large number of distinct mixture components $n^{*}$ (i.e., number of 
distinct $\mathcal{L}_i$ in the notation of Section \ref{subsec:postinfori}).
Given the number of covariates, one would expect that the common-atoms model requires
larger $n^{*}$. 
Indeed, the posterior median for $n^{*}$ is $8$, $12$, and $21$ under the common-weights, 
general, and common-atoms model, respectively; in fact, the common-atoms model did not produce 
a posterior draw for $n^{*}$ smaller than $10$. The relative inefficiency of the 
common-atoms model is also reflected in its larger penalty term for the posterior 
predictive loss criterion. As detailed in the Supplementary Material, the model comparison 
criterion essentially does not distinguish between the general and common-weights models, 
and inference results are overall similar under these two models. Here, we discuss results 
under the common-weights model.

\begin{figure}[t!]
\centering
\includegraphics[width=16cm,height=6cm]{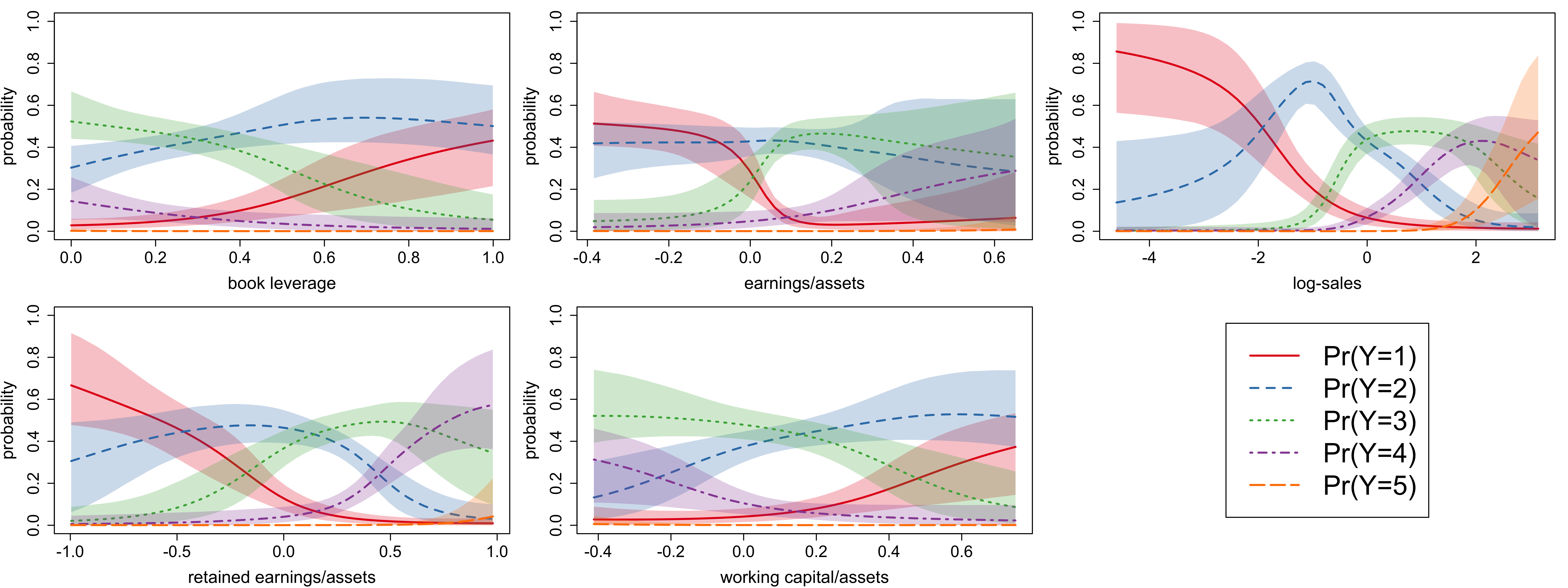}
\caption{
{\small Credit ratings data. Posterior mean (lines) and $95\%$ interval 
(shaded regions) estimates of probability response curves 
$\pi_{j}(x_{s})$. Estimates for all five response categories are displayed in 
a single panel for each covariate.}
}
\label{fig:1stordermargprob}
\end{figure}

We estimate first-order effects for each covariate $x_{s}$ (denoted by $\pi_{j}(x_{s})$, 
for $j=1,\ldots,5$), by computing posterior realizations for 
$\text{Pr}(\mathbf{Y}=j \mid G_{\mathbf{x}})$ in (\ref{eq:margcurvegen}) at a grid over 
the observed range for $x_{s}$, keeping the values of the other covariates fixed at 
their observed average. The resulting point and interval estimates are displayed in 
Figure \ref{fig:1stordermargprob}. The estimates reveal some interesting relationships 
between the firm's characteristics and its credit rating. For instance, debt may help to 
fuel growth of the firm, while uncontrolled debt levels can lead to credit downgrades. 
Hence, an important question pertains to the relevant debt to assets ratio. 
The substantial increase in the probability of the lowest credit rating when book 
leverage gets larger than 0.4 (top left panel of Figure \ref{fig:1stordermargprob}) 
suggests that the desirable ratio may not exceed $0.4$. Moreover, there is a 
positive association between standardized log-sales (a proxy for firm size) and the
firm's credit rating. The probability of the lowest credit rating decreases at a 
particular rate for low to moderate log-sales values, with the probability becoming 
exceedingly small for larger firms. The probabilities for ratings 2, 3 and 4 peak 
at increasingly larger log-sales values, and the probability of the highest rating 
is practically zero for low to moderate log-sales values and is increasing for the 
largest firms.

\begin{figure}[t!]
\centering
\includegraphics[width=16.5cm,height=4.3cm]{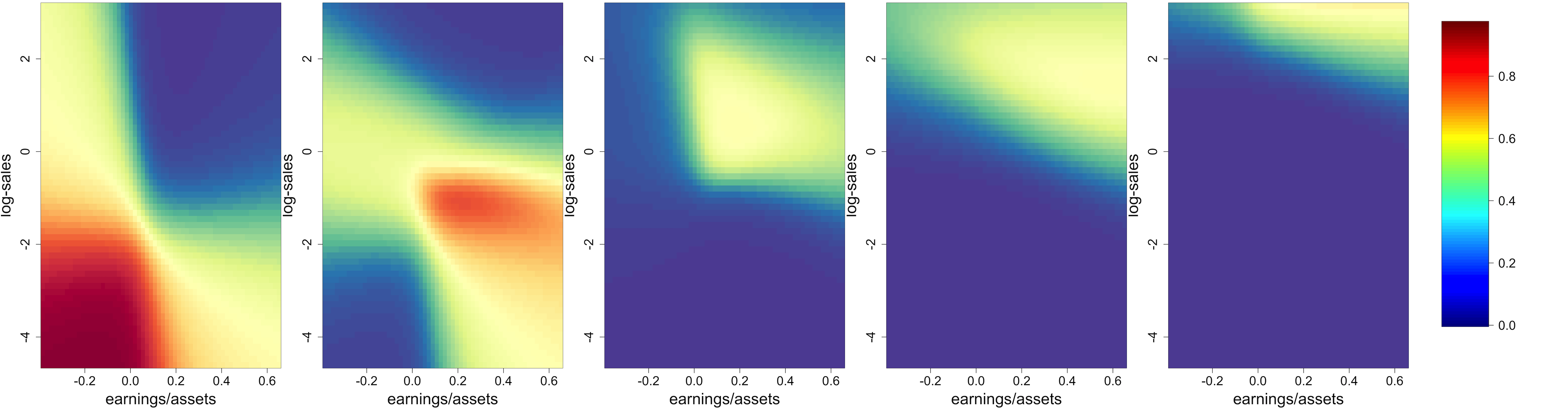}
\caption{
{\small
Credit ratings data. Posterior mean estimates of probability response surfaces 
$\pi_{j}(x_2,x_3)$, for $j=1,\ldots,5$ (from left to right).}
}
\label{fig:2ndordermargprob}
\end{figure}

Similarly to the first-order effects estimates, we can obtain inference for second-order
probability response surfaces for any pair of covariates $(x_{s},x_{s^{\prime}})$, 
denoted by $\pi_{j}(x_{s},x_{s^{\prime}})$, for $j=1,\ldots,5$. As an illustration of 
the model's capacity to accommodate interaction effects among the covariates, 
Figure \ref{fig:2ndordermargprob} plots posterior mean estimates for the second-order 
effects corresponding to earnings divided by total assets ($x_2$) and standardized 
log-sales ($x_3$).

The Supplementary Material includes additional results. 
In particular, we study how the posterior probability of obtaining investment grade 
(rating of 3 or higher) changes when each of the covariates changes its value from 
the 25th to the 75th observed percentile.

\subsection{Retinopathy data}
\label{subsec:retinopathy}

Problems from clinical research provide a broad application area for which the proposed modeling 
approach is particularly well-suited. For such problems, the severeness of a disease is often 
recorded in ordinal scale, and it is of interest to estimate effects of risk factors on disease 
status. This is a setting where it is natural to treat ordinal responses sequentially, 
from which conditional probability response relationships can be directly explored.

To illustrate the utility of our methodology in this context, we work with data set 
\texttt{retinopathy} from the R package ``catdata'' \citep{catdata}. The data set is 
from a 6-year follow-up study of type 1 diabetic patients; it contains 
information about 613 patients' retinopathy status, recorded as: no retinopathy 
(ordinal level 1), nonproliferative retinopathy (ordinal level 2), and advanced retinopathy 
or blind (ordinal level 3). Also available is information on four risk factors: 
smoking status (smoker/non-smoker), diabetes duration (years), glycosylated hemoglobin (percent), 
and diastolic blood pressure (mmHg). The primary scientific question pertains to association 
between retinopathy and smoking status, adjusted for the other risk factors.

The standard proportional odds regression model does not appear suitable for this data set.
Descriptive data analysis suggests that the odds of developing the retinopathy states are 
not proportional with respect to smoking; see \cite{Bender1998}. 
%
%
This is therefore a useful illustrative case for the LSBP mixture model, which, in contrast to 
any particular parametric model, supports essentially any collection of covariate-dependent 
ordinal response probabilities. We apply the general model to the data with the ordinal scale 
for the responses, focusing on the endpoints of ``at least nonproliferative retinopathy''
($\mathbf{Y} \geq 2$) and ``advanced retinopathy or blind'' ($\mathbf{Y} = 3$).

Since the main objective is to assess the relationship between smoking and retinopathy, we 
focus on inference results for the two retinopathy endpoint probabilities for smokers and 
non-smokers. The Supplementary Material provides additional results, including estimation 
of the effect of the other risk factors, 
and comparison between the different nonparametric models and the parametric 
continuation-ratio logits model.

\begin{figure}[t!]
\centering
\begin{subfigure}{0.475\textwidth}
  \centering 
  \includegraphics[width=7.55cm,height=4cm]{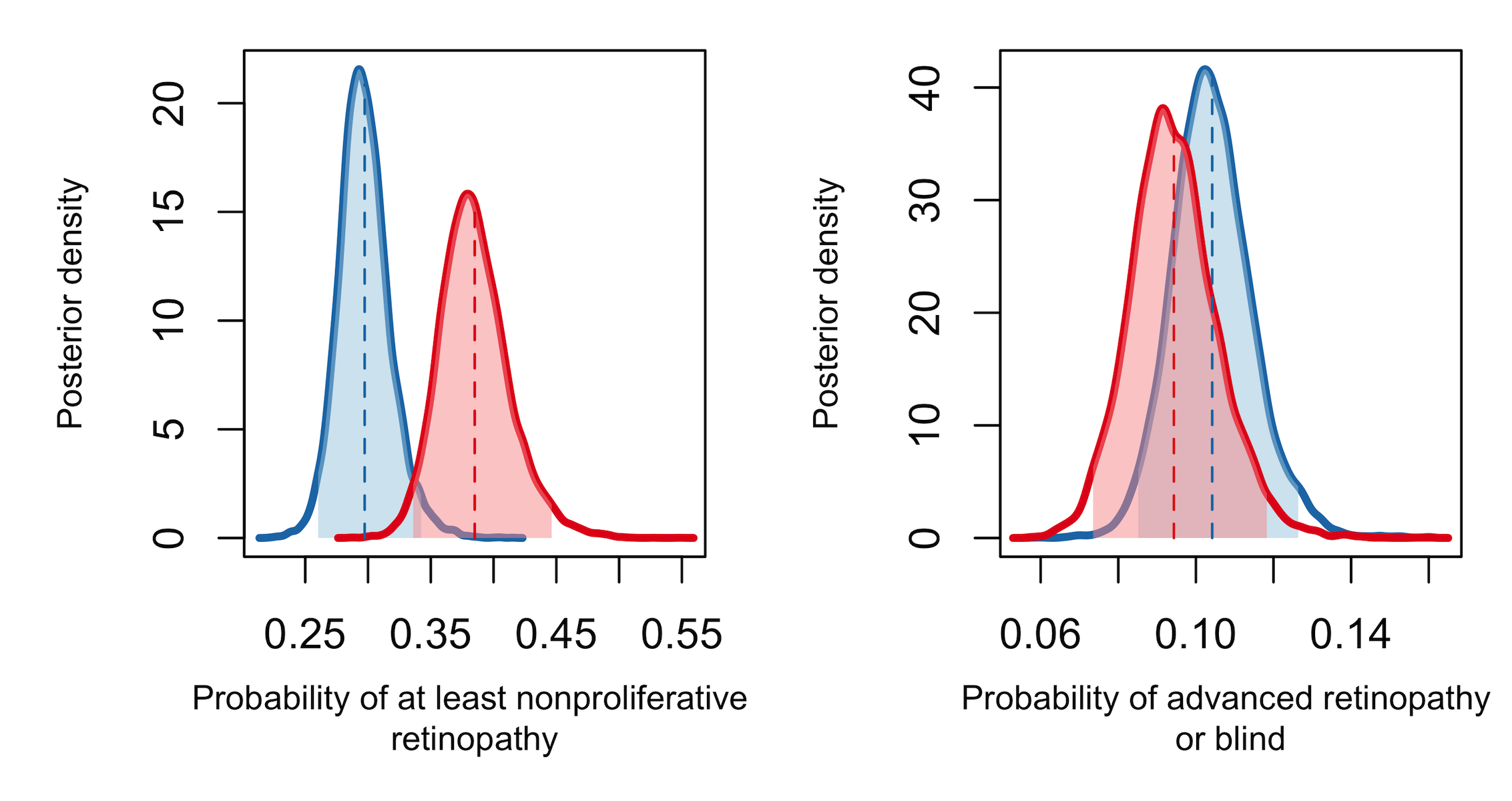}  
  \caption{\small Probability $\tilde{\pi}_{1s}$ (left) and $\tilde{\pi}_{2s}$ (right).}
  \label{subfig:retprob}
\end{subfigure}
\hfill
\begin{subfigure}{0.475\textwidth}
  \centering
  \includegraphics[width=7.55cm,height=4cm]{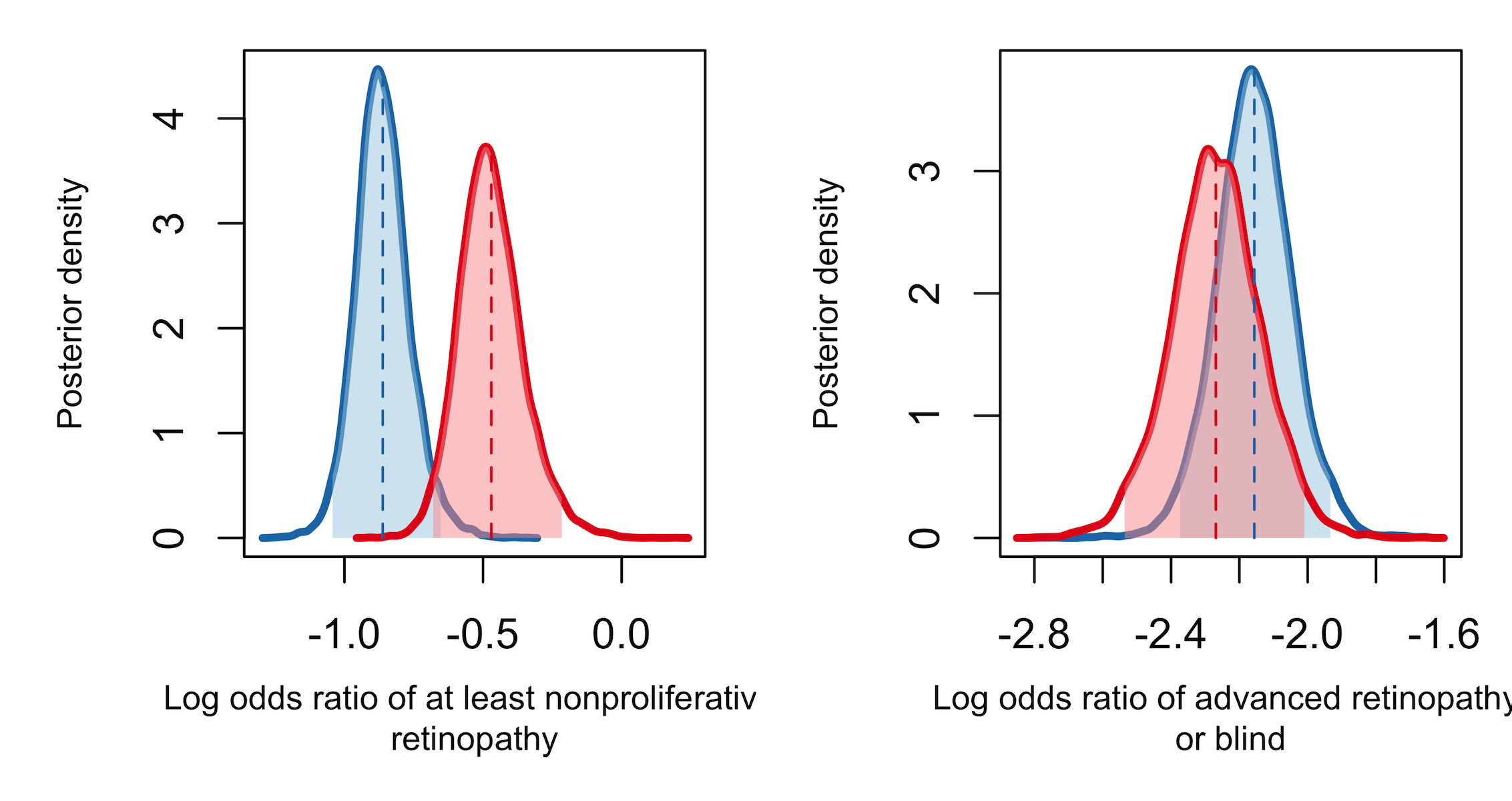}  
  \caption{\small Log odds ratio $r_{1s}$ (left) and $r_{2s}$ (right).}
  \label{subfig:retodd}
\end{subfigure}
\caption{\small Retinopathy data. Posterior densities for the probabilities and log odds 
ratios of the two retinopathy endpoints for smokers (in red) and non-smokers (in blue). 
The dashed line and shaded region correspond to the posterior mean and 95\% credible 
interval, respectively}
\label{fig:retprobodd}
\end{figure}

Keeping the values for the other risk factors fixed at their observed average, 
Figure \ref{subfig:retprob} displays the posterior densities for 
$\tilde{\pi}_{1s}=\text{Pr}(\mathbf{Y}\geq 2\mid G_{\mathbf{x}})$ and 
$\tilde{\pi}_{2s}=\text{Pr}(\mathbf{Y}=3\mid G_{\mathbf{x}})$, where the subscript 
$s=0,1$ indicates non-smokers and smokers, respectively. 
These results point to an adverse effect of smoking on the development of at least 
nonproliferative retinopathy for diabetic patients, whereas there is no clear suggestion of 
an effect on the terminal endpoint (advanced retinopathy of blindness). Indeed, the posterior 
mean and 95\% credible interval for $\tilde{\pi}_{11} - \tilde{\pi}_{10}$ are $0.088$ and 
$(0.042,0.144)$, whereas the corresponding estimates for $\tilde{\pi}_{21} - \tilde{\pi}_{20}$
are $-0.010$ and $(-0.031,0.012)$.

Consider the smoker/non-smoker log odds ratios for the two retinopathy endpoints, that is, 
$r_{1s}=$ $\log \{ \tilde{\pi}_{1s}/(1-\tilde{\pi}_{1s}) \}$ 
and $r_{2s}=$ $\log \{ \tilde{\pi}_{2s}/(1-\tilde{\pi}_{2s}) \}$, for $s=0,1$. 
The proportional odds regression model assumes
$\log \{ \text{Pr}(\mathbf{Y} \leq j)/\text{Pr}(\mathbf{Y} > j) \} =$
$\varkappa_{j} - \mathbf{x}^T\boldsymbol{\beta}$, for $j=1,2$, where the $\varkappa_{j}$ 
are the cut-off points in the notation of Section \ref{subsubsec:equivrep}, and 
$\mathbf{x}$ comprises the four risk factors. Hence, assuming proportional odds specifically 
with regard to the risk factor of smoking imposes the constraint $r_{11}-r_{10}=$ 
$r_{21}-r_{20}$. As discussed in \cite{Bender1998}, the proportional odds model identifies 
the diabetes duration, glycosylated hemoglobin, and blood pressure as significant risk factors, 
but estimates that the effect of smoking is negligible. \cite{Bender1998} question the 
proportional odds model assumption (the constraint above) based on descriptive data analysis,
and estimation results from fitting separate binary logistic regressions to the two 
retinopathy endpoints.

The results from the LSBP mixture model highlight the benefits of flexible nonparametric 
Bayesian modeling. Using a probability model for the ordinal response, we can identify 
the disease endpoint for which smoking has an adverse effect (Figure \ref{subfig:retprob}), 
as well as obtain clear evidence against the proportional odds structure with respect to 
the smoking risk factor (Figure \ref{subfig:retodd}). And, to reiterate, such model-based 
inferences arise from a prior probability model that does not impose restrictions on the 
ordinal regression relationships, making it practically useful for applications where it 
is difficult to check whether the assumptions of a specific parametric model are compatible 
with the data generating mechanism.

\section{Discussion}
\label{sec:summary}

We have developed Bayesian nonparametric mixture models for ordinal regression, 
modeling directly the discrete response distribution. The similarity between the logit 
stick-breaking prior and the continuation-ratio logits structure provides an elegant way of 
incorporating covariate effects in both the weights and the atoms of the mixture model, 
leading to the general model. To investigate the trade-off between model flexibility and 
complexity, we introduce two simpler models that retain covariate 
dependence only in the atoms (common-weights model) or only in the weights (common-atoms model). 
The methods yield a comprehensive toolbox that spans a wide range of flexibility 
in modeling ordinal regression relationships. Viewing the two simpler models as building 
blocks of the general model enables us to explore properties and develop inference algorithms 
under a unified framework. 
Full Kullback-Leibler support has been established as a key theoretical 
model property.
The practical advantage of the proposed models lies in the convenience in prior specification 
and the computationally efficient posterior simulation method. With regard to the latter, 
the key feature is the combination of the continuation-ratio logits representation for the 
mixture kernel with the Pólya-Gamma data augmentation technique.

A practical consideration is which model to apply to a specific problem. 
The data examples of Section \ref{sec:oriexample} were chosen to study different scenarios for 
suitability of the simplified models, as they pertain to the complexity 
of the probability response curves, the sample size, and the number of covariates. 
The common-weights model can not take advantage of the local adjustment offered by 
covariate-dependent weights, and this may be an issue for non-standard ordinal regression 
relationships. Among the two simpler model specifications, the common-atoms model is more 
suitable for complex covariate-response relationships. The caveat is that 
this model activates a large number of effective mixture components, thus increasing the 
computational cost and facing the potential risk of overfitting. 
Inheriting features from both of its building blocks, the general model offers the most 
versatile structure, especially for applications with sufficiently large amounts of data 
and non-standard regression relationships, as demonstrated by the 
synthetic data example of Section \ref{subsec:simstudy}. Nonetheless, in applications 
with small to moderate sample sizes and moderate to large number of response categories, 
the two simpler models are useful options to consider.


The continuation-ratio logits structure boosts computation in two ways. First, it implies 
conditional independence for category-specific parameters, allowing partial parallel 
computing across response categories. In addition, the MCMC algorithm can be 
replaced by a mean-field variational inference approach. 
Taking advantage of the Pólya-Gamma technique, the variational strategy for our models can 
be framed within the well-established exponential family setting, for which there exists a 
closed-form coordinate ascent variational inference algorithm \citep{Blei2017}. Therefore, 
there is potential to scale up the proposed models to handle ordinal regression 
problems with large amounts of data. 

The ordinal regression problem we have explored forms a building block for more general model 
settings involving ordinal responses. 
In fact, Proposition \ref{prop:lsbpeql} may widen the scope of the building block through 
alternative distributional assumptions for the latent variables.
A feature of the modeling framework is its modularity. For example, the model structure can be 
embedded in a hierarchical framework to develop nonparametric inference for longitudinal 
ordinal regression. Repeated measurements of ordinal responses are typically measured with covariates 
over time. A possible way to approach such problems could be built upon models that allow the ordinal 
regression relationships at each particular time point to be estimated in a flexible fashion, 
combined with a hyper-model for evolving temporal dynamics. 
In addition, variable selection can be incorporated into the model through the priors for the 
parameters of the mixture kernel and weights, adapting techniques used for local mixtures of 
normal densities \citep[e.g.,][]{Chung2009,MHAK_2022}.
We will report on such extensions in future work.
%

%
%

Finally, the methodology can also be applied to problems where the components of the ordinal response 
$\mathbf{Y}$ are not necessarily binary. A specific application area involves developmental toxicity 
studies. Here, the covariate is the level of a particular toxin, and, for each pregnant laboratory 
animal exposed to a specific toxin level, the typical data structure involves responses recorded for 
its offspring on embryolethality, malformation, and normal offspring. The modeling methods can be 
elaborated to extend the dependent DP mixture model in \cite{KF2013} for developmental toxicology data 
analysis.

\section*{Supplementary material}

The Supplementary Material includes: (i) MCMC algorithm details; (ii) proofs of the 
theoretical results; (iii) details about the prior specification strategy; 
(iv) results from MCMC diagnostics; (v) additional results for the data examples.  




\bibliographystyle{jasa3}
\bibliography{sample}

\pagebreak
\vspace*{50pt}
\begin{center}
\textbf{\LARGE Supplementary Material: Structured Mixture of Continuation-ratio Logits Models for Ordinal Regression}
\end{center}
\bigskip
\bigskip
\bigskip
\bigskip
\setcounter{equation}{0}
\setcounter{figure}{0}
\setcounter{table}{0}
\setcounter{page}{1}
\setcounter{section}{0}
\setcounter{proposition}{0}
\makeatletter
\renewcommand{\theequation}{S\arabic{equation}}
\renewcommand{\thefigure}{S\arabic{figure}}
\renewcommand{\thetable}{S\arabic{table}}
\renewcommand{\thesection}{S\arabic{section}}
\renewcommand{\theproposition}{S\arabic{proposition}}
\renewcommand{\bibnumfmt}[1]{[S#1]}
\renewcommand{\citenumfont}[1]{S#1}

\section{MCMC posterior simulation details}
\label{sec:smmcmcdetail}

\subsubsection*{The general model}

The development of the posterior simulation method for the general model (\ref{eq:hiergenmod}) relies heavily on effectively the same structure for the weights and atoms of the mixture model. The Pólya-Gamma data augmentation approach are used to update parameters defining both the weights and atoms, leading to conditionally conjugate update for all parameters. Denote the Pólya-Gamma distribution with shape parameter $b$ and tilting parameter $c$ by $PG(b,c)$. Specifically, for each $\mathbf{Y}_i$, we introduce two groups of Pólya-Gamma latent variables $\boldsymbol{\xi}_i=(\xi_{i1},\cdots,\xi_{i,L-1})$ and $\boldsymbol{\zeta}_{i}=(\zeta_{i1},\cdots,\zeta_{i,C-1})$, where $\xi_{i\ell}\stackrel{i.i.d.}{\sim}PG(1,0)$ and $\zeta_{ij}\stackrel{ind.}{\sim}PG(m_{ij},0)$. Proceeding to the joint posterior, the contribution from $\mathbf{Y}_i$ is given by
\begin{equation*}
    f(\mathbf{Y}_i \mid \{ \boldsymbol{\beta}_{j \ell} \},\mathcal{L}_i,\boldsymbol{\zeta}_i)\propto \prod_{j=1}^{C-1}\exp\{\frac{\zeta_{ij}}{2}(\mathbf{x}_i^T\boldsymbol{\beta}_{j\mathcal{L}_i}-\upsilon_{ij}/\zeta_{ij})^2\},
\end{equation*}
where $\upsilon_{ij}=Y_{ij}-\frac{m_{ij}}{2}$. Likewise, let $\iota_{i\ell}=\mathcal{L}_{i\ell}-\frac{1}{2}$, we can write the contribution from $\boldsymbol{\mathcal{L}}_i$ as
\begin{equation*}
    f(\boldsymbol{\mathcal{L}}_i \mid \{ \boldsymbol{\gamma}_{\ell} \},\boldsymbol{\xi}_i)\propto \prod_{\ell=1}^{L-1}\exp\{\frac{\xi_{i\ell}}{2}(\mathbf{x}_i^T\boldsymbol{\gamma}_{\ell}-\iota_{i\ell}/\xi_{i\ell})^2\}.
\end{equation*}
These expressions admit closed-form full conditional distributions for $\{ \boldsymbol{\beta}_{j \ell} \}$ and $\{ \boldsymbol{\gamma}_{\ell} \}$. 

We outline the MCMC sampling algorithm for the full augmented model. This process can be achieved entirely with Gibbs updates, by iterating the following steps. For notation simplicity, we let $(\phi\mid -)$ denote the posterior full conditional distribution for parameter $\phi$. 

\begin{description}
   \item[Step 1: update parameters in the atoms.] In this step, we update two sets of parameters, $\{\boldsymbol{\beta}_{j\ell}:j=1,\cdots,C-1,\ell=1,\cdots,L\}$ and $\{\zeta_{ij}:i=1,\cdots,n,j=1,\cdots,C-1\}$. Denote the set of distinct values of the configuration variables by $\{\mathcal{L}_r^*: r=1,\cdots,n^*\}$. Following \citet{Polson2013}, it can be done by $(\boldsymbol{\beta}_{j\ell}\mid -)\sim N(\tilde{\boldsymbol{\mu}}_{j\ell},\tilde{\Sigma}_{j\ell})$ and $(\zeta_{ij}\mid -)\sim PG(m_{ij},\mathbf{x}_i^T\boldsymbol{\beta}_{jL_i})$, where
   \begin{equation*}
       \left\{\begin{aligned}
       & \tilde{\boldsymbol{\mu}}_{j\ell}=\boldsymbol{\mu}_j,\,\,\,\tilde{\Sigma}_{j\ell}=\Sigma_j\,\,\, & \text{if}\,\, \ell\notin\{\mathcal{L}_r^*: r=1,\cdots,n^*\}\\
       & \tilde{\boldsymbol{\mu}}_{j\ell}=\tilde{\Sigma}_{j\ell}(X_{\ell}^T\boldsymbol{\upsilon}_{\ell}+\Sigma_j^{-1}\boldsymbol{\mu}_j),\,\,\,\tilde{\Sigma}_{j\ell}=(X_{\ell}^T\Omega_{\ell}X_{\ell}+\Sigma_j^{-1})^{-1}\,\,\, & \text{if}\,\, \ell\in\{\mathcal{L}_r^*: r=1,\cdots,n^*\}
       \end{aligned}\right..
   \end{equation*}
   Here $X_{\ell}$ is the matrix whose column vectors are given by $\{\mathbf{x}_i:\mathcal{L}_i=\ell\}$, $\Omega_{\ell}$ is the diagonal matrix with diagonal elements $\{\zeta_{ij}:\mathcal{L}_i=\ell\}$, and $\boldsymbol{\upsilon}_{\ell}$ is the vector of $\{\upsilon_{ij}:\mathcal{L}_i=\ell\}$. Notice that updating $\{\boldsymbol{\beta}_{j\ell}\}$ can be run in parallel across categories $j=1,\cdots, C-1$. 
   
   \item[Step 2: update parameters in the weights.] Similarly, we update $\{\boldsymbol{\gamma}_{\ell}:\ell=1,\cdots,L-1\}$ and $\{\xi_{i\ell}:i=1,\cdots,n,\ell=1,\cdots,L-1\}$ from $(\boldsymbol{\gamma}_{\ell}\mid -)\sim N(\tilde{\boldsymbol{\gamma}}_{\ell},\tilde{\Gamma}_{\ell})$ and $(\xi_{i\ell}\mid -)\sim PG(1,\mathbf{x}_i^T\boldsymbol{\gamma}_{\ell})$, where $\tilde{\boldsymbol{\gamma}}_{\ell}=\tilde{\Gamma}_{\ell}(X^T_{\ell}\boldsymbol{\iota}_{\ell}+\Gamma^{-1}_0\boldsymbol{\gamma}_0)$ and $\tilde{\Gamma}_{\ell}=(X^T_{\ell}\Xi_{\ell}X_{\ell}+\Gamma_0^{-1})^{-1}$. We denote the diagonal matrix formed by $\{\xi_{i\ell}:\mathcal{L}_i=\ell\}$ as $\Xi_{\ell}$, and the vector of$\{\iota_{i\ell}: \mathcal{L}_i=\ell\}$ as $\boldsymbol{\iota}_{\ell}$. 
   
   \item[Step 3: update configuration variables.]  Update $\mathcal{L}_i$, for $i=1,\cdots,n$ from
   \begin{equation*}
       P(\mathcal{L}_i=\ell\mid -)=\frac{p_{i\ell}\prod_{j=1}^{C-1}Bin(Y_{ij}\mid m_{ij},\varphi(\boldsymbol{x}_i^T\boldsymbol{\beta}_{j\ell}))}{\sum_{\ell=1}^Lp_{i\ell}\prod_{j=1}^{C-1}Bin(Y_{ij}\mid m_{ij},\varphi(\boldsymbol{x}_i^T\boldsymbol{\beta}_{j\ell}))}
   \end{equation*}
   where $\{p_{i\ell}: \ell=1,\cdots, L\}$ are calculated as $p_{i1}=\varphi(\mathbf{x}_i^T\boldsymbol{\gamma}_{1})$, $p_{i\ell}=\varphi(\mathbf{x}_i^T\boldsymbol{\gamma}_{\ell})\prod_{h=1}^{\ell-1}(1-\varphi(\mathbf{x}_i^T\boldsymbol{\gamma}_h))$, $\ell=2,\cdots,L-1$, and  $p_{iL}=\prod_{\ell=1}^{L-1}(1-\varphi(\mathbf{x}_i^T\boldsymbol{\gamma}_{\ell}))$.
   
   \item[Step 4: update hyperparameters.] By conjugacy, updating hyperparameters is standard. We update $\{\boldsymbol{\mu}_j\}$ and $\{\Sigma_j\}$ by $(\boldsymbol{\mu}_j\mid -)\sim N(\boldsymbol{\mu}^*_j,\Sigma_j/\kappa^*_j)$ and $(\Sigma_j\mid -)\sim IW(\nu^*_j,(\Lambda^*_j)^{-1})$, with the parameters given by
   \begin{equation*}
   \begin{split}
    &\boldsymbol{\mu}^*_j=\frac{\kappa_{0j}}{\kappa_{0j}+n^*}\boldsymbol{\mu}_{0j}+\frac{n^*}{\kappa_{0j}+n^*}\bar{\boldsymbol{\beta}}_j,\,\,\,\kappa^*_j=n^*+\kappa_{0j},\,\,\, \nu^*_j=n^*+\nu_{0j}\,\,\,\bar{\boldsymbol{\beta}}_j=\frac{1}{n^*}\sum_{r=1}^{n^*}\boldsymbol{\beta}_{j\mathcal{L}_r^*}, \\
    &\Lambda^*_j=\Lambda_{0j}+S_j+\frac{n^*\kappa_{0j}}{n^*+\kappa_{0j}}(\bar{\boldsymbol{\beta}}_j-\boldsymbol{\mu}_{0j})(\bar{\boldsymbol{\beta}}_j-\boldsymbol{\mu}_{0j})^T,\,\,\, S_j=\sum_{r=1}^{n^*}(\boldsymbol{\beta}_{j\mathcal{L}_r^*}-\bar{\boldsymbol{\beta}}_j)(\boldsymbol{\beta}_{j\mathcal{L}_r^*}-\bar{\boldsymbol{\beta}}_j)^T.   
   \end{split}
   \end{equation*}
\end{description}

We refer to the above process as the ``general process''. From the connection discussed in Section \ref{sec:specorimodel}, the Gibbs sampler for the two simpler models are straightforwardly adapted from the general process.

\subsubsection*{The common-weights model}

In the scenario that a common-weights model is adopted, the mixing weights and the configuration variables are determined by 
\begin{equation*}
\mathcal{L}_i\mid\boldsymbol{\omega}\sim\sum_{\ell=1}^L\omega_l\delta_{\ell}(\mathcal{L}_i),\,\,\boldsymbol{\omega}\mid\alpha\sim f(\boldsymbol{\omega}\mid\alpha),\,\,\alpha\sim Gamma(a_{\alpha},b_{\alpha}),
\end{equation*}
where $f(\boldsymbol{\omega}\mid\alpha)$ stands for a special case of the generalized Dirichlet distribution
\begin{equation*}
f(\boldsymbol{\omega}\mid\alpha)=\alpha^{L-1}\omega_L^{\alpha-1}(1-\omega_1)^{-1}(1-(\omega_1+\omega_2))^{-1}\cdots(1-\sum_{\ell=1}^{L-2}\omega_{\ell})^{-1},
\end{equation*}
while the atoms are the same as in the general model. Hence, we only need to introduce the group of Pólya-Gamma latent variables $\{\boldsymbol{\zeta}_i: i=1,\cdots,n\}$, which enable the same conjugate update in sampling atoms related parameters. We keep \textbf{Step 1} and \textbf{Step 4} in the general process, whereas the other two steps are replaced by:

\begin{description}
\item[Step $2^*$: update parameters in the weights.] The parameters to be updated in this step involve $\{\omega_{\ell}:\ell=1,\cdots,L-1\}$ and $\alpha$. From \citet{IshwaranJames2001}, it can be done by sample $V_{\ell}^*\stackrel{ind.}{\sim}Beta(1+M_{\ell},\alpha+\sum_{h=\ell+1}^LM_h)$ for $\ell=1,\cdots,L-1$. Then let $\omega_1=V_1^*$, $\omega_{\ell}=V_{\ell}^*\prod_{h=1}^{\ell-1}(1-V_h^*)$, $\ell=2,\cdots,L-1$ and $\omega_L=1-\sum_{\ell=1}^{L-1}\omega_{\ell}$. In addition, a new sample of $\alpha$ is obtained from $(\alpha\mid -)\sim Gamma(a_{\alpha}+L-1,b_{\alpha}-\sum_{\ell=1}^{L-1}\log(1-V_{\ell}^*))$.

\item[Step $3^*$: update configuration variables.] Update $\mathcal{L}_i$, $i=1,\cdots,n$, from
\begin{equation*}
   P(\mathcal{L}_i=\ell\mid -)=\frac{\omega_{\ell}\prod_{j=1}^{C-1}Bin(Y_{ij}\mid m_{ij},\varphi(\boldsymbol{x}_i^T\boldsymbol{\beta}_{j\ell}))}{\sum_{\ell=1}^L\omega_{\ell}\prod_{j=1}^{C-1}Bin(Y_{ij}\mid m_{ij},\varphi(\boldsymbol{x}_i^T\boldsymbol{\beta}_{j\ell}))}. 
\end{equation*}
\end{description}

\subsubsection*{The common-atoms model}

If one choose to fit the common-atoms model, the linear regression terms in the atoms are simplified by $\theta_{j\ell}$ with prior $\theta_{j\ell}\stackrel{ind.}{\sim} N(\mu_j,\sigma_j^2)$, $j=1,\cdots,C-1$ and $\ell=1,\cdots,L$. We replace \textbf{Step 1} and \textbf{Step 4} of the general process with the following alternatives, while the other steps remain the same. 

\begin{description}
\item[Step $1^*$: update parameters in the atoms.] The two sets of parameters $\{\theta_{j\ell}:j=1,\cdots,C-1,\ell=1,\cdots,L\}$ and $\{\zeta_{ij}:i=1,\cdots,n,j=1,\cdots,C-1\}$ are now updated by $(\theta_{j\ell}\mid -)\sim N(\tilde{\mu}_{j\ell},\tilde{\sigma}^2_{j\ell})$ and $(\zeta_{ij}\mid -)\sim PG(m_{ij},\theta_{j\mathcal{L}_i})$, where
\begin{equation*}
       \left\{\begin{aligned}
       & \tilde{\mu}_{j\ell}=\mu_j,\,\,\,\tilde{\sigma}^2_{j\ell}=\sigma^2_j,\,\,\, & \text{if}\,\, \ell\notin\{\mathcal{L}_r^*: r=1,\cdots,n^*\}
       \\
       & \tilde{\mu}_{j\ell}=\tilde{\sigma}_j^2(\sum_{\{i:\mathcal{L}_i=\ell\}}\upsilon_{ij}+\mu_j/\sigma_j^2),\,\,\,\tilde{\sigma}^2_{j\ell}=\sigma_j^2/(\sigma_j^2\sum_{\{i:\mathcal{L}_i=\ell\}}\zeta_{ij}+1),\,\,\, & \text{if}\,\, \ell\in\{\mathcal{L}_r^*: r=1,\cdots,n^*\}
       \end{aligned}\right..
\end{equation*}
\item[Step $4^*$: update hyperparameters.] That is, we update $\{\mu_j: j=1,\cdots, C-1\}$ and $\{\sigma_j^2: j=1,\cdots, C-1\}$ by $(\mu_j\mid -)\sim N(\mu^*_j,\sigma_j^2/\nu^*_j)$ and $(\Sigma_j\mid -)\sim IW(\nu^*_j,(\Lambda^*_j)^{-1})$, where
\begin{equation*}
\begin{split}
&\mu^*_j=\frac{\nu_{0j}\mu_{0j}+n^*\bar{\theta}_j}{\nu_{0j}+n^*},\,\,\,\nu^*_j=n^*+\nu_{0j},\,\,\, a^*_j=a_j+n^*/2\,\,\, \bar{\theta}_j=\frac{1}{n^*}\sum^{n^*}_{r=1}\theta_{jr},\\
&b_j^*=b_j+\frac{1}{2}\sum^{n^*}_{r=1}(\theta_{jr}-\bar{\theta}_j)^2+\frac{n^*\nu_{0j}}{n^*+\nu_{0j}}\frac{(\bar{\theta}_j-\mu_{0j})^2}{2}.
\end{split}
\end{equation*}
\end{description}

Finally, for notation consistency, we should also replace the terms $\mathbf{x}^T_i\boldsymbol{\beta}_{j\ell}$ with $\theta_{j\ell}$ in \textbf{Step 3}, while keeping the same updating mechanism. 

With the posterior samples for model parameters drawn by the MCMC mechanism described above, we can obtain full inference for any regression functional of interest. 
For instance, for any $j=1,\cdots,C$, posterior realizations for the marginal probability response curve, $\text{Pr}(\mathbf{Y}=j \mid G_{\mathbf{x}})$, can be computed over a grid in $\mathbf{x}$ via
\begin{equation*}
\sum_{\ell=1}^{L} \left\{ \varphi(\mathbf{x}^T\boldsymbol{\gamma}_{\ell}^{(t)})
\prod\nolimits_{h=1}^{\ell-1} (1-\varphi(\mathbf{x}^T\boldsymbol{\gamma}_{h}^{(t)})) \right\}
\, \left\{ \varphi(\mathbf{x}^T\boldsymbol{\beta}_{j\ell}^{(t)})
\prod\nolimits_{k=1}^{j-1}[1-\varphi(\mathbf{x}^T\boldsymbol{\beta}_{k\ell}^{(t)})] \right\}
\end{equation*}
where $\varphi(\mathbf{x}^T\boldsymbol{\gamma}_{L}^{(t)})=$
$\varphi(\mathbf{x}^T\boldsymbol{\beta}_{C\ell}^{(t)})\equiv 1$, and the superscript $(t)$ 
indicates the $t$th posterior sample for the model parameters.

\section{Proofs of theoretical results}
\label{sec:smproofs}

\subsection{Proof of Proposition \ref{prop:seqbreaklogit}}
\label{subsec:pfseqbreak}

\begin{proof}
Under the augmented model, we have
\begin{equation*}
\text{Pr}(\mathbf{Y}=j,\boldsymbol{\mathcal{Z}}\mid\boldsymbol{\theta}) = 
\text{Pr}(\mathbf{Y}=j \mid \boldsymbol{\mathcal{Z}}) \,
f(\boldsymbol{\mathcal{Z}}\mid\boldsymbol{\theta}) =
\mathbf{1}(\boldsymbol{\mathcal{Z}}\in\mathcal{R}_j) \, 
\prod_{j=1}^{C-1}\mathfrak{L}(\mathcal{Z}_j\mid\theta_j).
\end{equation*}
Integrating out $\boldsymbol{\mathcal{Z}}$, we obtain
\begin{equation*}
\begin{split}
\text{Pr}(\mathbf{Y}=j\mid \boldsymbol{\theta}) & = 
\int \mathbf{1}(\boldsymbol{\mathcal{Z}}\in\mathcal{R}_j)
\prod_{j=1}^{C-1}\mathfrak{L}(\mathcal{Z}_j\mid\theta_j) \,
\text{d}\boldsymbol{\mathcal{Z}} =
\int_{\mathcal{R}_j}\prod_{j=1}^{C-1}\mathfrak{L}(\mathcal{Z}_j\mid\theta_j) \,
\text{d}\boldsymbol{\mathcal{Z}} \\
& = \left(  \int_0^{\infty} \mathfrak{L}(\mathcal{Z}_j\mid\theta_j) \, \text{d}\mathcal{Z}_j
\right) \, 
\prod_{k=1}^{j-1} \left(
\int_{-\infty}^0\mathfrak{L}(\mathcal{Z}_k\mid\theta_k) \, \text{d}\mathcal{Z}_k
\right) =
\varphi(\theta_j) \, \prod_{k=1}^{j-1} \{ 1-\varphi(\theta_k)\},
\end{split}
\end{equation*}
for $j=2,\cdots,C-1$. Similarly, $\text{Pr}(\mathbf{Y}=1\mid \boldsymbol{\theta})=\varphi(\theta_1)$,
and $\text{Pr}(\mathbf{Y}=C\mid \boldsymbol{\theta})=$
$\prod_{k=1}^{C-1} \{ 1-\varphi(\theta_k) \}$. Therefore, 
$\mathbf{Y}\mid \boldsymbol{\theta}\sim K(\mathbf{Y}\mid \boldsymbol{\theta})$, i.e., the ordinal 
response distribution is the multinomial with the continuation-ratio logits parameterization.
\end{proof}

\subsection{Proof of Proposition \ref{prop:lsbpeql}}
\label{subsec:pflsbpeql}

\begin{proof}
In this scenario, under the augmented model,
\begin{equation*}
\text{Pr}(\mathbf{Y}=j,\boldsymbol{\mathcal{Z}}\mid G_{\mathbf{x}}) = 
\mathbf{1}(\boldsymbol{\mathcal{Z}}\in\mathcal{R}_j) \,
\sum_{\ell=1}^{\infty}\omega_{\ell}(\mathbf{x})
\prod_{j=1}^{C-1}\mathfrak{L}(\mathcal{Z}_j\mid\theta_{j\ell}(\mathbf{x})).
\end{equation*}
Integrating out $\boldsymbol{\mathcal{Z}}$, the probability for the $j$-th 
response category becomes
\begin{equation*}
\begin{split}
\text{Pr}(\mathbf{Y}=j\mid G_{\mathbf{x}}) & =
\int_{\mathcal{R}_j}\sum_{\ell=1}^{\infty}\omega_{\ell}(\mathbf{x})
\prod_{j=1}^{C-1}\mathfrak{L}(\mathcal{Z}_j\mid\theta_{j\ell}(\mathbf{x})) \,
\text{d}\boldsymbol{\mathcal{Z}} \\
& = 
\sum_{\ell=1}^{\infty}\omega_{\ell}(\mathbf{x})\int_{\mathcal{R}_j}
\prod_{j=1}^{C-1}\mathfrak{L}(\mathcal{Z}_j\mid\theta_{j\ell}(\mathbf{x})) \,
\text{d}\boldsymbol{\mathcal{Z}} \\
& = \sum_{\ell=1}^{\infty}\omega_{\ell}(\mathbf{x})
\left\{
\left( \int^{+\infty}_0\mathfrak{L}(\mathcal{Z}_j \mid \theta_{j\ell}(\mathbf{x})) \, 
\text{d}\mathcal{Z}_j \right) \, 
\prod_{k=1}^{j-1} \left( \int_{-\infty}^0\mathfrak{L}(\mathcal{Z}_k\mid\theta_{k\ell}(\mathbf{x})) 
\, \text{d}\mathcal{Z}_k \right)
\right\} \\
& = \sum_{\ell=1}^{\infty}\omega_{\ell}(\mathbf{x})
\left\{ \varphi(\theta_{j\ell}(\mathbf{x})) \, 
\prod_{k=1}^{j-1} [1-\varphi(\theta_{k\ell}(\mathbf{x}))] \right\},
\end{split} 
\end{equation*}
for $j=2,\cdots,C-1$. 
The function under integration and countable summation in the first line takes non-negative values, 
and we can thus switch the order of the two operations. 
Similarly, we obtain $\text{Pr}(\mathbf{Y}=1\mid G_{\mathbf{x}})=$
$\sum_{\ell=1}^{\infty}\omega_{\ell}(\mathbf{x}) \, \varphi(\theta_{1\ell}(\mathbf{x}))$,
and $\text{Pr}(\mathbf{Y}=C \mid G_{\mathbf{x}})=$
$\sum_{\ell=1}^{\infty}\omega_{\ell}(\mathbf{x}) 
\left\{ \prod_{k=1}^{C-1}[1-\varphi(\theta_{k \ell}(\mathbf{x}))] \right\}$. 
Hence, $\mathbf{Y} \mid G_{\mathbf{x}}\sim \sum_{\ell=1}^{\infty} \omega_{\ell}(\mathbf{x}) \,
K(\mathbf{Y} \mid \boldsymbol{\theta}_{\ell}(\mathbf{x}))$, i.e., the multinomial 
LSBP mixture model.
\end{proof}

\subsection{Proof of Lemma \ref{lem:klsupport}}
\label{subsec:pflemma}

\begin{proof}
Consider the set of probability densities 
$\{f^0_{\mathbf{x}}(\mathbf{z}):\mathbf{x}\in\mathcal{X}\}$ for 
$\mathbf{Z}\in\mathbb{R}^{C-1}$. 
Let $\mathfrak{F}$ be the set of all distributions defined on $\mathbb{R}^{C-1}$. A prior $\mathcal{F}_{\mathbf{x}}=\{F_{\mathbf{x}}(w,B):\mathbf{x}\in\mathcal{X}\}$ on $\mathfrak{F}^{\mathcal{X}}$ is a stochastic process on an appropriate probability space $(\mathcal{W},\mathscr{F},\Pi)$, such that for every $\mathbf{x}\in\mathcal{X}$ and almost every $w\in\mathcal{W}$, $F_{\mathbf{x}}(w,\cdot)\in\mathfrak{F}$. The set of densities $\{f^0_{\mathbf{x}}(\mathbf{z}):\mathbf{x}\in\mathcal{X}\}$ having KL property relative to $\mathcal{F}_{\mathbf{x}}$ refers to
\begin{equation}
\Pi \left\{
w \in \mathcal{W}: \int f_{\mathbf{x}_t}^0(\mathbf{z})
\log(f_{\mathbf{x}_t}^0(\mathbf{z})/f_{\mathbf{x}_t}(\mathbf{z}))\,
\text{d}\mathbf{z}<\epsilon, \,\,\,\, t=1,\ldots,T \right\} >0, 
\label{eq:KLcontinuous}
\end{equation}
for any $\epsilon>0$, $\mathbf{x}_1,\ldots,\mathbf{x}_T\in\mathcal{X}$, $T\in\mathbb{N}^+$.

Now consider the set of probability masses $\{p_{\mathbf{x}}(y):\mathbf{x}\in\mathcal{X}\}$ for 
ordinal variable $Y$ with $C$ categories. Denote the set of all distributions on $\{1,\ldots,C\}$ 
by $\mathfrak{P}$. Let $\{\mathcal{R}_1,\ldots,\mathcal{R}_C\}$ be a partition of $\mathbb{R}^{C-1}$. 
To connect with the distribution of continuous variable, we consider the mapping 
from $\mathfrak{F}^{\mathcal{X}}$ to $\mathfrak{P}^{\mathcal{X}}$, given by
\begin{equation}
f_{\mathbf{x}}\mapsto p_{\mathbf{x}}(y)=\int_{\mathcal{R}_y}f_{\mathbf{x}}(\mathbf{z}) 
\, \text{d}\mathbf{z}
,\quad y=1,\ldots,C.
\label{eq:contmaptodisc}
\end{equation}
This mapping induces a prior on $\mathfrak{P}^{\mathcal{X}}$ from $\mathcal{F}_{\mathbf{x}}$. 
The prior, denoted by $\mathcal{P}_{\mathbf{x}}$, is a $\mathfrak{P}$-valued stochastic process 
on probability space $(\mathcal{W},\mathscr{P},\Pi)$. Additionally, let $p_{\mathbf{x}}^0(y)$ 
denote the discrete distribution induced by $f_{\mathbf{x}}^0(\mathbf{z})$. 
Following the definition of KL property for continuous distributions, we say 
$\{p_{\mathbf{x}}^0(y):\mathbf{x}\in\mathcal{X}\}$ possesses the KL property with respect 
to $\mathcal{P}_{\mathbf{x}}$ if
\begin{equation}
\Pi \left\{
w\in\mathcal{W}: \sum_{y=1}^C p_{\mathbf{x}_t}^0(y) 
\log(p_{\mathbf{x}_t}^0(y)/p_{\mathbf{x}_t}(y))<\epsilon, \,\,\,\, t=1,\ldots,T 
\right\} > 0
\label{eq:KLdiscrete}
\end{equation}
for any $\epsilon>0$, $\mathbf{x}_1,\ldots,\mathbf{x}_T\in\mathcal{X}$, $T\in\mathbb{N}^+$.

To prove the lemma, it is adequate to show that every $w$ that satisfies the criterion 
in (\ref{eq:KLcontinuous}) also satisfies the criterion in (\ref{eq:KLdiscrete}).  

The proof relies on the following inequality of KL divergence, 
\begin{equation}
    \int_{\mathcal{A}}g_1(\mathbf{u})\log\left(\frac{g_1(\mathbf{u})}{g_2(\mathbf{u})}\right)\,\text{d}\mathbf{u}
    \geq\int_{\mathcal{A}}g_1(\mathbf{u})\,\text{d}\mathbf{u}\times \log\left(\frac{\int_{\mathcal{A}}g_1(\mathbf{u})\,\text{d}\mathbf{u}}{\int_{\mathcal{A}}g_2(\mathbf{u})
    \,\text{d}\mathbf{u}}\right),
    \label{eq:propertyKL}
\end{equation}
where $g_r(\mathbf{u})$ is a density of $\mathbf{u}\in\mathbb{R}^s$, $r=1,2$ and $\mathcal{A}$ is a generic subset of $\mathbb{R}^s$. To show this inequality, let $\mathcal{H}_r=\int_{\mathcal{A}} g_r(\mathbf{u})\,\text{d}\mathbf{u}$, such that $h_r(\mathbf{u})=g_r(\mathbf{u})/\mathcal{H}_r$, $r=1,2$, are densities on $\mathcal{A}$. The left-hand-side of (\ref{eq:propertyKL}) can be expressed as $\mathcal{H}_1\int_{\mathcal{A}}h_1(\mathbf{u})\log(\frac{\mathcal{H}_1h_1(\mathbf{u})}{\mathcal{H}_2h_2(\mathbf{u})})\,\text{d}\mathbf{u}=\mathcal{H}_1\log(\frac{\mathcal{H}_1}{\mathcal{H}_2})+\mathcal{H}_1\int_{\mathcal{A}}h_1(\mathbf{u})\log(\frac{h_1(\mathbf{u})}{h_2(\mathbf{u})})\,\text{d}\mathbf{u}\geq \mathcal{H}_1\log(\frac{\mathcal{H}_1}{\mathcal{H}_2})$, 
because $\int_{\mathcal{A}}h_1(\mathbf{u})\log(\frac{h_1(\mathbf{u})}{h_2(\mathbf{u})})\,\text{d}\mathbf{u}$ is the KL divergence between densities $h_1$ and $h_2$.

For any set $\mathcal{R}_y$ in the partition of $\mathbb{R}^{C-1}$, from (\ref{eq:propertyKL}) we obtain
\begin{equation*}
\int_{\mathcal{R}_y}f_{\mathbf{x}_t}^0(\mathbf{z})
\log(f_{\mathbf{x}_t}^0(\mathbf{z})/f_{\mathbf{x}_t}(\mathbf{z}))\,\text{d}\mathbf{z}\geq \int_{\mathcal{R}_y}f_{\mathbf{x}_t}^0(\mathbf{z})\,\text{d}\mathbf{z}\times \log\left(\frac{\int_{\mathcal{R}_y}
f_{\mathbf{x}_t}^0(\mathbf{z})\,\text{d}\mathbf{z}}{\int_{\mathcal{R}_y}f_{\mathbf{x}_t}(\mathbf{z})\,\text{d}\mathbf{z}}\right)
=p^0_{\mathbf{x}_t}(y)\log\left(\frac{p^0_{\mathbf{x}_t}(y)}{p_{\mathbf{x}_t}(y)}\right),
\end{equation*}
for $y=1,\ldots,C$. Consider any $w\in\mathcal{W}$ satisfying $\int f_{\mathbf{x}_t}^0(\mathbf{z}) \log(f_{\mathbf{x}_t}^0(\mathbf{z})/f_{\mathbf{x}_t}(\mathbf{z}))\,\text{d}\mathbf{z}<\epsilon$, 
for $t=1,\ldots,T$. Then, we have
\begin{equation*}
    \begin{split}
        \epsilon&>\int f_{\mathbf{x}_t}^0(\mathbf{z}) \log(f_{\mathbf{x}_t}^0(\mathbf{z})/f_{\mathbf{x}_t}(\mathbf{z}))\,\text{d}\mathbf{z}=\sum_{y=1}^C \int_{\mathcal{R}_y}f_{\mathbf{x}_t}^0(\mathbf{z})
\log(f_{\mathbf{x}_t}^0(\mathbf{z})/f_{\mathbf{x}_t}(\mathbf{z}))\,\text{d}\mathbf{z}\\
&\geq \sum_{y=1}^C p^0_{\mathbf{x}_t}(y)\log\left(\frac{p^0_{\mathbf{x}_t}(y)}{p_{\mathbf{x}_t}(y)}\right).
    \end{split}
\end{equation*}
We have thus obtained that $w$ also satisfies 
$\sum_{y=1}^C p^0_{\mathbf{x}_t}(y)\log({p^0_{\mathbf{x}_t}(y)}/{p_{\mathbf{x}_t}(y)})<\epsilon$, 
for $t=1,\ldots,T$, which completes the argument for the proof.
\end{proof}

\subsection{Proof of Theorem \ref{thm:KLgeneralmodel}}
\label{subsec:pfklgeneral}

\begin{proof}
Based on Proposition \ref{prop:lsbpeql}, the multinomial LSBP mixture model can be formulated
in terms of latent continuous responses as follows:
$\mathbf{Y}\mid \boldsymbol{\mathcal{Z}}\sim \boldsymbol{1}(\mathbf{Y}=j\Longleftrightarrow \boldsymbol{\mathcal{Z}}\in\mathcal{R}_j)$, for $j=1,\ldots,C$, and 
\begin{equation}
\begin{split}
&\boldsymbol{\mathcal{Z}} \mid G_{\mathbf{x}}\sim \sum_{\ell=1}^{\infty}\omega_{\ell}(\mathbf{x})\prod_{j=1}^{C-1}\mathfrak{L}(\mathcal{Z}_j\mid\theta_{j\ell}(\mathbf{x}))=\int \prod_{j=1}^{C-1}\mathfrak{L}(\mathcal{Z}_j\mid\theta_{j}(\mathbf{x}))\,\text{d}G_{\mathbf{x}}(\boldsymbol{\theta}),\\
& \omega_{1}(\mathbf{x}) \, = \, \varphi(\mathbf{x}^T\boldsymbol{\gamma}_1), \,\,\,
\omega_{\ell}(\mathbf{x}) \, = \,
\varphi(\mathbf{x}^T\boldsymbol{\gamma}_{\ell}) 
\prod_{h=1}^{\ell-1} (1-\varphi(\mathbf{x}^T\boldsymbol{\gamma}_h)), \,\,\, \ell \geq 2,\\
&\theta_{j\ell}(\mathbf{x}) \, = \, \mathbf{x}^T\boldsymbol{\beta}_{j\ell}
\mid \boldsymbol{\mu}_j, \Sigma_j \, \stackrel{ind.}{\sim} \, 
N(\mathbf{x}^T\boldsymbol{\mu}_j,\mathbf{x}^T\Sigma_j\mathbf{x}), \,\,\,\,\,
j=1,\ldots,C-1, \,\,\, \ell \geq 1.
\end{split}
\label{eq:LSBPcontinuous}
\end{equation}
where $\{\mathcal{R}_j:j=1,\ldots,C\}$ is the partition of $\mathbb{R}^{C-1}$ defined 
in equation (\ref{eq:partition}) of the main paper.

Let $\mathcal{F}_{\mathbf{x}}$ be the above LSBP mixture prior for continuous random 
vector $\boldsymbol{\mathcal{Z}} \in \mathbb{R}^{C-1}$. The original multinomial LSBP mixture
prior on ordinal distributions is denoted by $\mathcal{P}_{\mathbf{x}}$. Consider the 
set of probability masses $\{p^0_{\mathbf{x}}:\mathbf{x}\in\mathcal{X}\}$ on 
$\{1,\ldots,C\}$. From Lemma \ref{lem:klsupport}, to show that 
$\{p^0_{\mathbf{x}}:\mathbf{x}\in\mathcal{X}\}$ has the KL property with respect 
to $\mathcal{P}_{\mathbf{x}}$, we can utilize the result regarding the KL support 
of $\mathcal{F}_{\mathbf{x}}$.

Suppose the probability densities $\{f^0_{\mathbf{x}}:\mathbf{x}\in\mathcal{X}\}$ on 
$\mathbb{R}^{C-1}$ satisfy the regularity conditions (v) to (viii) in 
\citet[][Theorem~5]{Barrientos2012}. We prove that 
$\{f^0_{\mathbf{x}}:\mathbf{x}\in\mathcal{X}\}$ possesses the KL property relative 
to $\mathcal{F}_{\mathbf{x}}$. The proof consists of two parts. We first show that
$\mathcal{F}_{\mathbf{x}}$ falls in the scheme of dependent stick-breaking process 
(DSBP) priors \citep[][Definition~4]{Barrientos2012}. Then, we confirm that the 
mixture kernel of the model for $\boldsymbol{\mathcal{Z}}$ satisfies the conditions 
in  \citet[][Theorem~10]{Barrientos2012}.

For the LSBP prior discussed in this paper, we introduce the marginal 
distributions $\mathscr{V}_{\mathcal{X}}^{V_{\ell}}$, $G_{\mathcal{X}}^0$, and the 
copula functions $\mathscr{C}_{\mathcal{X}}^{V_{\ell}}$, 
$\mathscr{C}_{\mathcal{X}}^{\theta}$, $\ell=1,2,\ldots$, defined as follows:
\begin{equation*}
\begin{split}
&\mathscr{V}_{\mathcal{X}}^{V_{\ell}}=\{N(\mathbf{x}^T\boldsymbol{\gamma}_0,\mathbf{x}^T\Gamma_0\mathbf{x}):\mathbf{x}\in\mathcal{X}\},\quad G_{\mathcal{X}}^0=\{\prod_{j=1}^{C-1}N(\mathbf{x}^T\boldsymbol{\beta}_j,\mathbf{x}^T\Sigma_j\mathbf{x}):\mathbf{x}\in\mathcal{X}\},\\
    &\mathscr{C}_{\mathcal{X}}^{V_{\ell}}=\{C_{\mathbf{x}_1,\ldots,\mathbf{x}_d}(u_1,\ldots,u_d)=\Phi_{S(\mathbf{x}_1,\ldots,\mathbf{x}_d)}(\Phi^{-1}(u_1),\ldots,\Phi^{-1}(u_d)),\mathbf{x}_1,\ldots,\mathbf{x}_d\in\mathcal{X}\},\\
    &\mathscr{C}_{\mathcal{X}}^{\theta}=\{C_{\mathbf{x}_1,\ldots,\mathbf{x}_d}(u_1,\ldots,u_d)=\prod_{j=1}^{C-1}\Phi_{R_j(\mathbf{x}_1,\ldots,\mathbf{x}_d)}(\Phi^{-1}(u_1),\ldots,\Phi^{-1}(u_d)),\mathbf{x}_1,\cdots,\mathbf{x}_d\in\mathcal{X}\},
\label{eq:genlsbpcopula}    
\end{split}    
\end{equation*}
where $\Phi$ is the c.d.f. of the standard normal distribution. In addition,
$\Phi_{S(\mathbf{x}_1,\cdots,\mathbf{x}_d)}$ and  $\Phi_{R_j(\mathbf{x}_1,\cdots,\mathbf{x}_d)}$ 
denote the c.d.f. of a $d-$variate normal distribution with mean 0, variance 1,
and correlation matrix $S(\mathbf{x}_1,\cdots,\mathbf{x}_d)$, $R_j(\mathbf{x}_1,\cdots,\mathbf{x}_d)$, whose $(s,t)$-entry is given respectively by
\begin{equation*}
S(\mathbf{x}_1,\cdots,\mathbf{x}_d)_{(s,t)}=\frac{\mathbf{x}_s^T\Gamma_0\mathbf{x}_t}{\sqrt{\mathbf{x}_s^T\Gamma_0\mathbf{x}_s}\sqrt{\mathbf{x}_t^T\Gamma_0\mathbf{x}_t}},\quad R_j(\mathbf{x}_1,\cdots,\mathbf{x}_d)_{(s,t)}=\frac{\mathbf{x}_s^T\Sigma_j\mathbf{x}_t}{\sqrt{\mathbf{x}_s^T\Sigma_j\mathbf{x}_s}\sqrt{\mathbf{x}_t^T\Sigma_j\mathbf{x}_t}}
\end{equation*}
Let $\mathscr{T}$ denote the transformation induced by the standard logistic function, i.e. $x\mapsto\varphi(x)$, which is strictly increasing, and define $\mathscr{V}_{\mathcal{X}}^{\eta_{\ell}}\coloneqq\mathscr{T}\circ\mathscr{V}_{\mathcal{X}}^{V_{\ell}}$, $\mathscr{C}_{\mathcal{X}}^{\eta_{\ell}}\coloneqq \mathscr{C}_{\mathcal{X}}^{V_{\ell}}$. Consequently, it is easy to check that the LSBP prior fits in the definition of DSBP prior given in \citet{Barrientos2012}. Specifically, the LSBP prior can be written as $DSBP(\mathscr{C}_{\mathcal{X},\mathbb{N}}^{\eta},\mathscr{C}_{\mathcal{X}}^{\theta},\mathscr{V}_{\mathcal{X},\mathbb{N}}^{\eta},G^0_{\mathcal{X}})$.

Now, consider the mixing kernel of (\ref{eq:LSBPcontinuous}), given by
\begin{equation*}
\chi(\mathbf{z}\mid \boldsymbol{\theta})=\prod_{j=1}^{C-1}\frac{e^{-(z_j-\theta_j)}}{(1+e^{-(z_j-\theta_j)})^2}.
\end{equation*}
We check that the kernel satisfies conditions (i) $-$ (iv) of \citet[][Theorem~5]{Barrientos2012}. 
Letting $\chi_j(z)=\frac{e^{-(z-\theta_j)}}{(1+e^{-(z-\theta_j)})^2}$, we have 
\begin{equation}
\chi_j^{\prime}(z)=-\frac{e^{-(z-\theta_j)}(1-e^{-(z-\theta_j)})}{(1+e^{-(z-\theta_j)})^3} .
\label{eq:derivlogit}
\end{equation}
Obviously, $\chi(\mathbf{z}\mid \boldsymbol{\theta})$ is continuous and strictly 
positive on $\mathbb{R}^{C-1}$. It is bounded above by $\frac{1}{4^{C-1}}$ because from (\ref{eq:derivlogit}), each $\chi_j(z)$ takes its maximum $\frac{1}{4}$ at $z=\theta_j$. 
Condition (i) is satisfied. In addition, since $\chi_j(z)$ decreases as $z>\theta_j$, 
we can choose $l_1=\sqrt{\sum_{j=1}^{C-1}\theta_j^2}$ such that 
$\chi(\mathbf{z}\mid \boldsymbol{\theta})$ decreases as $\mathbf{z}$ outside the 
ball $\{\mathbf{z}: \|\mathbf{z}\|<l_1\}$. Condition (ii) is also satisfied. 
As for condition (iii), it is satisfied because $\chi_j^{\prime}(z)/\chi(z)\to -1$ 
as $z\to \infty$, and $\chi_j^{\prime}(z)/\chi_j(z)\to 1$ as $z\to-\infty$. Finally, 
condition (iv) is obviously satisfied. Moreover, $\mathscr{C}_{\mathcal{X}}^{\eta_{\ell}}$ 
and $\mathscr{C}_{\mathcal{X}}^{\theta}$ are copulas with positive density on the 
appropriate unitary hyper-cubes. Based on \citet[][Theorem~10]{Barrientos2012}, 
$\{f^0_{\mathbf{x}}:\mathbf{x}\in\mathcal{X}\}$ possess the KL property relative 
to $\mathcal{F}_{\mathbf{x}}$. Consequently, by Lemma \ref{lem:klsupport}, the 
induced distributions on discrete space $\{p^0_{\mathbf{x}}:\mathbf{x}\in\mathcal{X}\}$ 
possess the KL property relative to $\mathcal{P}_{\mathbf{x}}$.

To complete the proof, we finally consider the required regularity conditions such that 
$p^0_{\mathbf{x}}$ is in the KL support of $\mathcal{P}_{\mathbf{x}}$. We notice that the 
regularity conditions for continuous distributions are not necessary in our ordinal regression
setting. In fact, the only condition we need for $p^0_{\mathbf{x}}$ is that it comprises
strictly positive probabilities for all response categories. To obtain this result, we first 
show that there exists a specific $f^0_{\mathbf{x}}$ that enables the connection with 
$p^0_{\mathbf{x}}$ given in (\ref{eq:contmaptodisc}). Then, we show that such $f^0_{\mathbf{x}}$
satisfies the regularity conditions (v) to (viii) of \citet[][Theorem~5]{Barrientos2012}.

For a generic probability mass $p^0_{\mathbf{x}}$ on $\{1,\ldots,C\}$, without loss of generality,
we assume $p^0_{\mathbf{x}}$ is positive hereinafter. We define
\begin{equation}
    f_{\mathbf{x}}^0(\mathbf{z})=
    \sum_{y=1}^C\frac{p_{\mathbf{x}}^0(y)\mathbf{1}_{\mathcal{R}_y}
    (\mathbf{z})\phi(\mathbf{z})}
    {\int_{\mathcal{R}_y}\phi(\mathbf{z})\,\text{d}\mathbf{z}},
    \label{eq:existcont}
\end{equation}
where $\phi(\mathbf{z})$ denotes the p.d.f. of the standard $C-1$ dimensional normal 
distribution. It is straightforward to check that $\int_{\mathcal{R}_y}f_{\mathbf{x}}^0(\mathbf{z})\,\text{d}\mathbf{z}=p_{\mathbf{x}}^0(y)$, 
for $y=1,\ldots,C$. Hence, the relationship defined in (\ref{eq:contmaptodisc}) is satisfied.

We now show that this specific $f_{\mathbf{x}}^0(\mathbf{z})$ satisfies the regularity 
conditions (v) to (viii) of \citet[][Theorem~5]{Barrientos2012}. To proceed, we notice 
that because $\phi(\mathbf{z})$ is symmetric around the origin, we have
\begin{equation}
    \frac{1}{2^{C-1}}\leq \int_{\mathcal{R}_y}\phi(\mathbf{z})\,\text{d}\mathbf{z}\leq \frac{1}{2},\quad y=1,\ldots,C
    \label{eq:fzbound}
\end{equation}
where $\{\mathcal{R}_y:y=1,\ldots,C\}$ is the partition defined in equation (10) of the 
main paper. Hence, condition (v) follows directly from (\ref{eq:fzbound}). For condition (vi), 
because $\log(f_{\mathbf{x}}^0(\mathbf{z}))$ is also bounded, we have
\begin{equation*}
    \int f_{\mathbf{x}}^0(\mathbf{z})\log(f_{\mathbf{x}}^0(\mathbf{z}))\,\text{d}\mathbf{z}=\sum_{y=1}^c\frac{p_{\mathbf{x}}^0(y)}{\int_{\mathcal{R}_y}\phi(\mathbf{z})\,\text{d}\mathbf{z}}\int_{\mathcal{R}_y}\log(f_{\mathbf{x}}^0(\mathbf{z}))\phi(\mathbf{z})\,\text{d}\mathbf{z}<\infty.
\end{equation*}
To show that condition (vii) holds, let $\mathcal{B}$ denote the set of boundaries for the 
partition $\{\mathcal{R}_y:y=1,\cdots,C\}$. The set $\mathcal{B}$ has measure 0, and outside 
$\mathcal{B}$ the function $\log(f_{\mathbf{x}}^0(\mathbf{z})/h_{\delta}(\mathbf{z}))$ is bounded, 
where $h_{\delta}(\mathbf{z})=\inf_{\|\mathbf{z}^{\prime}-\mathbf{z}\|<\delta}f_{\mathbf{x}}^0(\mathbf{z}^{\prime})$. Therefore, we can obtain
\begin{equation*}
\begin{split}
\int f_{\mathbf{x}}^0 \log(f_{\mathbf{x}}^0(\mathbf{z})/h_{\delta}(\mathbf{z}))\,\text{d}\mathbf{z}&=\int_{\mathcal{B}}f_{\mathbf{x}}^0 \log(f_{\mathbf{x}}^0(\mathbf{z})/h_{\delta}(\mathbf{z}))\,\text{d}\mathbf{z}+\int_{\mathcal{B}^{C}}f_{\mathbf{x}}^0 \log(f_{\mathbf{x}}^0(\mathbf{z})/h_{\delta}(\mathbf{z}))\,\text{d}\mathbf{z}\\
        &=\sum_{y=1}^C\frac{p_{\mathbf{x}}^0(y)}{\int_{\mathcal{R}_y}\phi(\mathbf{z})\,\text{d}\mathbf{z}}
        \int_{\mathcal{B}^{C}\cap \mathcal{R}_y}\log(f_{\mathbf{x}}^0(\mathbf{z})/h_{\delta}(\mathbf{z}))\phi(\mathbf{z})\,\text{d}\mathbf{z}<\infty .
    \end{split}
\end{equation*}
Finally, condition (viii) can be verified using the fact that the tails of $\log(\chi(\mathbf{z}))$ 
behave like $|\mathbf{z}|$.

Hence, we have proved that the set of generic, positive probability mass functions
$\{p_{\mathbf{x}}^0:\mathbf{x}\in\mathcal{X}\}$ on $\{1,\ldots,C\}$ possesses the 
KL property with respect to the proposed nonparametric prior model.

Similarly, we can establish the KL property for the nonparametric prior induced by 
the two simpler models. First, it is straightforward to check that the common-weights 
and the common-atoms nonparametric priors fall in the wDDP \citep[][Definition~2]{Barrientos2012} 
and $\theta$DSBP \citep[][Definition~4]{Barrientos2012} framework, respectively. 
In addition, the mixing kernel is the same as in Theorem \ref{thm:KLgeneralmodel} of the 
main paper, and therefore the relevant regularity conditions are satisfied as shown 
earlier. Hence, by Theorem 5 and Theorem 10 of \citet{Barrientos2012}, and 
Lemma \ref{lem:klsupport} of the main paper, we obtain the following corollary.

\begin{corollary}
    Denote by $\text{w}\mathcal{P}_{\mathbf{x}}$ the common-weights LSBP mixture prior 
    discussed in Section \ref{subsec:lddpmodel},
    and $\theta\mathcal{P}_{\mathbf{x}}$ the common-atoms LSBP mixture prior proposed in Section \ref{subsec:comatomlsbpmodel}. Consider $\{p^0_{\mathbf{x}}:\mathbf{x}\in\mathcal{X}\}$, a generic collection of covariate-dependent probabilities for an ordinal response with $C$ categories. Assume that the probability of each response category is strictly positive. Then, the mass functions $\{p^0_{\mathbf{x}}:\mathbf{x}\in\mathcal{X}\}$ are in the KL support of $\text{w}\mathcal{P}_{\mathbf{x}}$ and $\theta\mathcal{P}_{\mathbf{x}}$. 
\end{corollary}

\end{proof}

\subsection{Proof of Proposition \ref{prop:equdiscont}}
\label{subsec:pfeqdiscount}

\begin{proof}
Denote by $\text{Pr}(Y=j \mid \mathcal{M}_1)$ and $\text{Pr}(Y=j \mid \mathcal{M}_2)$, for $j=1,...,C$, 
the $j$-th category response probability under the continuation-ratio logits model and the cumulative 
logit model, respectively. The parameter vector for the former model is $(\theta_1,...,\theta_{C-1})$, 
whereas for the latter it is $(\vartheta,\varkappa_2,...,\varkappa_{C-1})$. Recall that, for 
identifiability, $\varkappa_1=0$.

For the first response probability, we have $\text{Pr}(Y=1\mid \mathcal{M}_1) =$
$e^{\theta_1} / (1+e^{\theta_1})$, and $\text{Pr}(Y=1\mid \mathcal{M}_2) =$
$\text{Pr}(Z \leq 0 \mid \mathcal{M}_2) =$ $e^{-\vartheta} / (1+e^{-\vartheta})$, from which 
we obtain $\vartheta = -\theta_1$. The equality for these two parameters holds true for any value 
of $C$, but this step also establishes the result for the simplest case of $C=2$

For $C=3$, from the argument above, we have $\vartheta = - \theta_1$. Now, under the 
continuation-ratio logits model
\begin{equation}
\text{Pr}(Y=3 \mid \mathcal{M}_1) = \frac{1}{1+e^{\theta_1}} \, \frac{1}{1+e^{\theta_2}},
\label{eq:probcrl3}
\end{equation}
while under the cumulative logit model
\begin{equation}
\text{Pr}(Y=3 \mid \mathcal{M}_2) = \text{Pr}(Z > \varkappa_2 \mid \mathcal{M}_2) = 
\frac{1}{1+e^{\varkappa_2-\vartheta}} = \frac{1}{1+e^{\varkappa_2 + \theta_{1}}}.
\label{eq:probdllv3}
\end{equation}
Setting (\ref{eq:probcrl3}) and (\ref{eq:probdllv3}) equal, and solving for $\varkappa_2$, 
we have
\begin{equation*}
\varkappa_2 = \log(1+e^{\theta_2}+e^{\theta_2-\theta_1}) =
\log(e^{\varkappa_1}+e^{\varkappa_1+\theta_2}+e^{\theta_2-\theta_1}),
\end{equation*}
which establishes the result for $C=3$.

To prove the proposition by induction, assume the correspondence between the parameters of the 
two models holds true for an ordinal response with $C-1$ categories, that is, $\vartheta=-\theta_1$ 
and $\varkappa_j=$ $\log(e^{\varkappa_{j-1}}+e^{\varkappa_{j-1}+\theta_j}+e^{\theta_j-\theta_1})$, 
for $j=2,\ldots,C-2$.

%
%

Now, assume that the ordinal response has $C$ categories. To complete the argument, we work 
with the probability for category $C$. For the continuation-ratio logits model:
\begin{equation*}
\text{Pr}(Y=C \mid \mathcal{M}_1) = \frac{1}{1+e^{\theta_{C-1}}} 
\prod_{k=1}^{C-2} \frac{1}{1+e^{\theta_k}} 
= \frac{1}{1+e^{\theta_{C-1}}} \, \frac{1}{1+e^{\varkappa_{C-2} + \theta_1}}
\label{eq:probcrlC}
\end{equation*}
where the equality $\prod_{k=1}^{C-2} 1 / (1+e^{\theta_k}) =$
$1/(1+e^{\varkappa_{C-2} + \theta_1})$ is obtained by starting from the right-hand side 
and using recursively (for $j = C-2,C-3,...,2$) the induction assumption 
$\varkappa_j=$ $\log(e^{\varkappa_{j-1}}+e^{\varkappa_{j-1}+\theta_j}+e^{\theta_j-\theta_1})$.
On the other hand,  
$$
\text{Pr}(Y=C\mid\mathcal{M}_2) = \text{Pr}(Z > \varkappa_{C-1} \mid\mathcal{M}_2) =
\frac{1}{1+e^{\varkappa_{C-1} - \vartheta}} = \frac{1}{1+e^{\varkappa_{C-1} + \theta_{1}}}.
$$ 
Setting the two probabilities equal to each other, we can solve for $\varkappa_{C-1}$, resulting in 
\begin{equation*}
\varkappa_{C-1}=\log(e^{\varkappa_{C-2}}+e^{\varkappa_{C-2}+\theta_{C-1}}+e^{\theta_{C-1}-\theta_1}),
\end{equation*}
thus completing the induction argument.


Note that we can write $\varkappa_j=$ $\log( A \, e^{\varkappa_{j-1}} + B)$, where 
$A=$ $1 + e^{\theta_j} > 1$, and $B=$ $e^{\theta_j-\theta_1} > 0$, from which 
we can easily confirm the order restriction for the cut-off points, 
$\varkappa_j>\varkappa_{j-1}$, for $j=2,\cdots,C-1$.
\end{proof}

\section{Prior specification strategy}
\label{sec:smpriorspec}

Here, we elaborate on the prior specification example presented at the 
end of Section \ref{subsec:prispecori}. We propose a general prior specification strategy that 
can force the prior expected regression curves to have specific patterns (especially monotonicity)
that reflect available information. The strategy relies on the bounds provided in 
Proposition \ref{prop:logitnormalexpbound}. 

\begin{figure}[t!]
\centering
\begin{subfigure}{.5\textwidth}
  \centering
  \includegraphics[width=8cm,height=4cm]{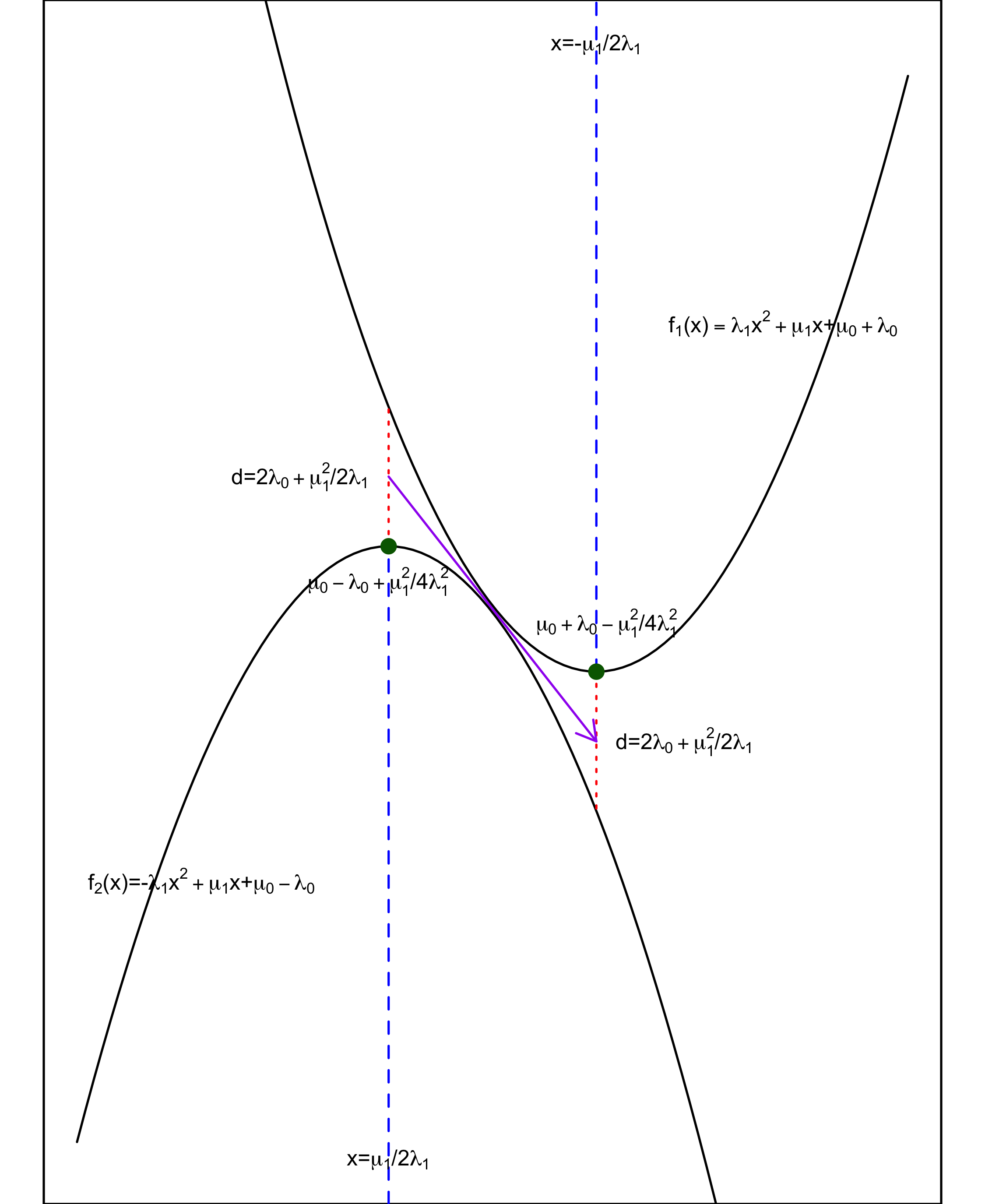}
  \caption{Monotonically decreasing function.}
  \label{fig:subdecprior}
\end{subfigure}%
\begin{subfigure}{.5\textwidth}
  \centering
  \includegraphics[width=8cm,height=4cm]{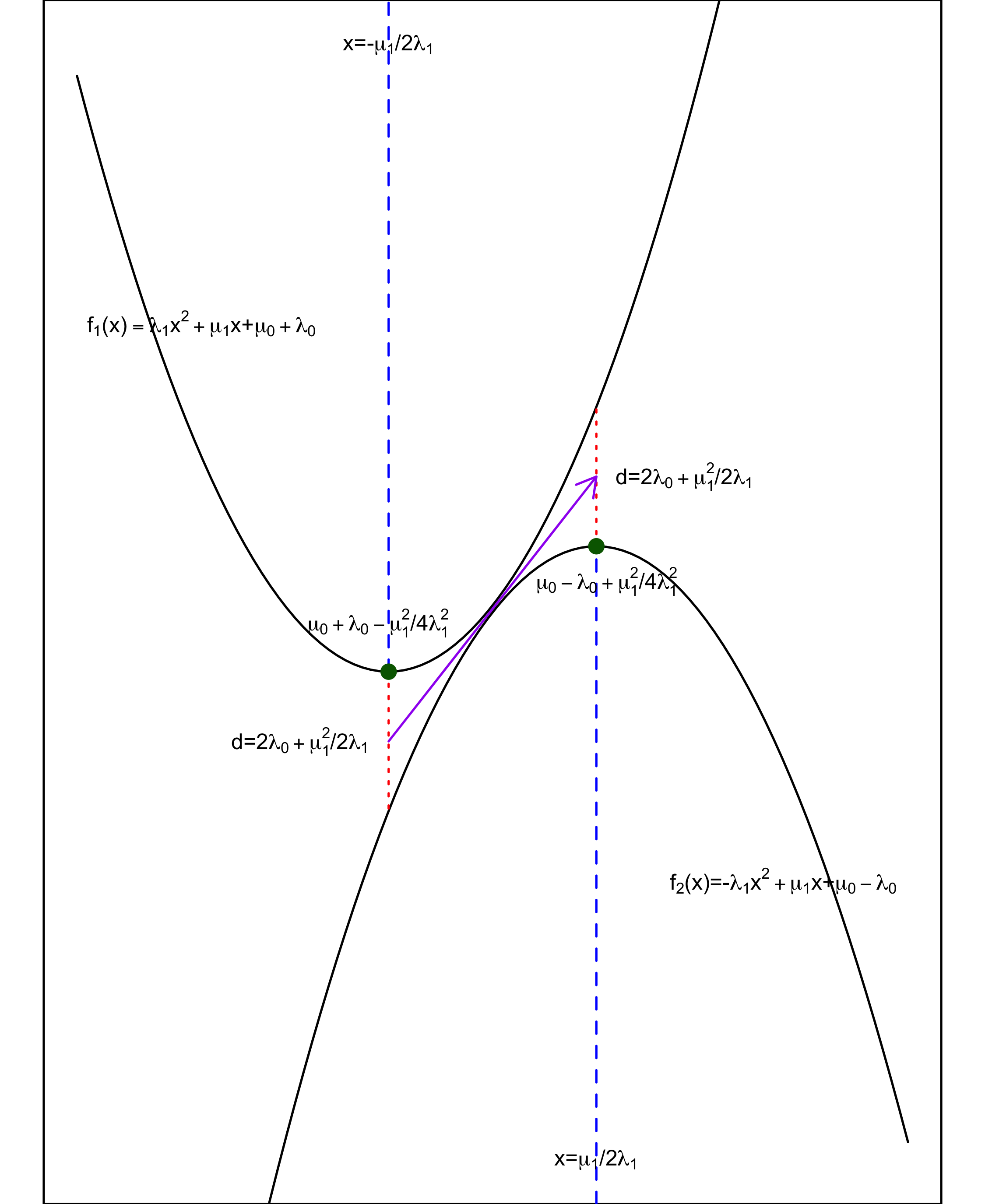}
  \caption{Monotonically increasing function.}
  \label{fig:subincprior}
\end{subfigure}
\caption{\small Illustration of how the two bounds can be used to set the monotonic pattern of the prior expected probability response curve.}
\label{fig:priorexpillustrate}
\end{figure}

As an illustrative example, consider the case when the covariates vector is $\mathbf{x}=(1,x)^T$ and 
the information is available for the first probability response curve. Suppose the prior hyperparameters 
are $\boldsymbol{\mu}_{01}=(\mu_{01,0},\mu_{01,1})^T$ and $\Lambda_{01}=diag(\lambda_{01,0},\lambda_{01,1})$. In such a case, the prior expected first probability response curve $\text{E}(\text{Pr}(\mathbf{Y}=1\mid G_{\mathbf{x}}))=\text{E}(\varphi(\mathbf{x}^T\boldsymbol{\beta}_1))$, where $\mathbf{x}^T\boldsymbol{\beta}_1\sim N(\mu_{01,0}+\mu_{01,1}x,(\kappa_{01}+1)/(\kappa_{01}(\nu_{01}-p-1))(\lambda_{01,0}+\lambda_{01,1}x^2))$. For notation simplicity, let us denote $\mu_s=\mu_{01,s},\lambda_s=(\kappa_{01}+1)/(2\kappa_{01}(\nu_{01}-p-1))\lambda_{01,s}$, $s=0,1$. Then from Proposition \ref{prop:logitnormalexpbound}, $\text{E}(\text{Pr}(\mathbf{Y}=1\mid G_{\mathbf{x}}))$ is bounded by
\begin{equation*}
    \varphi(-\lambda_1x^2+\mu_1x+\mu_0-\lambda_0)\leq \text{E}(\text{Pr}(\mathbf{Y}=1\mid G_{\mathbf{x}}))\leq \varphi(\lambda_1x^2+\mu_1x+\mu_0+\lambda_0)
\end{equation*}
Because the logistic function preserves monotonicity, it is helpful to study the relative position 
of the two parabolas inside. Indeed, we can choose the prior hyperparameters such that the two bounds squeeze a small region. The first prior expected probability response curve pinches through that region, possessing certain monotonicity, illustrated in Figure \ref{fig:priorexpillustrate}.   

Specifically, suppose the prior guess for the first probability response curve is a decreasing 
function with respect to $x$. As shown in Figure \ref{fig:subdecprior}, we can put the range of $x$ inside the two axes of symmetry. In addition, the quantity $d=2\lambda_0+\mu_1^2/2\lambda_1$ determines the maximum difference of the two bounds. The two vertices determine the prior mean at the minimum and maximum value of $x$. To summarize, the parameters $\mu_0,\mu_1,\lambda_0,\lambda_1$ can be specified by the equations
\begin{equation}
    \left\{\begin{aligned}&\frac{\mu_1}{2\lambda_1}=a_1,\quad -\frac{\mu_1}{2\lambda_1}=-a_1\\
    &2\lambda_0+\frac{\mu_1^2}{2\lambda_1}=a_2\\
    &\mu_0+\lambda_0-\frac{\mu_1^2}{4\lambda_1}=-a_3\\
    &\mu_0-\lambda_0+\frac{\mu_1^2}{4\lambda_1}=a_4\end{aligned}
    \right.\Longleftrightarrow \left\{\begin{aligned}&\mu_0=\frac{a_4-a_3}{2}\\
    &\mu_1=-\frac{a_2+a_3+a_4}{2a_1}\\
    &\lambda_0=\frac{a_2-a_3-a_4}{4}\\
    &\lambda_1=\frac{a_2+a_3+a_4}{4a_1^2}\end{aligned}
    \right.  
\label{eq:specpriordec}    
\end{equation}
with positive numbers $a_1,a_2,a_3,a_4$ chosen based on the prior information. 
Note that $\lambda_0$ should be positive, so it imposes the constraint $a_2>a_3+a_4$ on the choice of these four numbers. Using (\ref{eq:specpriordec}), we can specific the prior hyperparameters $\mu_{01,0},\mu_{01,1},\lambda_{01,0},\lambda_{01,1}$. The same strategy can be extended for 
the monotonically increasing case. 

To specify $\boldsymbol{\mu}_{0j}$ and $\Lambda_{0j}$ for $j>1$, we can sequentially implement this strategy. Furthermore, if the dimension of covariates $p>2$, it becomes more difficult to specify hyperparameters, but the same strategy can be applied by considering each covariate $x_s,s=1,\cdots,p$ marginally while fixing $x_{s^{\prime}},s^{\prime}\neq s$.

We specify the prior hyperparameters of the illustrative example in Section \ref{subsec:prispecori} by the proposed strategy. Suppose the prior information we want to incorporate is that the probability response curve for the first category decreases from 1 to 0 while the probability response curve for the second category increases from 0 to 1 in the region $(-10,10)$. For the first decreasing probability curve, we set $a_1=a_2=10,a_3=6,a_4=2$ to specify $\boldsymbol{\mu}_{01}$ and $\Lambda_{01}$. 
As for the second probability curve, since 
$\text{E}(\text{Pr}(\mathbf{Y}=2\mid G_{\mathbf{x}}))=[1-\text{E}(\text{Pr}(\mathbf{Y}=1\mid G_{\mathbf{x}}))]\text{E}[\varphi(\mathbf{x}^T\boldsymbol{\beta}_2)]$ and utilizing the 
specified monotonicity for $\text{E}(\text{Pr}(\mathbf{Y}=1\mid G_{\mathbf{x}}))$, 
we focus on $\text{E}[\varphi(\mathbf{x}^T\boldsymbol{\beta}_2)]$. To force a increasing trend, we further choose $\boldsymbol{\mu}_{02}$ and $\Lambda_{02}$ by applying the strategy for the increasing case with same setting on $a_1$ to $a_4$. After simple algebra, we obtain the following prior hyperaprameters:
\begin{equation*}
    \boldsymbol{\mu}_{01}=(-2,-0.9)^T,\,\,\boldsymbol{\mu}_{02}=(-2,0.9)^T,\,\,\Lambda_{01}=\Lambda_{02}=\begin{pmatrix} 0.8 & 0 \\ 0 & 0.072\end{pmatrix}
\end{equation*}
This set of prior hyperparameters leads to Figure \ref{fig:priorpiest} in the main paper.

\section{MCMC diagnostics}
\label{sec:mcmcdiag}

Here, we provide some results from assessing convergence and performance of the MCMC algorithm. 
For simplicity in the presentation of results, we focus on the general LSBP mixture model for 
the first synthetic data example (the one presented in the main paper), with sample size $n=800$. 
For this example, a total of 4000 MCMC samples were taken from a chain of length 30000. The first 
10000 were discarded as burn in, and the remaining draws were thinned to reduce autocorrelation. 

\begin{figure}[t!]
\centering
\includegraphics[width=16cm,height=4cm]{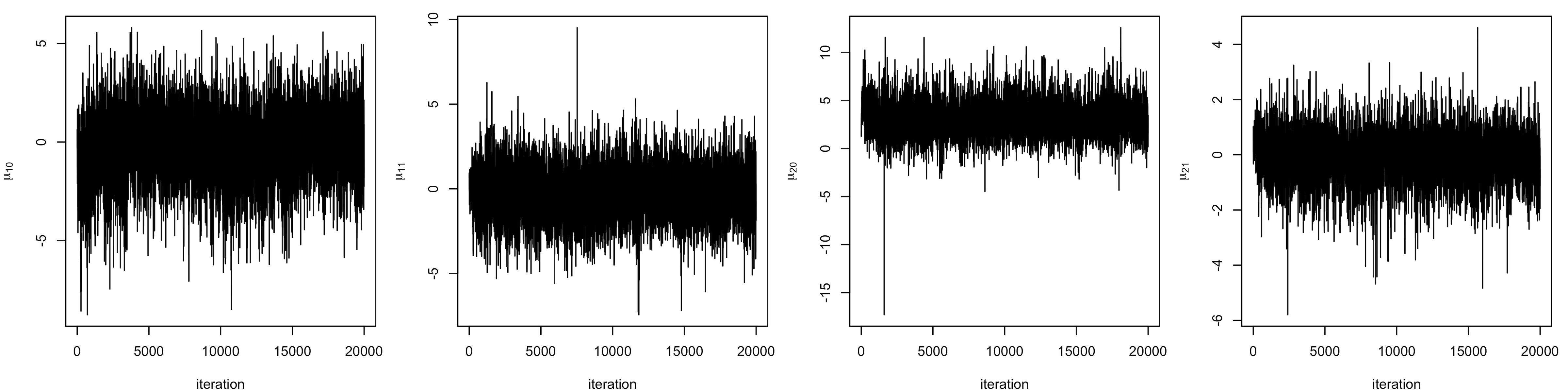}  
\caption{\small MCMC diagnostics for the first synthetic data example. Trace plots of 
the first 20000 posterior samples for the elements of $\boldsymbol{\mu}_j$, $j=1,2$.}
\label{fig:MCMCdiagmu}
\end{figure}

\begin{figure}[t!]
\centering
\includegraphics[width=16cm,height=8cm]{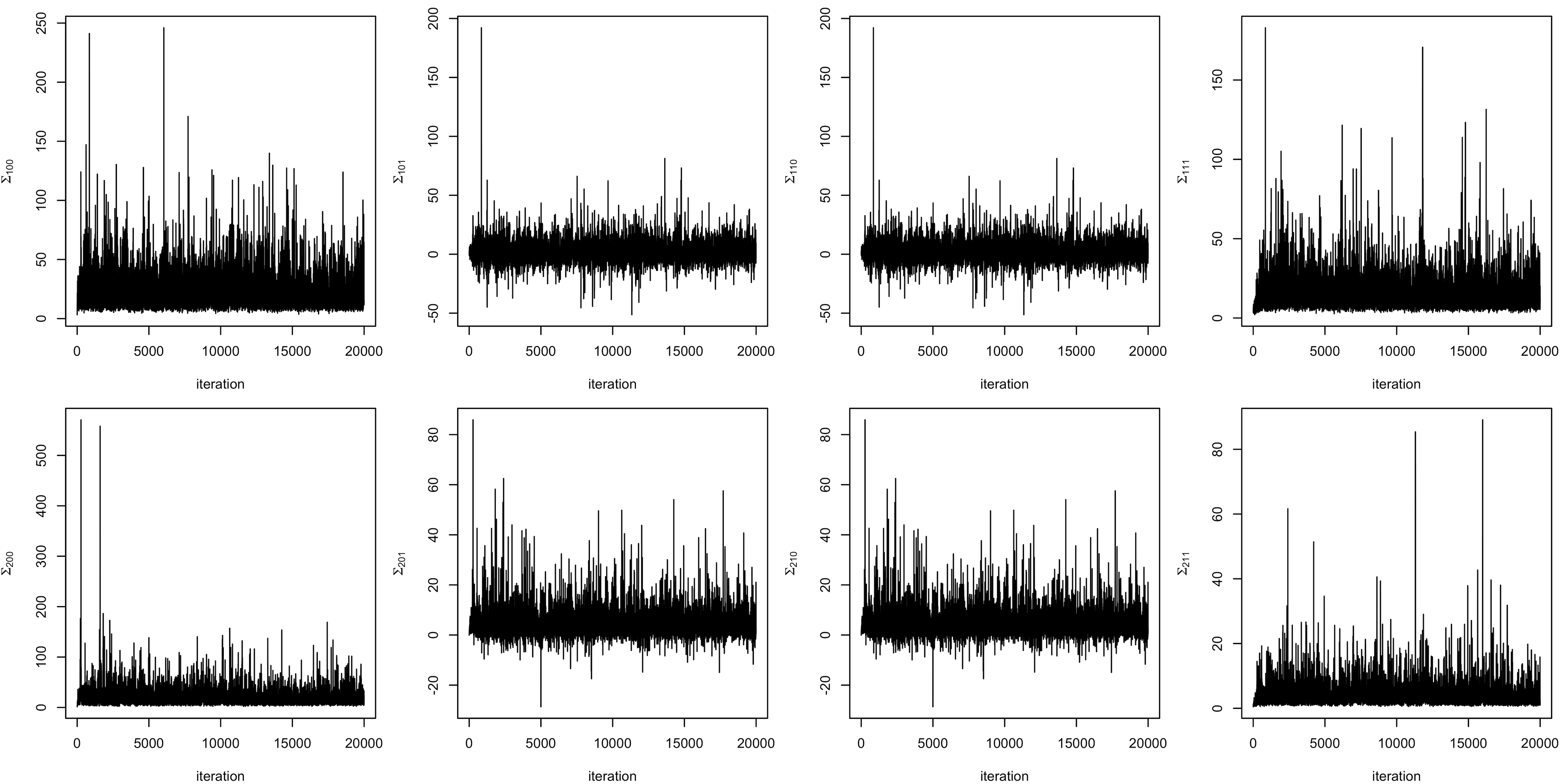}  
\caption{\small MCMC diagnostics for the first synthetic data example. Trace plots of 
the first 20000 posterior samples for the elements of $\Sigma_j$, $j=1,2$.}
\label{fig:MCMCdiagSigma}
\end{figure}

Figure \ref{fig:MCMCdiagmu} and Figure \ref{fig:MCMCdiagSigma} show the trace plots of the 
first 20000 posterior samples for the elements of $\boldsymbol{\mu}_j$ and $\Sigma_j$, $j=1,2$, 
respectively. The plots suggest that convergence is achieved quickly for these parameters.

\begin{figure}[t!]
\centering
\includegraphics[width=16cm,height=4cm]{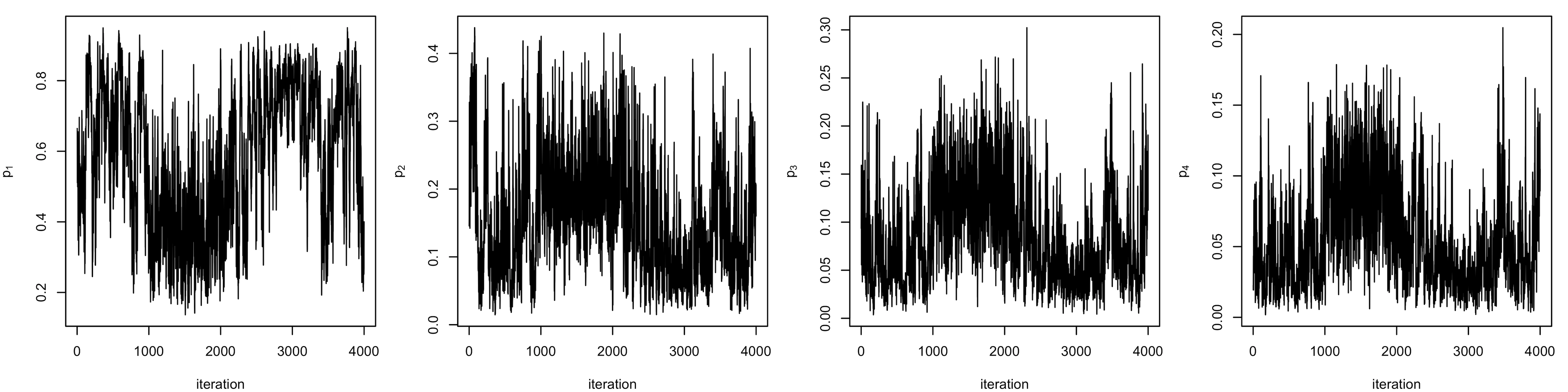}  
\caption{\small MCMC diagnostics for the first synthetic data example. Trace plots of the 4000 remaining posterior samples for the four largest elements of the mixing weight vector.}
\label{fig:MCMCdiagweight}
\end{figure}

The remaining trace plots focus on the 4000 MCMC samples collected after burn in and 
thinning. Figure \ref{fig:MCMCdiagweight} shows the trace plots for the largest 4 elements 
of the mixing weight vector $\{p_{i\ell}:\ell=1,\ldots,L\}$, for a randomly selected $i$. 
Specifically, at each MCMC iteration the largest element of the vector of mixture weights 
is found, and this corresponds to $p_1$ in the plot. Similarly, 
the second largest element at each iteration is found, and this corresponds to $p_2$ in the plot,
with $p_3$ and $p_4$ obtained in the same fashion.

\begin{figure}[t!]
\centering
\includegraphics[width=16cm,height=4cm]{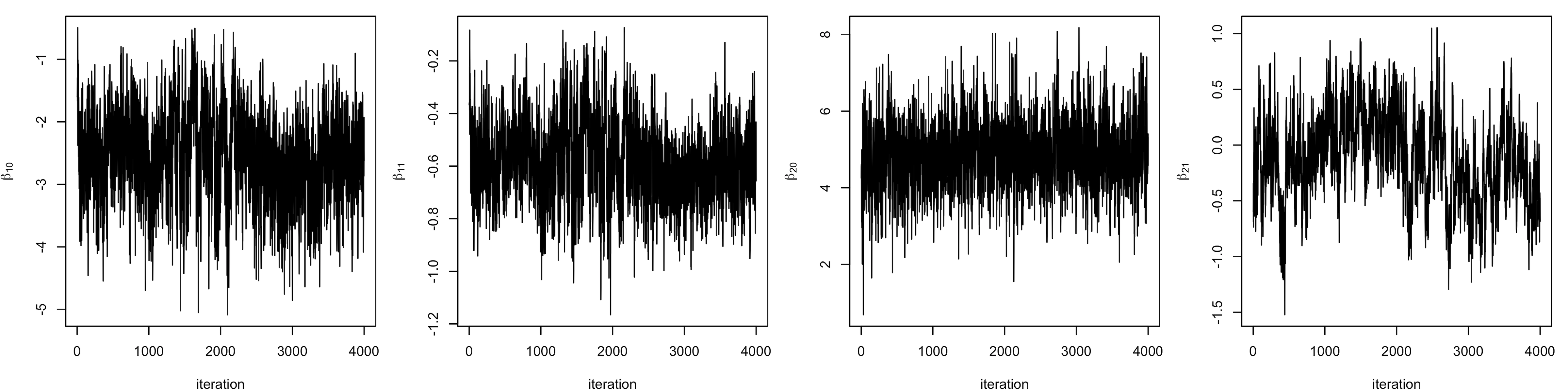}  
\caption{\small MCMC diagnostics for the first synthetic data example. Trace plots of the 4000 
remaining posterior samples for the label-invariant functions of $\boldsymbol{\beta}_{j\ell}$.}
\label{fig:MCMCdiagbeta}
\end{figure}

\begin{figure}[t!]
\centering
\includegraphics[width=16cm,height=4cm]{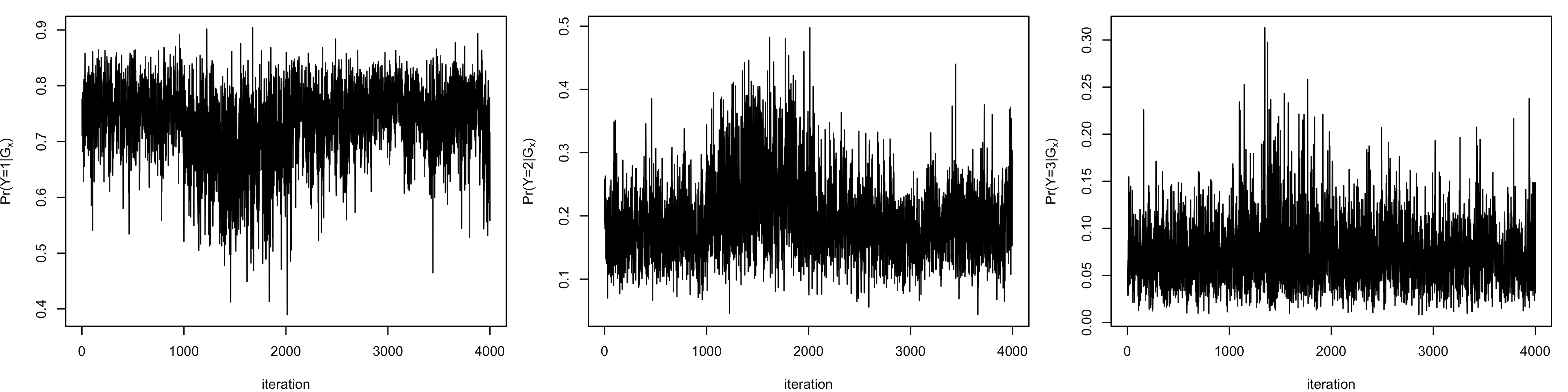}  
\caption{\small MCMC diagnostics for the first synthetic data example. Trace plots of the 4000 
remaining posterior samples for the response probabilities $\text{Pr}(Y=j\mid G_{\mathbf{x}})$, 
$j=1,2,3$, corresponding to a randomly selected $\mathbf{x}$.}
\label{fig:MCMCdiagprob}
\end{figure}

Because of the label switching problem in mixture models, we turn to label-invariant 
functions of parameters $\boldsymbol{\beta}_{j\ell}$ to diagnose convergence and mixing for these 
parameters. We use $\sum_{i=1}^n\mathbf{\beta}_{j\mathcal{L}_i}/n$, where 
$\mathcal{L}_i$ represents the configuration variable corresponding to subject $i$. This quantity 
can be interpreted as the average of the coefficient vectors to which the observations are 
assigned at a particular MCMC iteration, or a weighted average of the $\boldsymbol{\beta}_{j\ell}$
parameters. It is commonly used to assess convergence in mixture models (see, e.g., Antoniano-Villalobo and Walker, 2016). 
Figure \ref{fig:MCMCdiagbeta} shows the trace plots for 
the components of the label-invariant functions of $\boldsymbol{\beta}_{j\ell}$, which also 
suggests convergence of the Markov chain. 
Moreover, Figure \ref{fig:MCMCdiagprob} displays the trace plots for the estimated response 
probabilities $\text{Pr}(Y=j\mid G_{\mathbf{x}})$, $j=1,2,3$, for a randomly chosen $\mathbf{x}$. 

Finally, to provide information on computing time, we report the time in seconds for producing 
1000 posterior samples of all the model parameters. We consider both sample sizes for the first 
synthetic data example (the one presented in the main paper), truncation level $L=30$ or $L=50$, 
and all three LSBP mixture models. The models were programmed in R and run on a laptop with Apple 
M1 pro 3.2 GHz. The results are summarized in Table \ref{tab:computingtime}. Note that the true 
probability response curves have non-standard shapes under this simulation scenario, resulting 
in more effective mixing components under the common-atoms model. As a consequence, computing
time is longer for the common-atoms model because of the additional calculations in the MCMC 
algorithm for sampling a mixing component that has been assigned to some subjects in the data. 

\begin{table}[t!] \centering
\small
\caption{First synthetic data example. Computing time (in seconds) for obtaining 1000 posterior 
samples under each scenario.} 
\label{tab:computingtime}
\begin{tabular}{ccccc}
\hline
\hline
\multirow{2}{*}{Truncation level} & \multirow{2}{*}{Sample Size} & \multicolumn{3}{c}{Model} \\
\cline{3-5} & & Common-weights & Common-atoms & General \\
\hline
\hline
\multirow{2}{*}{$L=30$} & $n=200$ & 33.64 & 83.92 & 56.32 \\
& $n=800$ & 124.12 & 289.73 & 174.28 \\
\hline
\multirow{2}{*}{$L=50$} & $n=200$ & 51.50 & 138.29 & 89.24 \\
& $n=800$ & 189.26 & 492.34 & 294.21  \\
\hline
\hline
\end{tabular}
\end{table}

\section{Additional results for data examples}
\label{sec:smdataexample}

\subsection{First synthetic data example}
\label{subsec:SMfirst}

\begin{figure}[t!]
\centering
\begin{subfigure}{\textwidth}
  \centering 
  \includegraphics[width=16cm,height=3cm]{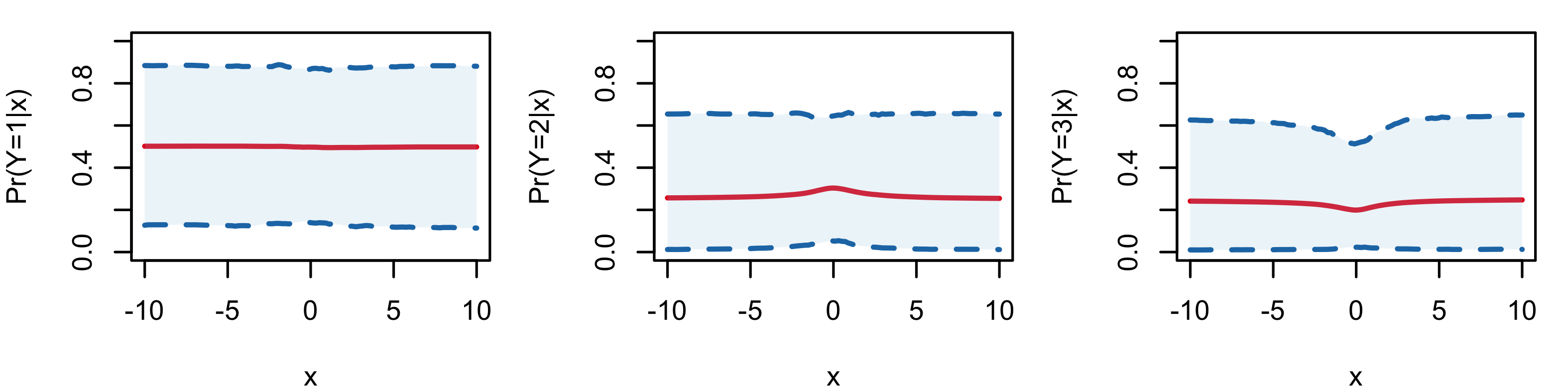}  
  \caption{Common-weights model.}
  \label{subfig:sim2priorspeclddp}
\end{subfigure}
\begin{subfigure}{\textwidth}
  \centering 
  \includegraphics[width=16cm,height=3cm]{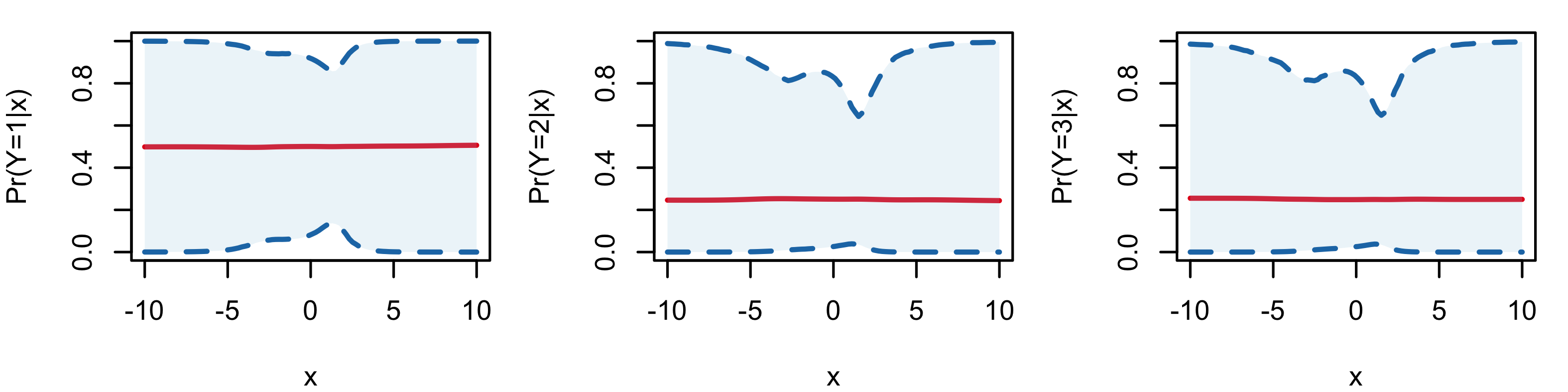}  
  \caption{Common-atoms model.}
  \label{subfig:sim2priorspeclddp}
\end{subfigure}
\begin{subfigure}{\textwidth}
  \centering
  \includegraphics[width=16cm,height=3cm]{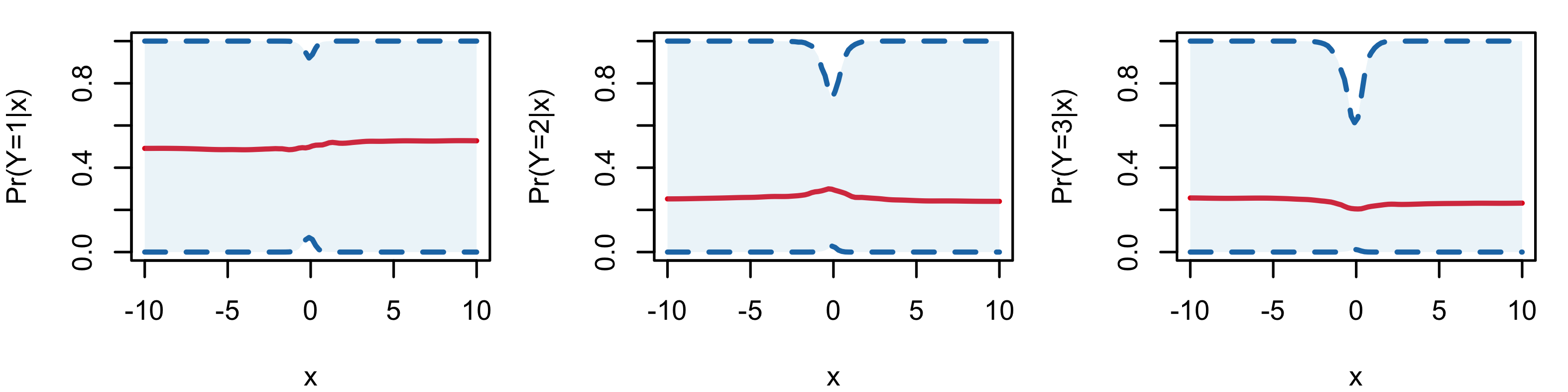}   
  \caption{General model.}
  \label{subfig:sim2priorspecgen}
\end{subfigure}
\caption{\small First simulation example. In each panel, the red solid line corresponds 
to the prior expected probability response curve, the blue dashed lines and shaded region 
indicate the prior 95\% interval estimate.}
\label{fig:sim2prior}
\end{figure}

The prior  hyperparameters are specified to yield a fairly noninformative prior such 
that the three nonparametric models provide similar prior point and interval estimates 
of the probability response curves. As a consequence, the difference in posterior estimates 
should be driven by the difference in the mixing structure. More specifically, we set the 
LSBP prior hyperparameters $(\boldsymbol{\gamma}_0,\Gamma_0)$ to favor a priori enough 
mixture components over the covariate space. The hyperparameters $(a_{\alpha},b_{\alpha})$ 
in the common-weights model are set accordingly to favor a comparable number of mixing 
components. In addition, the hyperparameters corresponding to the atoms are set as the 
baseline choice. Figure \ref{fig:sim2prior} displays the prior point and interval estimates 
under the proposed models. The three subfigures display the same pattern: the prior mean 
estimates are flat, and the prior $95\%$ interval estimates span a substantial portion 
of the unit interval.

Under the displayed prior, the posterior estimates are shown in Figure \ref{fig:sim2postgeneral} 
of the main paper. 
To further investigate how the proposed models behave in capturing non-standard probability 
response curves, we conduct a formal model comparison using the posterior predictive loss 
criterion \citep{GelfandGhosh1998}. The criterion contains a goodness-of-fit term and a penalty 
term. Since the response variable $\mathbf{Y}$ is multivariate, we consider the posterior predictive 
loss for every entry of it. Specifically, let $\mathbf{Y}_i^*$ denote the replicate response drawn 
from the posterior predictive distribution. Then, the goodness-of-fit term is defined as
$G_j(\mathcal{M})=$ $\sum_{i=1}^n[\mathbf{Y}_{ij}-\text{E}^{\mathcal{M}}
(\mathbf{Y}^*_{ij} \mid \text{data})]^2$, whereas the penalty term is defined 
as $P_j(\mathcal{M})=\sum_{i=1}^n\text{Var}^{\mathcal{M}}(\mathbf{Y}^*_{ij} \mid \text{data})$, 
for $j=1,\ldots,C$. The results are summarized in Table \ref{tab:sim2summary}. The two 
models with covariate-dependent weights outperform the common-weights model. The common-atoms 
model and the general model are comparable in terms of goodness of fit. However, the 
common-atoms model activates more components to compensate for the constant atoms, 
resulting in a larger penalty.   



\begin{table}[t!]
\centering 
\caption{First simulation example. Summary of model comparison using the posterior 
predictive loss criterion. The values corresponding to the best model are given in bold.} 
\label{tab:sim2summary}
\begin{tabular}{ccccccc} 
\hline \hline 
Model & $G_1(\mathcal{M})$ & $P_1(\mathcal{M})$ & $G_2(\mathcal{M})$ & $P_2(\mathcal{M})$ & $G_3(\mathcal{M})$ & $P_3(\mathcal{M})$\\ 
\hline 
Common-weights & 132.59 & 141.27 & 72.76 & 83.54 & 136.40 & 141.91\\
Common-atoms & 90.52 & 103.84 & 65.79 & 82.25 & 89.45 & 107.64 \\
General & \textbf{89.96} & \textbf{96.25} & \textbf{64.14} & \textbf{72.56} & \textbf{88.24} & \textbf{94.39} \\
\hline 
\hline 
\end{tabular} 
\end{table}

\subsection{Second synthetic data example}
\label{subsec:SMsecond}

We generate $n=100$ responses from a probit model, that is, we first sample normally distributed
latent continuous variables $\tilde{y}_i$, and then discretize the $\tilde{y}_i$ with cut-off points 
to get the ordinal responses $\mathbf{Y}_i$, for $i=1,\ldots,n$. The covariates, $(1,x_i)^{T}$, 
with the $x_i$ sampled from the $Unif(-10,10)$ distribution, enter through the mean of the 
normal distribution for the latent variables. The objective is to study how 
the different models handle the challenge of recovering standard regression relationships for 
which the nonparametric mixture model structure is not necessary.

\begin{figure}[t!]
\centering
\begin{subfigure}{\textwidth}
  \centering 
  \includegraphics[width=15.5cm,height=3cm]{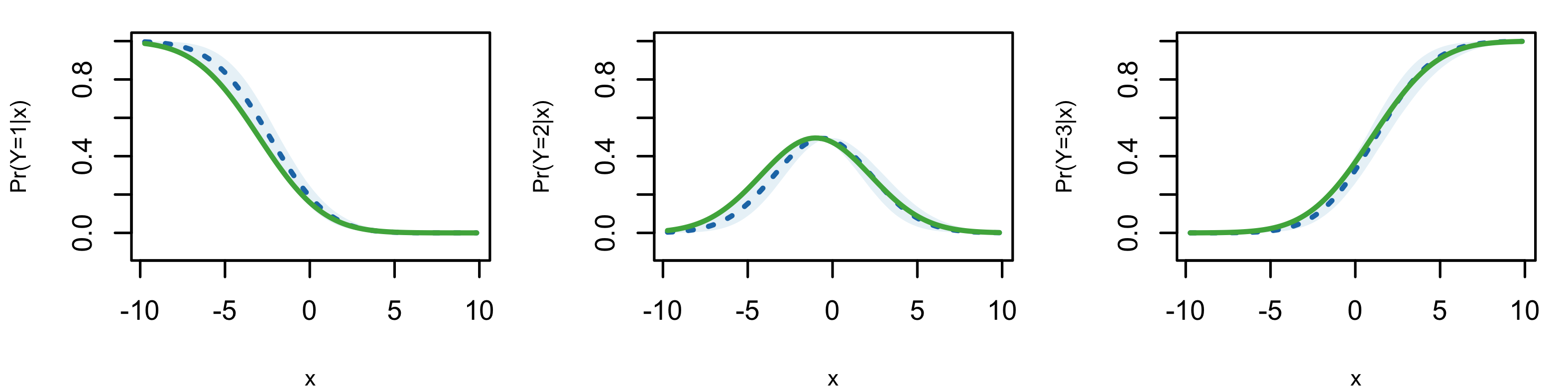}  
  \caption{Probit model.}
  \label{subfig:sim1lddppost}
\end{subfigure}
\begin{subfigure}{\textwidth}
  \centering 
  \includegraphics[width=15.5cm,height=3cm]{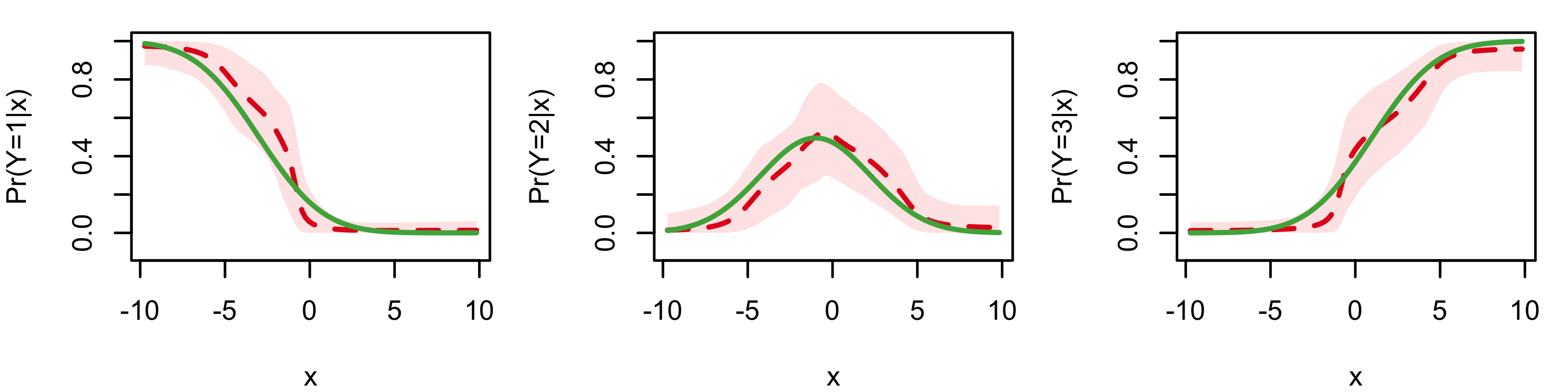}  
  \caption{Common-weights model.}
  \label{subfig:sim1lddppost}
\end{subfigure}
\begin{subfigure}{\textwidth}
  \centering
  \includegraphics[width=15.5cm,height=3cm]{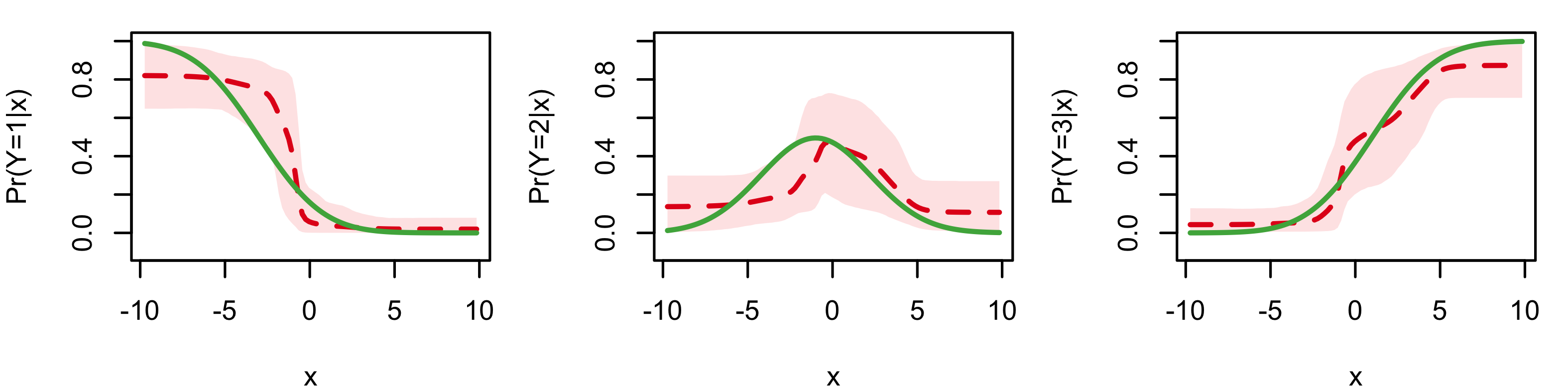}  
  \caption{Common-atoms model.}
  \label{subfig:sim1lsbppost}
\end{subfigure}
\begin{subfigure}{\textwidth}
  \centering
  \includegraphics[width=15.5cm,height=3cm]{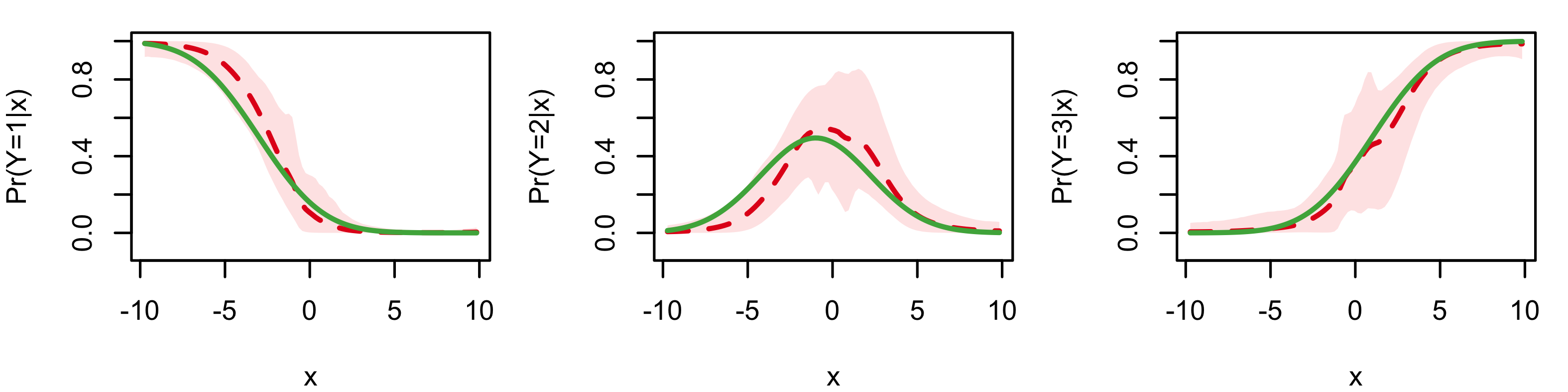}   
  \caption{General model.}
  \label{subfig:sim1generalpost}
\end{subfigure}
\caption{
{\small Second simulation example. Inference results for the marginal probability 
response curves. 
In each panel, the dashed line and shaded region correspond to the posterior mean and 
$95\%$ credible interval estimates, whereas the (green) solid line denotes the true 
regression function.}
}
\label{fig:sim1post}
\end{figure}

The nonparametric mixture models are applied to the data, using the (non-informative)
baseline prior for their hyperparameters. Figure \ref{fig:sim1post} plots posterior point 
and interval estimates for the marginal probability response curves, including, as a 
reference point, estimates under the parametric probit model used to generate the data. 
As expected, the nonparametric models result in wider posterior uncertainty bands than the parametric model. In terms of recovering the underlying regression curves, the common-atoms 
model is less effective than the common-weights and the general model. As discussed in 
Section \ref{subsec:comatomlsbpmodel}, this can be explained from the common-atoms model 
property that the regression curve shapes are adjusted essentially only through the mixture weights. The findings from the graphical comparison are supported by results from formal comparison, using the aforementioned posterior predictive loss criterion. The results, summarized in Table \ref{tab:sim1summary}, suggests comparable 
performance for the common-weights and general models, whereas both outperform the common-atoms model.

\begin{table}[t!]
\centering 
\caption{Second simulation example. Summary of model comparison using the posterior predictive loss criterion. The values correspond to the best model are given in bold.} 
\label{tab:sim1summary}
\begin{tabular}{ccccccc} 
\hline \hline 
Model & $G_1(\mathcal{M})$ & $P_1(\mathcal{M})$ & $G_2(\mathcal{M})$ & $P_2(\mathcal{M})$ & $G_3(\mathcal{M})$ & $P_3(\mathcal{M})$\\ 
\hline 
Common-weights & \textbf{6.94} & 7.94 & 12.95 & 13.59 & \textbf{8.41} & 9.00\\
Common-atoms & 7.35 & 9.73 & 13.76 & 15.43 & 8.73 & 11.99 \\
General & 7.26 & \textbf{7.38} & \textbf{12.94} & \textbf{12.78} & 8.51 & \textbf{8.84} \\
\hline 
\hline 
\end{tabular} 
\end{table}

%
%

\begin{figure}[t!]
\centering
\begin{subfigure}{\textwidth}
  \centering 
  \includegraphics[width=15.5cm,height=3cm]{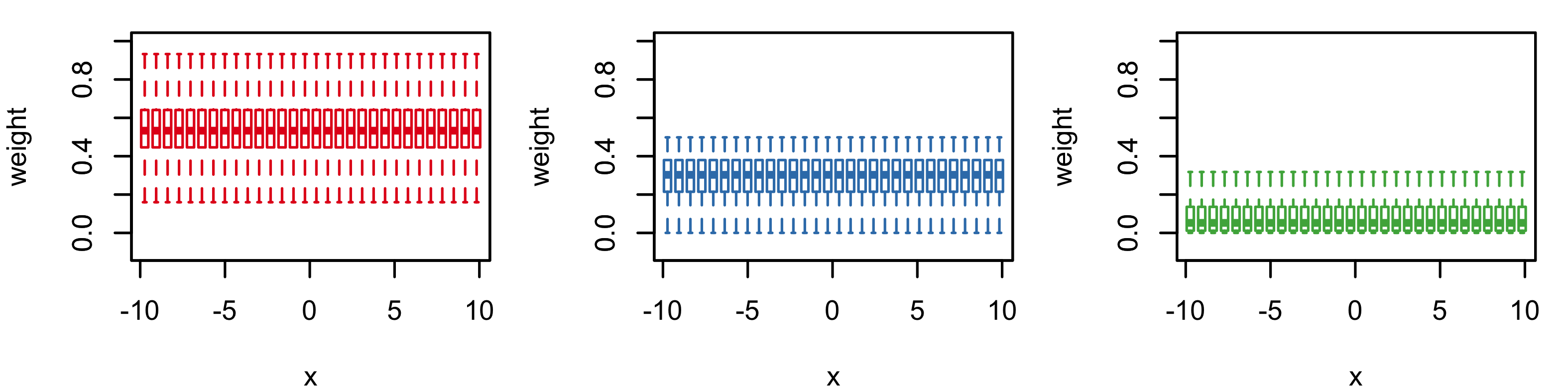}  
  \caption{Common-weights model.}
  \label{subfig:sim1lddpweight}
\end{subfigure}
\begin{subfigure}{\textwidth}
  \centering
  \includegraphics[width=15.5cm,height=3cm]{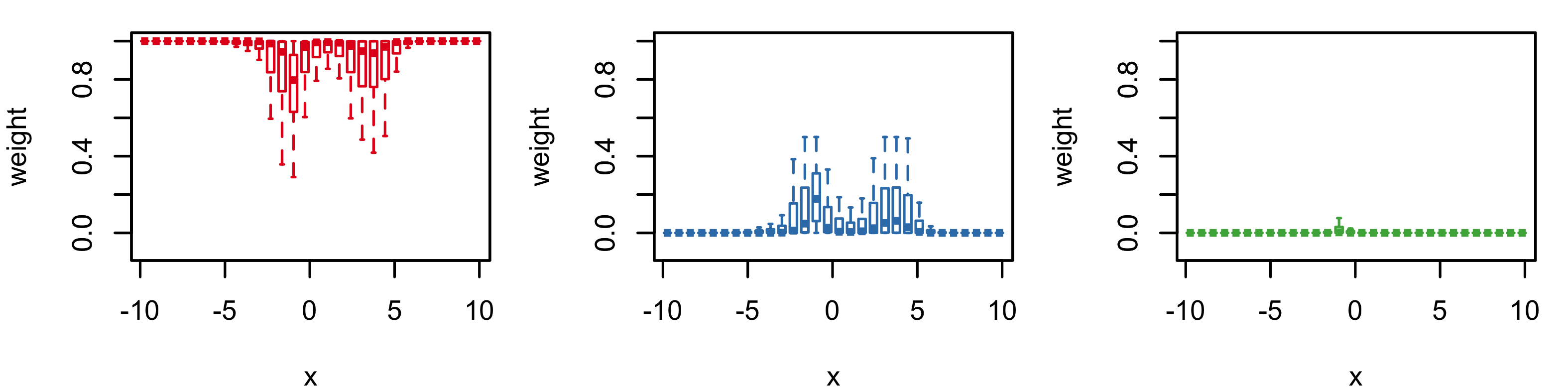}  
  \caption{Common-atoms model.}
  \label{subfig:sim1lsbpweight}
\end{subfigure}
\begin{subfigure}{\textwidth}
  \centering
  \includegraphics[width=15.5cm,height=3cm]{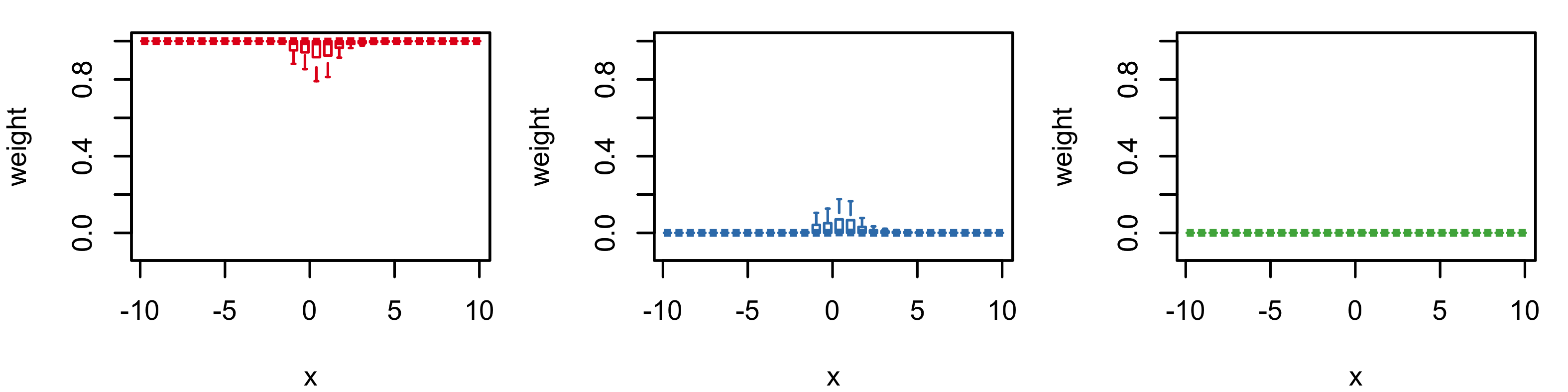}   
  \caption{General model.}
  \label{subfig:sim1generalweight}
\end{subfigure}
\caption{
{\small Second simulation example. Box plots of the posterior samples for the three largest mixture weights under each of the nonparametric models.}
}
\label{fig:sim1postweight}
\end{figure}

To further explore how the different nonparametric models utilize the mixture structure, 
Figure \ref{fig:sim1postweight} shows the posterior distributions of the three largest 
mixture weights across covariate values. 
The general model is the most efficient in terms of the number of effective mixture 
components, using a second component (with small weight) only for covariates values
around $0$. This is to be expected, since it is those covariate values that result in 
practically relevant differences between the probit regression function (used to generate 
the data) and the logistic regression kernel. The common-atoms model activates effectively 
one extra component for covariate values where the regression functions are not flat. Compared
to the general model, it places larger weights on the second component to account 
for the constant atoms. On the other hand, the mixture weights can not change with the 
covariates for the common-weights model. Hence, to recover the probit regression 
function, this model utilizes effectively three mixture components, with the second and 
third assigned larger (global) weight than the other two models.

\begin{figure}[t!]
\centering
\begin{subfigure}{\textwidth}
  \centering 
  \includegraphics[width=16cm,height=3cm]{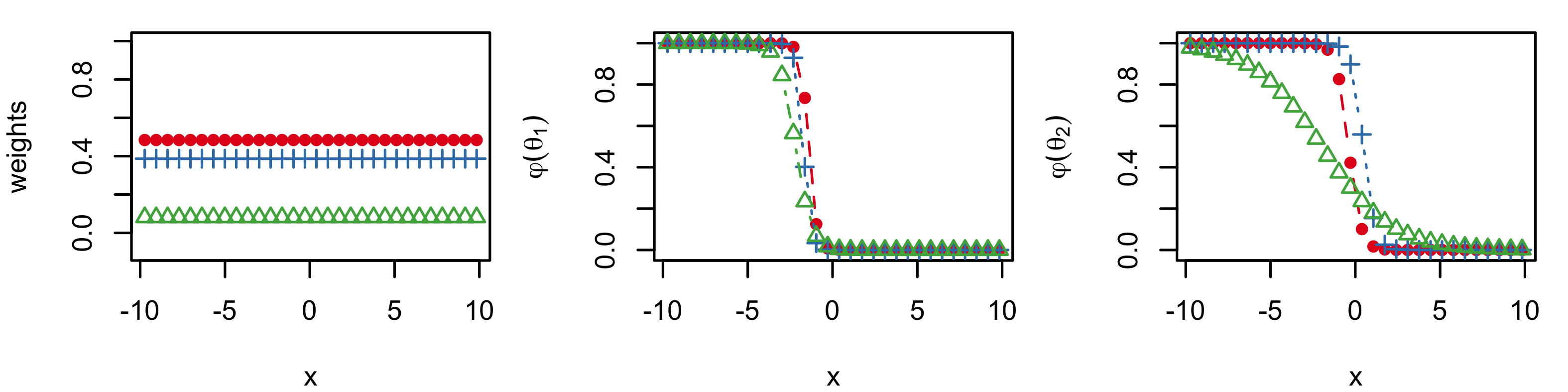}  
  \caption{Common-weigths model.}
  \label{subfig:lddpwasim1}
\end{subfigure}
\begin{subfigure}{\textwidth}
  \centering
  \includegraphics[width=16cm,height=3cm]{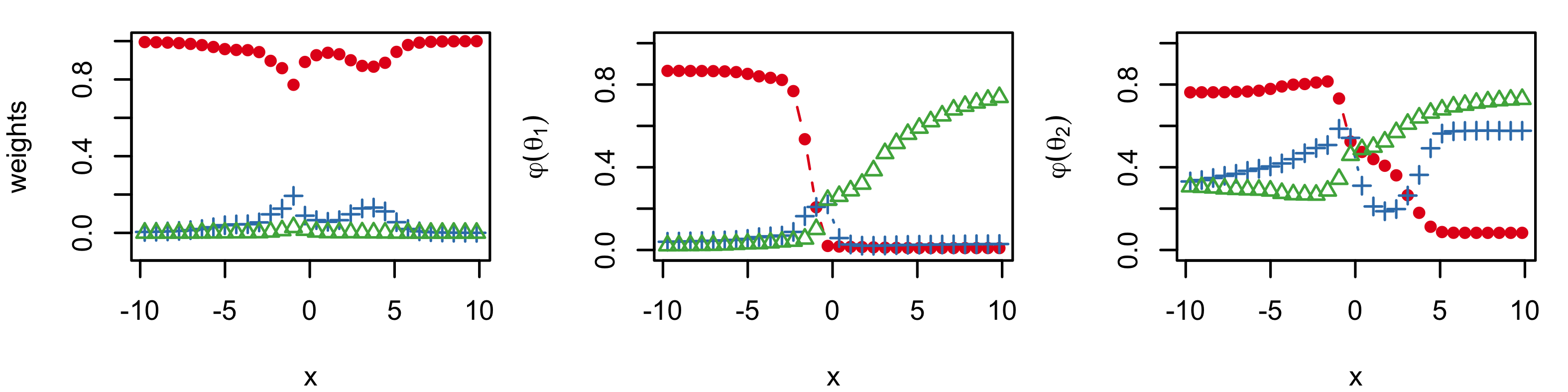}  
  \caption{Common-atoms model.}
  \label{subfig:lsbpwasim1}
\end{subfigure}
\begin{subfigure}{\textwidth}
  \centering
  \includegraphics[width=16cm,height=3cm]{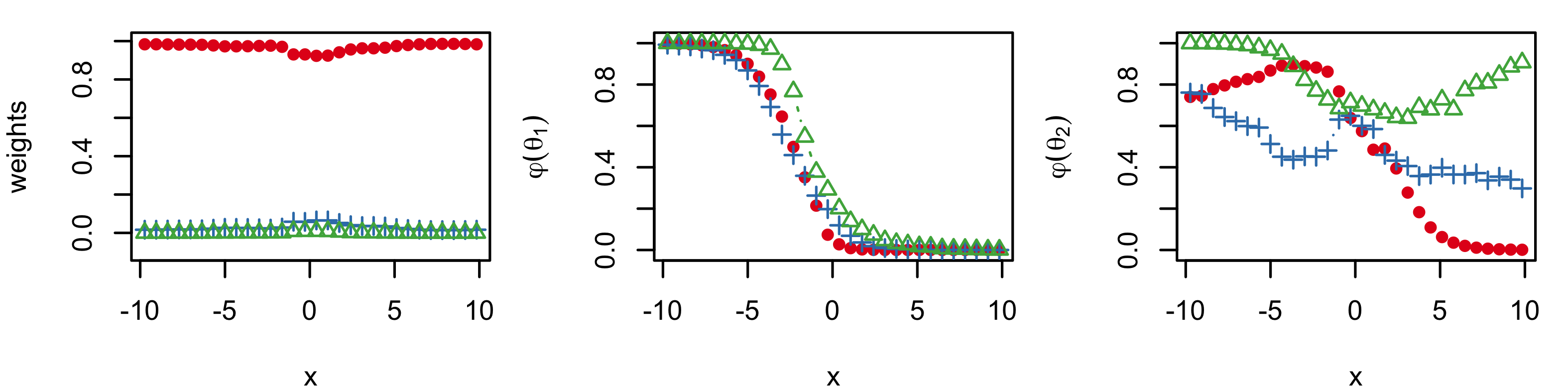}   
  \caption{General model.}
  \label{subfig:lsbpgeneralwasim1}
\end{subfigure}

\caption{\small Second simulation example. Posterior mean estimates of the three largest 
mixture weights and atoms. The red circle, blue plus, and green triangle correspond to the 
first, second, and third largest weights, respectively.}
\label{fig:ordweights}
\end{figure}

We also plot the posterior mean of the three largest weights and the corresponding 
atoms $\varphi(\theta_1)$ and $\varphi(\theta_2)$ in Figure \ref{fig:ordweights}. Combining 
with the posterior predictive loss criterion for each model, we can diagnose how the three models 
estimate the probability response curves. It appears that all three models are dominated by the 
mixing component with the largest weight, whose shape is similar to the truth. (The common-weights 
model favors two mixing components, but the two components are close to each other.) 
Regarding differences, the common-atoms model can only adjust the shape of the regression functions 
through the mixing weights. It thus uses more effective mixing components with shapes 
differing dramatically, yielding larger goodness-of-fit and penalty terms. The general model 
is overall the most effective in capturing the actual shape. It uses fewer and similar effective 
mixing components, leading to smaller penalty terms. 

%
%

\begin{figure}[t!]
\centering
\begin{subfigure}{\textwidth}
  \centering 
  \includegraphics[width=16cm,height=3cm]{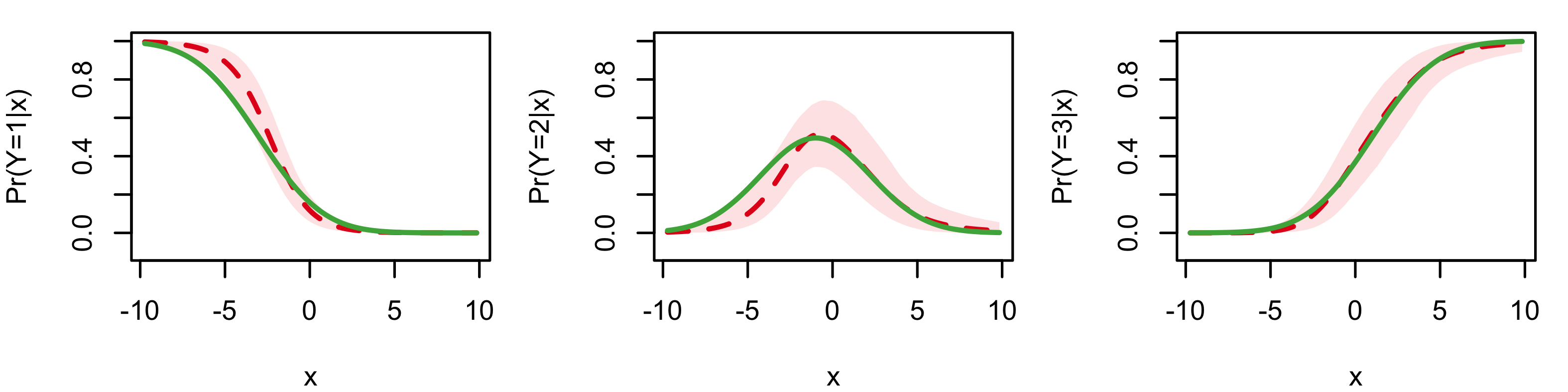}  
  \caption{Common-weights model.}
  \label{subfig:sim1lddppostspec}
\end{subfigure}
\begin{subfigure}{\textwidth}
  \centering
  \includegraphics[width=16cm,height=3cm]{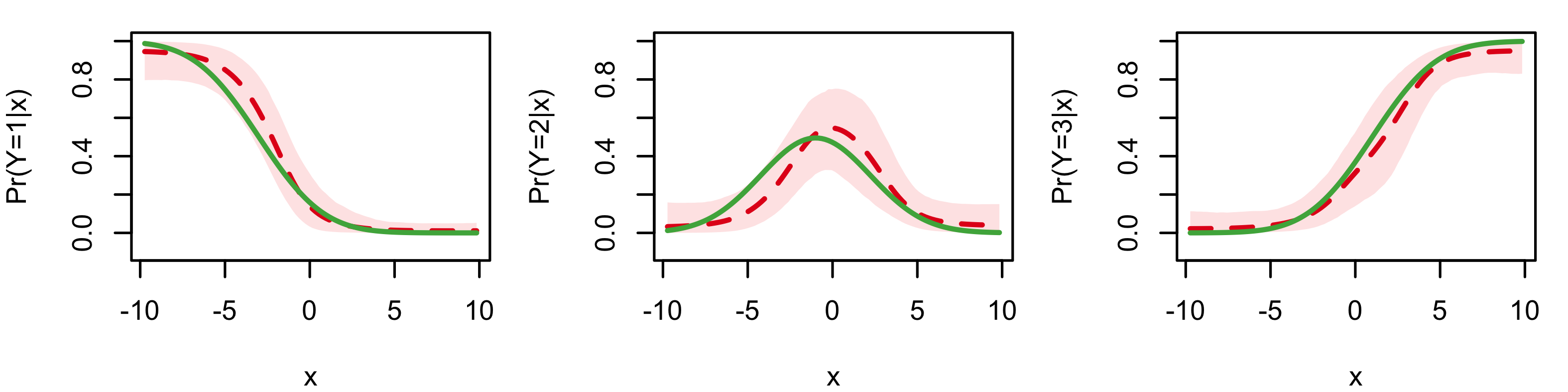}  
  \caption{Common-atoms model.}
  \label{subfig:sim1lsbppostspec}
\end{subfigure}
\begin{subfigure}{\textwidth}
  \centering
  \includegraphics[width=16cm,height=3cm]{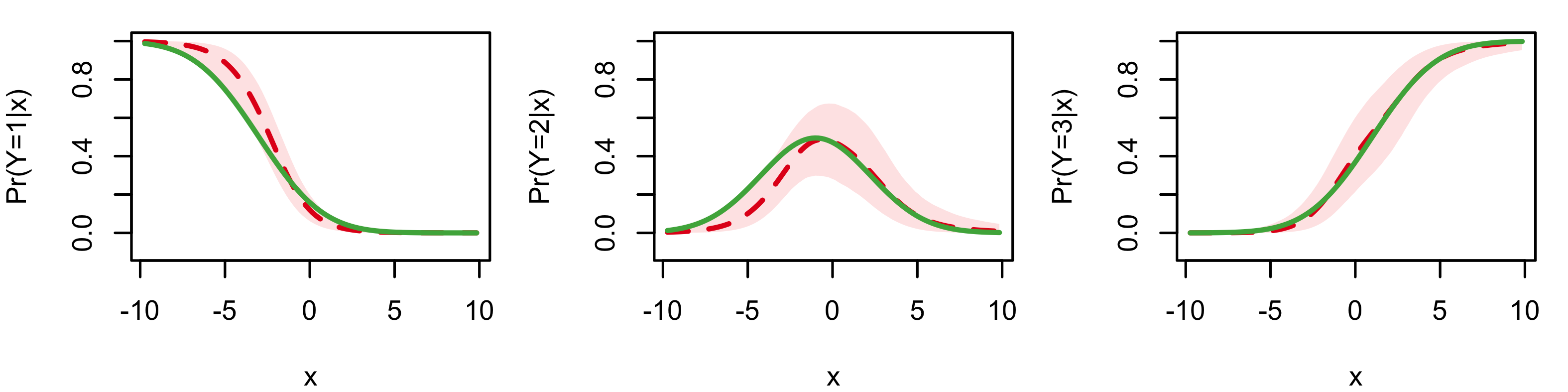}   
  \caption{General model.}
  \label{subfig:sim1generalpostspec}
\end{subfigure}
\caption{\small Second simulation example. Inference results for the marginal probability 
response curves, under the informative prior specification. In each panel, the dashed line 
and shaded region correspond to the posterior mean and $95\%$ credible interval estimates, 
whereas the (green) solid line denotes the true regression function.}
\label{fig:sim1postspec}
\end{figure}

The sample size for this example was intentionally taken to be relatively small, in order 
to study sensitivity to the prior choice, as well as to demonstrate the practical utility 
of a more focused prior specification approach. If the monotonicity of two of the regression 
functions was in fact available as prior information, such information can be incorporated 
into the model, as discussed in Section \ref{subsec:prispecori} of the main paper. 
Indeed, we consider a more information prior choice to reflect a decreasing shape 
for the first probability response function, and an increasing trend for the third response
probability function. Using the aforementioned prior specification strategy, the prior 
hyperparameters for the general model are
\begin{equation*}
    \boldsymbol{\mu}_{0j}=(-2,-0.9)^T,\,\,\Lambda_{0j}=\begin{pmatrix} 0.8 & 0 \\ 0 & 0.072\end{pmatrix},\,\,j=1,2 
\end{equation*}
and we set $\boldsymbol{\gamma}_0=(-2.5,0)$ and $\Gamma_0=diag(10,1)$ to favor a relatively large 
number of distinct components a priori. The prior hyperparameters of the simplified models are 
specified accordingly. This set of prior hyperparameters leads to the posterior estimates 
shown in Figure \ref{fig:sim1postspec}. We note the more accurate posterior mean estimates 
and the reduction in the width of the posterior uncertainty bands, the improvement being 
particularly noteworthy for the common-atoms model.

\subsection{Third synthetic data example}
\label{subsec:SMthird}

\begin{figure}[t!]
\centering
\includegraphics[width=15cm,height=5cm]{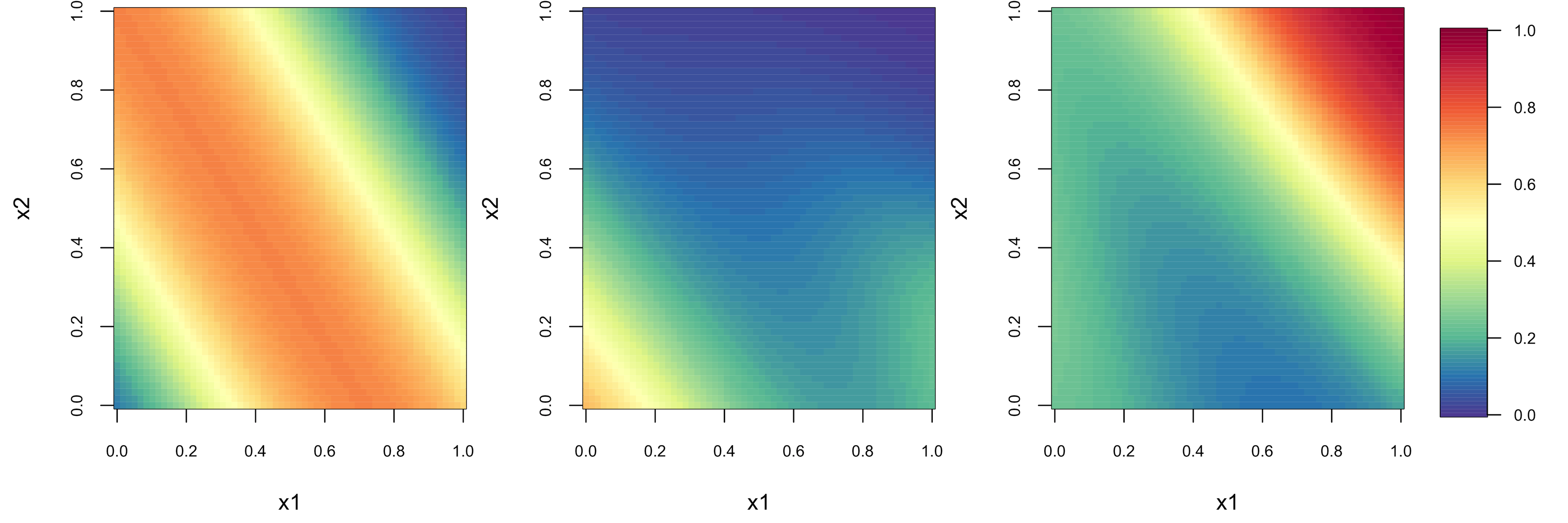}
\caption{
{\small Third simulation example. The true probability response surface 
$\pi_j(x_1,x_2)$, for $j=1,2,3$ (from left to right).}
}
\label{fig:sim3truesurface}
\end{figure}

The purpose of the third simulation example is to investigate the effectiveness of the proposed 
models in capturing the joint effect of covariates. We consider two covariates and sample their 
values as $x_{is}\stackrel{i.i.d.}{\sim}Unif(0,1)$, for $s=1,2$. The responses are sampled from 
the multinomial distribution with the continuation-ratio logits parameterization, 
where the $\theta_j(\mathbf{x})$, for $j=1,2$, are non-linear functions of the
covariates. Specifically, we take $\theta_1(\mathbf{x})=$ $c_{11}+c_{12}\sin(a_{11}x_1+a_{12}x_2)$, 
and $\theta_2(\mathbf{x})=$ $c_{21}+c_{22}\exp(a_{21}x_1+a_{22}x_2)$. The covariate effects are 
non-linear and non-additive, resulting in non-standard probability response surfaces (displayed in 
Figure \ref{fig:sim3truesurface}). We fit the general LSBP mixture model, as well as its two 
simplified versions. Note that covariates enter the mixture model structure linearly and 
additively, through the weights (common-atoms model), the atoms (common-weights model), or 
both (general model). It is therefore of interest to examine how the proposed models capture 
the non-standard probability response surfaces 
through the mixing of linear combinations of covariates. We take a fairly large sample 
size ($n=3000$) to ensure the data is representative of the underlying data generating mechanism.

\begin{figure}[t!]
\centering
\begin{subfigure}{0.475\textwidth}
  \centering 
  \includegraphics[width=8cm,height=2.67cm]{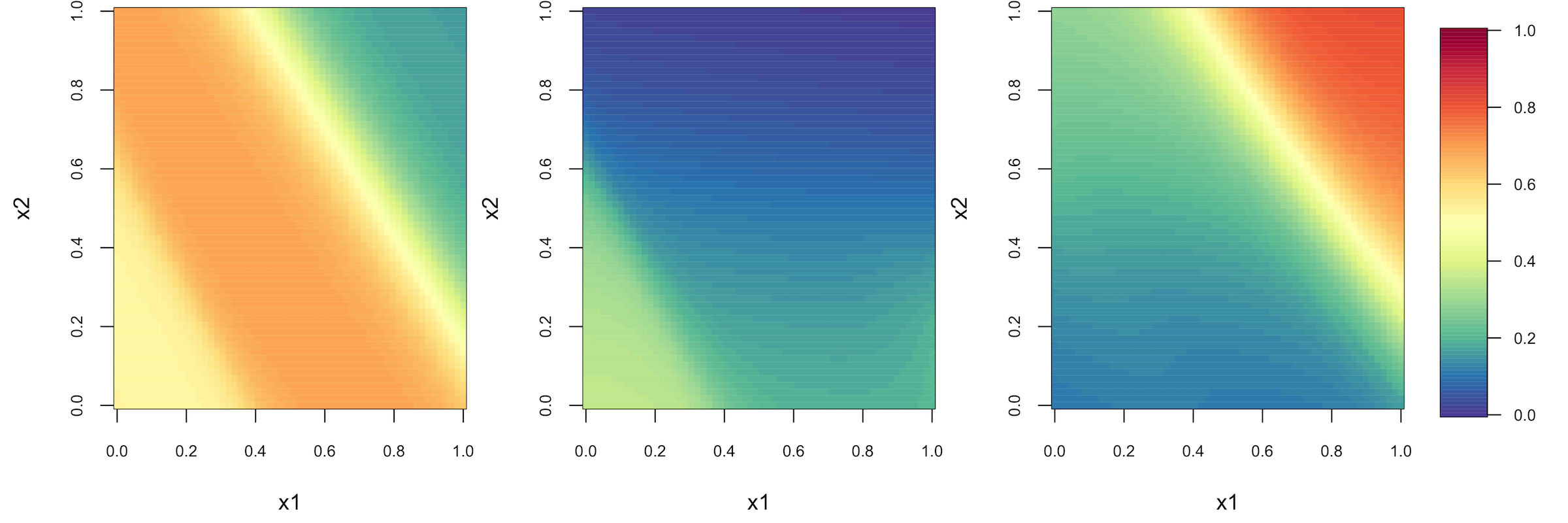}  
  \caption{Common-weights model.}
  \label{subfig:sim3lddp}
\end{subfigure}
\hfill
\begin{subfigure}{0.475\textwidth}
  \centering
  \includegraphics[width=8cm,height=2.67cm]{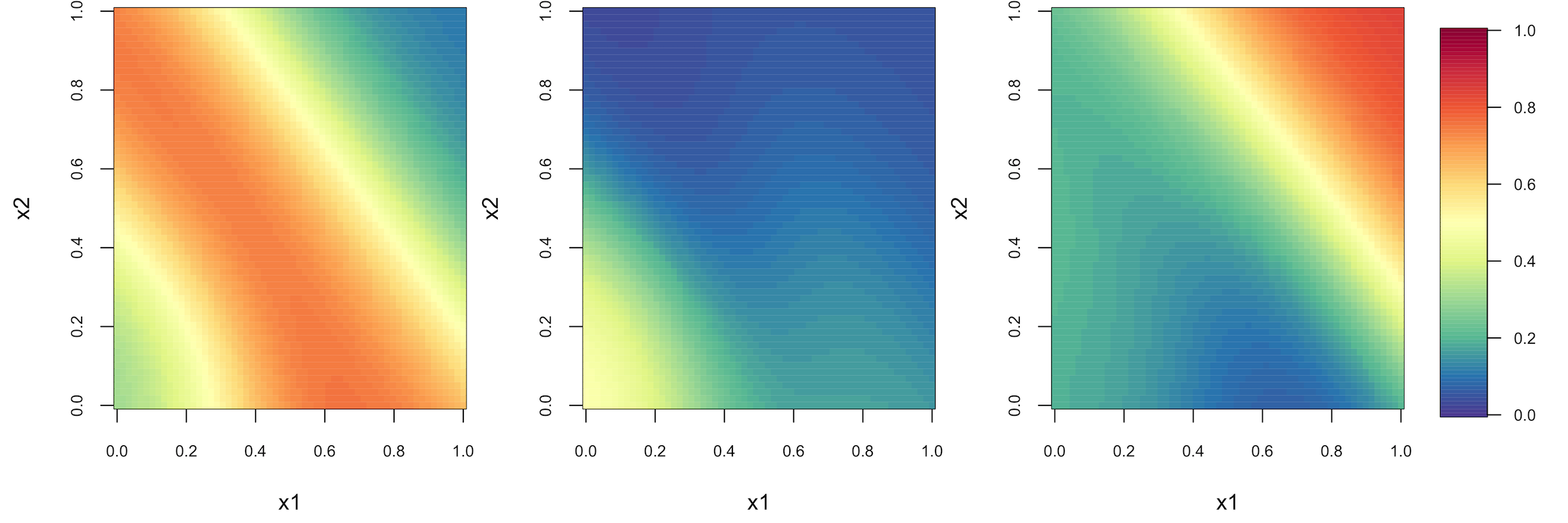}  
  \caption{Common-atoms model.}
  \label{subfig:sim3lsbp}
\end{subfigure}
\vskip\baselineskip
\begin{subfigure}{0.475\textwidth}
  \centering
  \includegraphics[width=8cm,height=2.67cm]{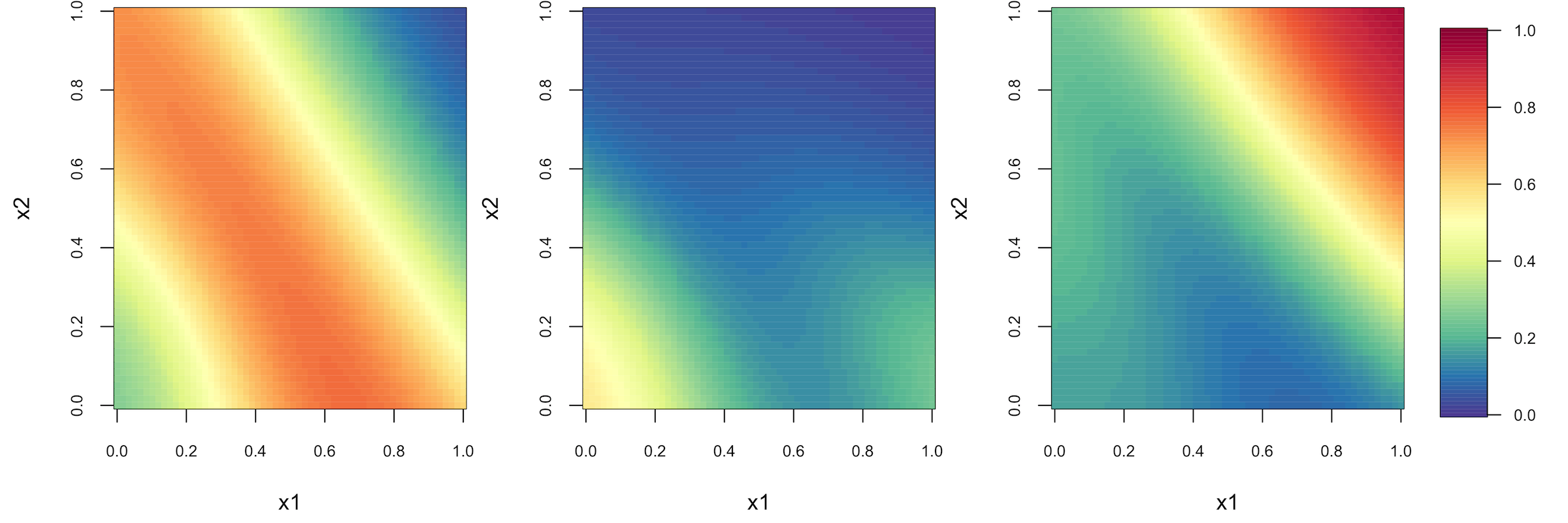}  
  \caption{General model.}
  \label{subfig:sim3gen}
\end{subfigure}
\hfill
\begin{subfigure}{0.475\textwidth}
  \centering
  \includegraphics[width=8cm,height=2.67cm]{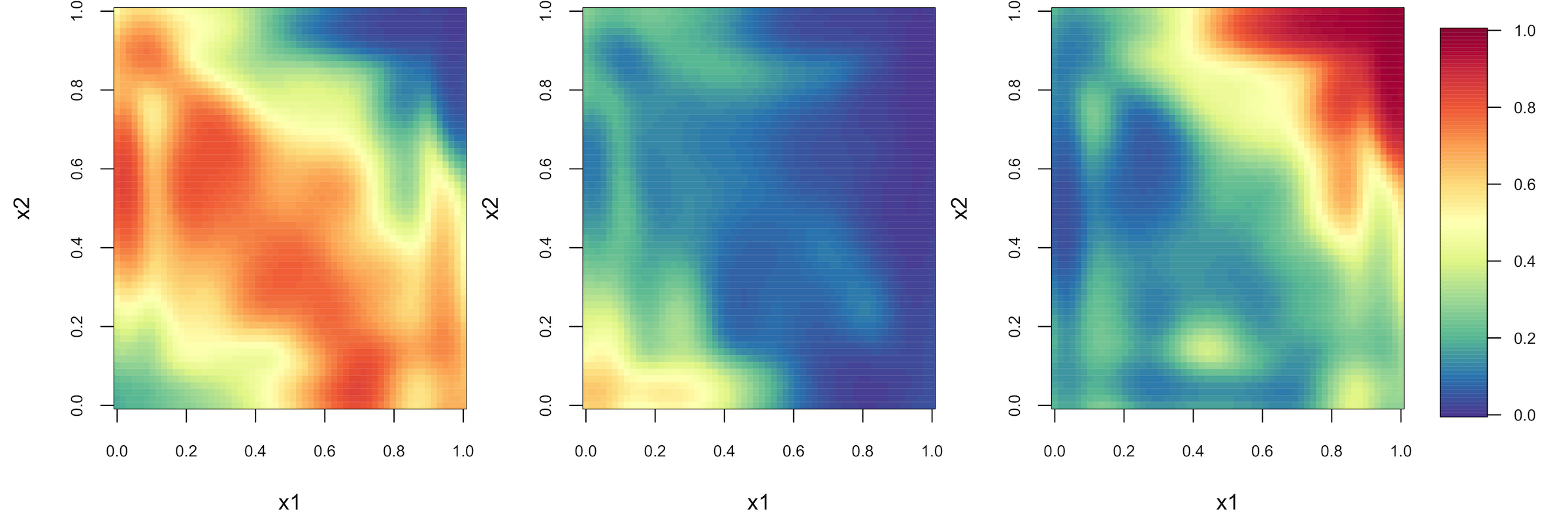}  
  \caption{Density regression model.}
  \label{subfig:sim3comp}
\end{subfigure}
\caption{\small Third simulation example. Posterior mean estimates of 
$\pi_j(x_1,x_2)$, for $j=1,2,3$ (from left to right).}
\label{fig:sim3postestimate}
\end{figure}

We set the prior hyperparameters for the atoms according to the baseline choice, while the 
hyperparameters for the weights are set to encourage a relatively large number of effective 
mixture components. The truncation level is set as $L=50$. Figure panels \ref{subfig:sim3lddp}, \ref{subfig:sim3lsbp}, and 
\ref{subfig:sim3gen} present the posterior mean estimates of the probability response surfaces 
under the common-weights model, the common-atoms model, and the general model, respectively. 
Although none of the models include non-linear or interaction terms for the covariates, the 
general model captures the non-linear joint effect particularly well, and the common-atoms 
model also demonstrates good estimation performance.
These two models involve covariate-dependent weights, which allow for local adjustment in 
the regression surface estimates. As illustrated by this example, such local adjustment 
is beneficial, especially when the covariate effects are expected to be non-standard.

\begin{figure}[t!]
\centering
\begin{subfigure}{0.475\textwidth}
  \centering 
  \includegraphics[width=8cm,height=2.67cm]{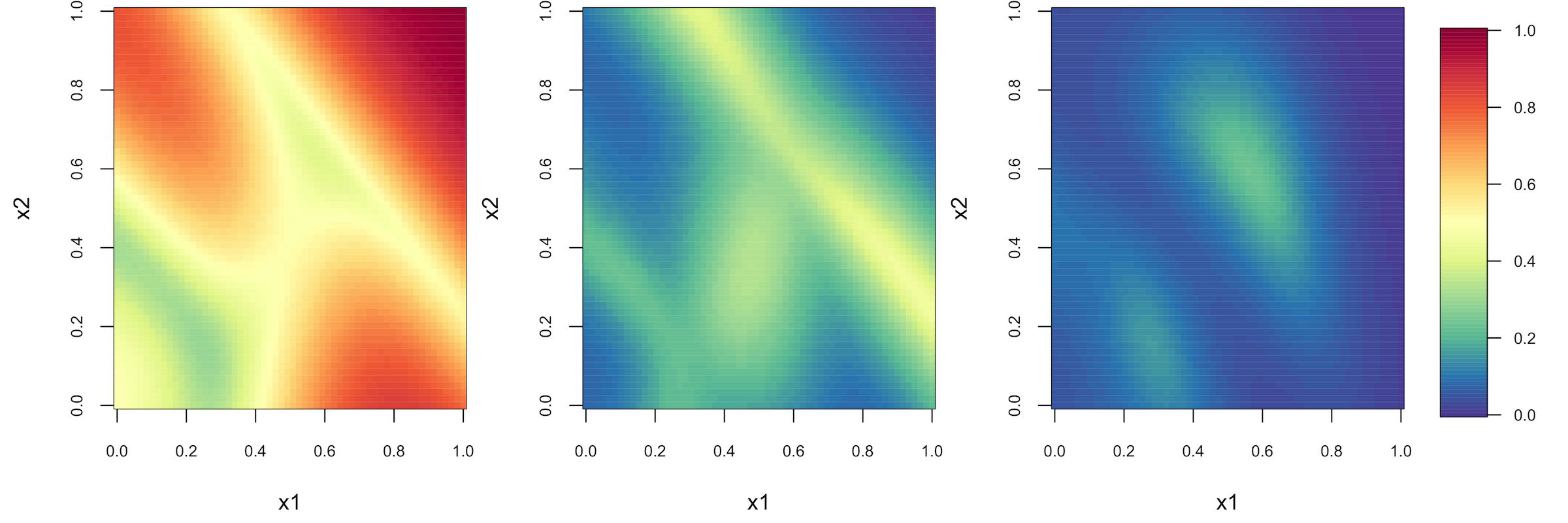}  
  \caption{Common-atoms model.}
  \label{subfig:sim3lsbpweight}
\end{subfigure}
\hfill
\begin{subfigure}{0.475\textwidth}
  \centering
  \includegraphics[width=8cm,height=2.67cm]{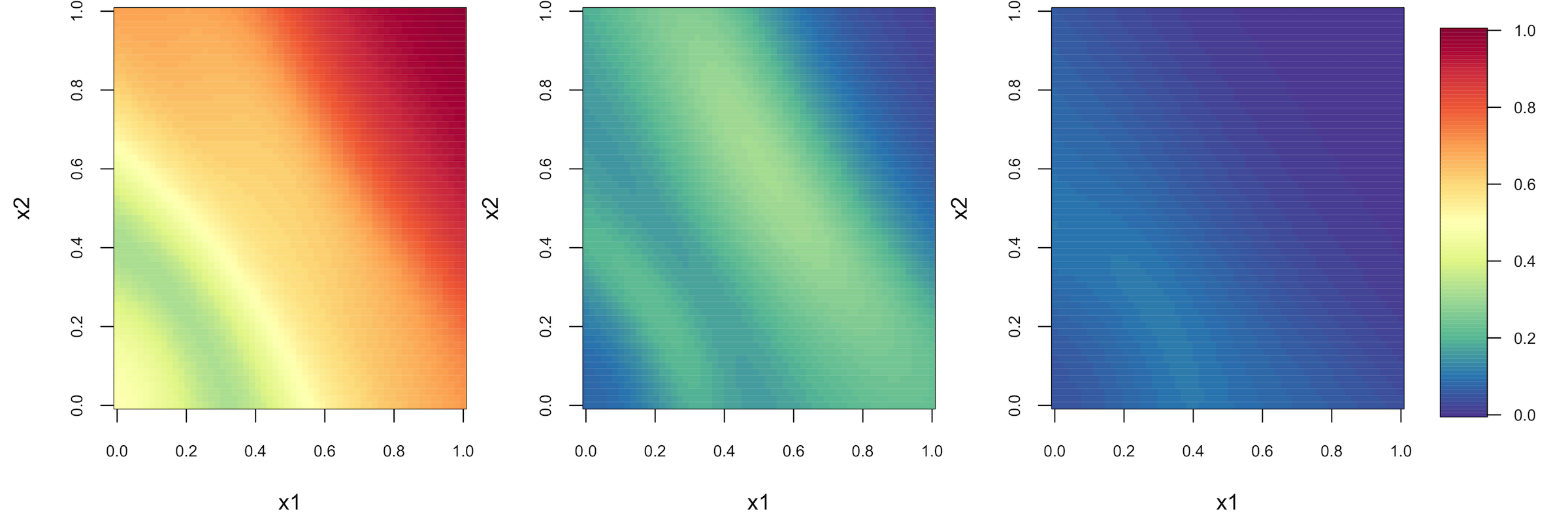}  
  \caption{General model.}
  \label{subfig:sim3generalweight}
\end{subfigure}
\caption{\small Third simulation example. Posterior mean of the three largest mixture 
weights for the common-atoms and general LSBP mixture models.}
\label{fig:sim3weights}
\end{figure}

Focusing on the two LSBP mixture models with covariate-dependent weights, 
Figure \ref{fig:sim3weights} plots the posterior mean estimates for the three largest weights 
over a grid in the covariate space. The common-atoms model has more pronounced local changes, 
which is to be expected because it can adjust the shape of the regression surfaces only through 
the mixture weights. The general model exhibits more smooth estimated weight surfaces.

As discussed in the Introduction section of the main paper, the literature contains a relatively 
small collection of fully nonparametric Bayesian methods for ordinal regression. Here, we include
comparison with the density regression model from \cite{DeYoreoKottas2018}, for which the R code 
to implement the MCMC posterior simulation algorithm is available (from the online supplemental 
material of the journal article). Under the density regression modeling approach, the joint 
distribution of the two covariates and the latent continuous response is modeled with a DP mixture
of trivariate normal densities, mixing on both the mean and the covariance matrix. The ordinal 
response probabilities emerge by discretizing the implied conditional distribution for the 
latent response given the covariates. The conditioning results in a model structure for 
the ordinal response distribution that can be interpreted as a mixture of probit regressions
with covariate-dependent mixture weights.

The posterior mean estimates for the probability response surfaces 
are plotted in Figure \ref{subfig:sim3comp}. The density regression model captures the 
general trends, but it overestimates the local changes of the probability surfaces. This is 
likely due to both the different model structure (modeling the joint covariate-response 
distribution rather than the conditional response distribution) and to the underlying truth, 
which is much more structured than the density regression model. The results in 
Figure \ref{subfig:sim3comp} are similar under different prior choices, in particular, under
priors for the DP precision parameter that favor both large and 
fairly small number of distinct mixture components. 

We further compare models based on their performance in estimating the probability response 
surfaces $\pi_j(x_1,x_2)$, for $j=1,2,3$. We consider three metrics: the rooted mean 
square error (RMSE); the average length of the 95\% posterior credible interval; and, the ratio 
at which the 95\% credible interval covers the true probability. More specifically, for $N$ 
grid points on the covariate space, the RMSE is calculated by 
$$
\bar{E}_j \, = \, N^{-1} \, 
\sqrt{\sum_{i=1}^N \{ \pi^*_j(x_{1i},x_{2i})-\hat{\pi}_j(x_{1i},x_{2i}) \}^2}
$$
where $\pi^*_j(x_{1i},x_{2i})$ and $\hat{\pi}_j(x_{1i},x_{2i})$ denote respectively the 
posterior mean estimate and the true value of the $j$-th category response probability at 
covariate values $(x_{1i},x_{2i})$. In addition, the average 95\% posterior credible interval 
length regarding the $j$-th category is obtained as $\bar{L}_j=$
$N^{-1} \, \sum_{i=1}^N \{ \pi^{\text{U}}_j(x_{1i},x_{2i}) - \pi^{\text{L}}_j(x_{1i},x_{2i}) \}$, 
with $\pi^{\text{U}}_j(x_{1i},x_{2i})$ and $\pi^{\text{L}}_j(x_{1i},x_{2i})$ denoting the 
$97.5$th and $2.5$th percentiles of the posterior samples. Finally, the coverage percentage 
of the 95\% posterior credible interval is calculated by 
$$
\bar{R}_j \, = \, N^{-1} \, 
\sum_{i=1}^N \mathbf{1} \{ \pi^{\text{L}}_j(x_{1i},x_{2i}) \leq \hat{\pi}_j(x_{1i},x_{2i})
\leq\pi^{\text{U}}_j(x_{1i},x_{2i}) \}.
$$
Table \ref{tab:sim3compare} reports the metrics values for the four models. 
Among the LSBP mixture models, the general model performs better with respect to essentially
all metrics, followed by the common-atoms model. These results reinforce the findings 
from the graphical comparison of the posterior mean estimates for the probability response 
surfaces. Also consistent with the graphical comparison in Figure \ref{fig:sim3postestimate},
the general and common-atoms LSBP mixture models outperform the density regression model
in terms of RMSE. This is also the case with respect to the average 95\% posterior credible 
interval length. The density regression model achieves the best results
for the coverage criterion, with the general LSBP mixture model a fairly close second. 
Overall, the general LSBP mixture model yields the best performance under the 
particular simulation scenario.

\begin{table}[t!]
\centering 
\caption{Third simulation example. Summary of model comparison results, using the RMSE 
$\bar{E}_j$, average $95\%$ posterior credible interval length $\bar{L}_j$, and the 
coverage of the $95\%$ posterior credible interval $\bar{R}_j$, for $j=1,2,3$. The 
values that correspond to the best model are given in bold.} 
\label{tab:sim3compare}
\begin{tabular}{cccccccccc} 
\hline \hline 
Model & $\bar{E}_1$ & $\bar{L}_1$ & $\bar{R}_1$ & $\bar{E}_2$ & $\bar{L}_2$ 
& $\bar{R}_2$ & $\bar{E}_3$ & $\bar{L}_3$ & $\bar{R}_3$\\ 
\hline 
Common-weights & 4.075 & 0.088 & 0.594 & 2.243 & 0.029 & 0.932 & 2.586 & 0.092 & 0.736 \\
Common-atoms & 2.260 & 0.058 & 0.785 & 1.686 & 0.044 & 0.735 & 2.149 & 0.071 & 0.884 \\
General & \textbf{1.629} & \textbf{0.038} & 0.905 & \textbf{0.827} & \textbf{0.019} & 0.998 & \textbf{1.310} & \textbf{0.043} & \textbf{0.923} \\
Density regression & 3.823 & 0.069 & \textbf{0.951} & 2.091 & 0.062 & \textbf{0.968} 
& 3.579 & 0.099 & \textbf{0.923} \\
\hline 
\hline 
\end{tabular} 
\end{table}

\subsection{Credit ratings of U.S. firms}
\label{subsec:SMcreditdata}

We compare the posterior point estimates of the first-order marginal probability 
curves $\pi_j(x_s)$, $j=1,\ldots,5$ and $s=1,\ldots,5$, obtained by the proposed nonparametric 
models and their parametric backbone. The results are shown in Figure \ref{fig:compareallmodel}. 
The continuation-ratio logits regression model contains fewer parameters than the nonparametric models, 
leading to reduced flexibility. As shown in Figure \ref{subfig:conratiomargin}, the estimated curves 
have a standard shape. In contrast, the flexible nature of the nonparametric models enables more 
complex regression relationships to be extracted from the data. To discuss the difference in the 
estimated regression trends regarding a certain covariate, consider the standardized log-sales 
variable as an example. For low to moderate log-sales values, the probability of the lowest rating 
level decreases at about the same rate under the continuation-ratio logits model, while the 
decreasing rate varies under the three nonparametric models. 

\afterpage{
\begin{figure}[!htbp]
\centering
\begin{subfigure}{\textwidth}
  \centering 
  \includegraphics[width=16cm,height=4.32cm]{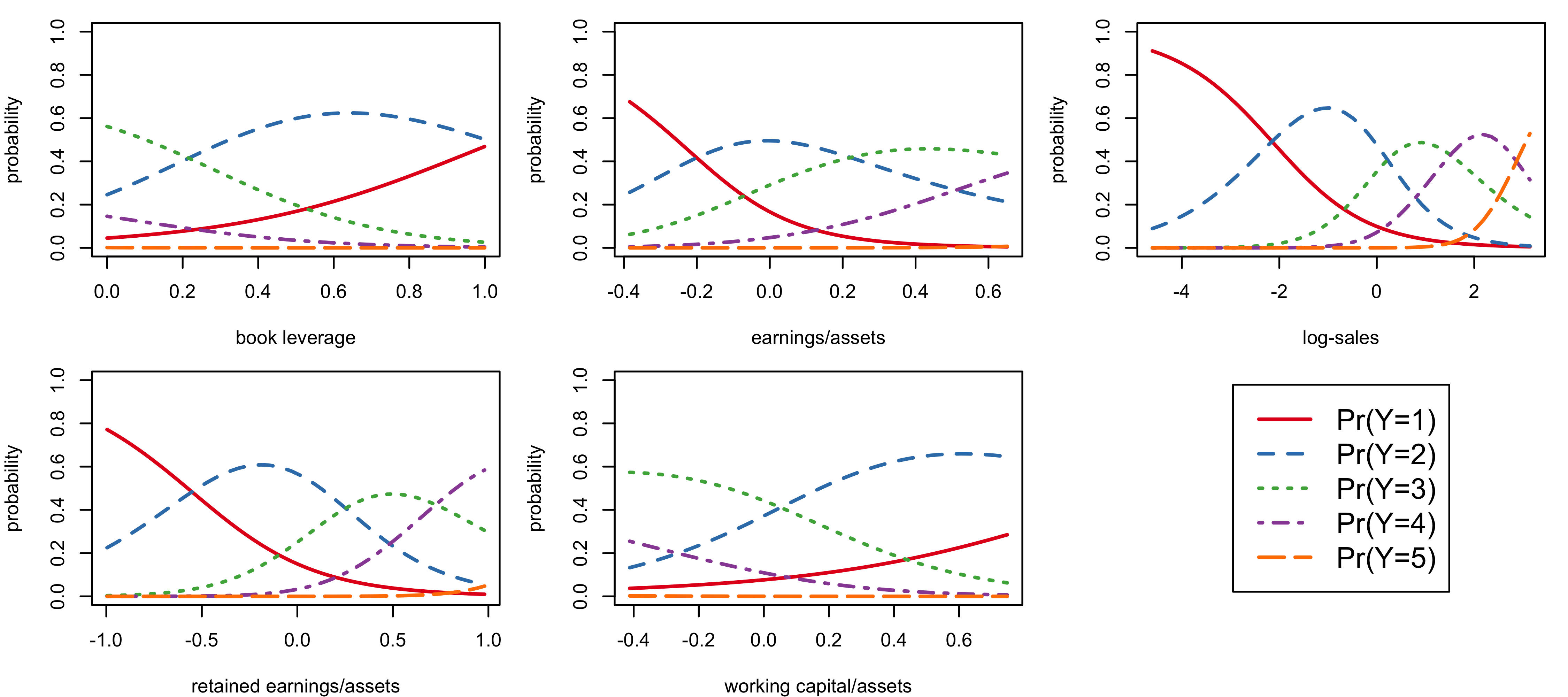}  
  \caption{The continuation-ratio logits regression model.}
  \label{subfig:conratiomargin}
\end{subfigure}
\begin{subfigure}{\textwidth}
  \centering
  \includegraphics[width=16cm,height=4.32cm]{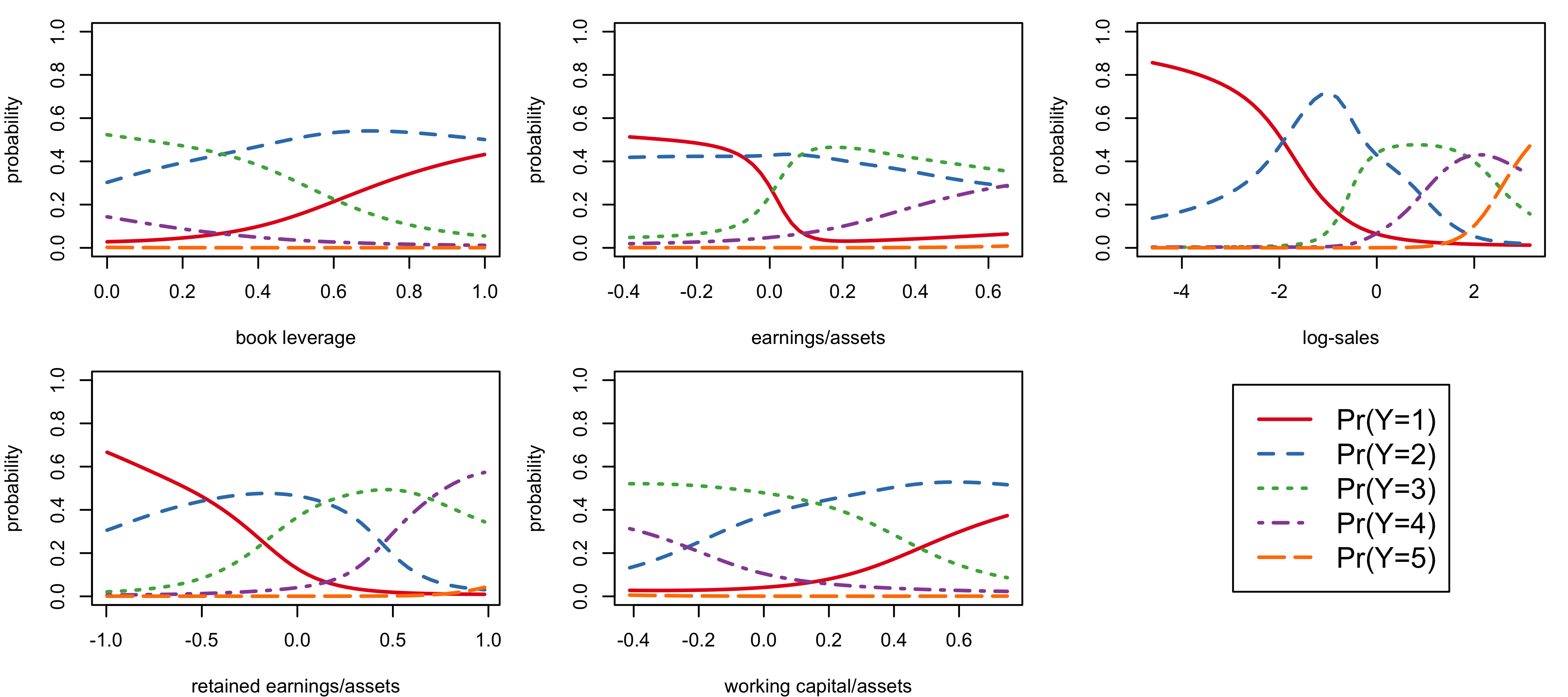}  
  \caption{The common-weights model.}
  \label{subfig:lddpmargin}
\end{subfigure}
\begin{subfigure}{\textwidth}
  \centering
  \includegraphics[width=16cm,height=4.32cm]{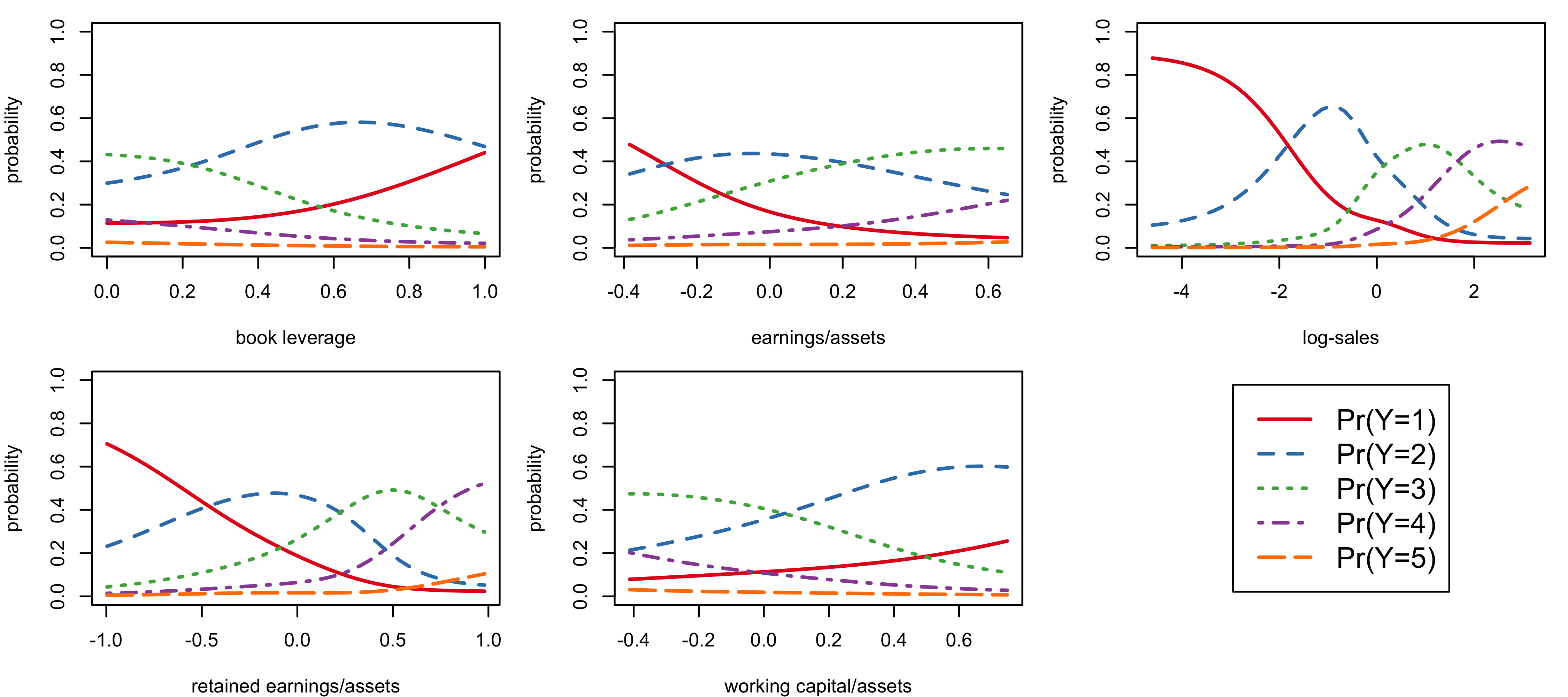}  
  \caption{The common-atoms model.}
  \label{subfig:lsbpmargin}
\end{subfigure}
\begin{subfigure}{\textwidth}
  \centering
  \includegraphics[width=16cm,height=4.32cm]{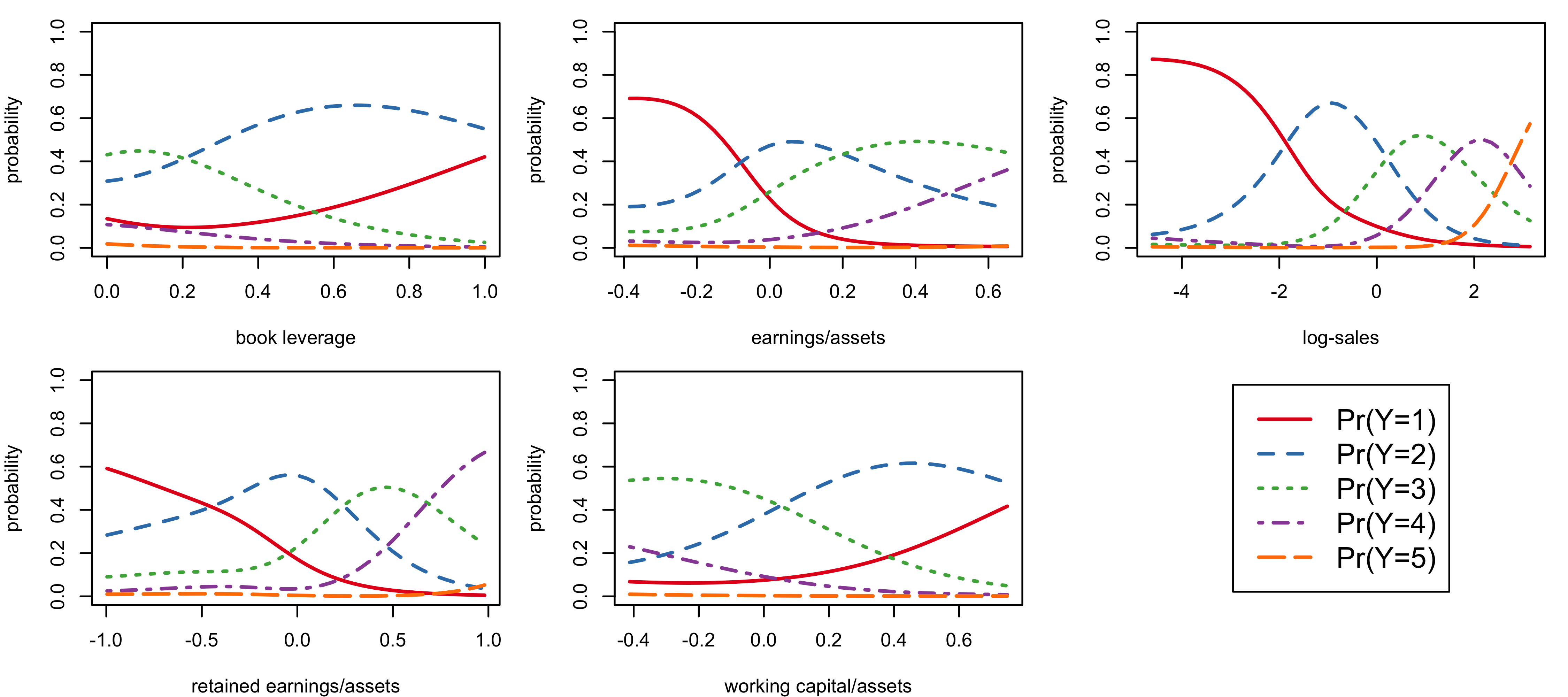}  
  \caption{The general model.}
  \label{subfig:generalmargin}
\end{subfigure}
\caption{\small Credit ratings data. Posterior mean of $\pi_j(x_s)$, for $s=1,\ldots, 5$ 
and $j=1,\ldots,5$. All five ordinal response curves are displayed in a single panel for each covariate.}
\label{fig:compareallmodel}
\end{figure}
}

A formal model comparison based on the posterior predictive loss criterion (definition provided 
in Section \ref{subsec:SMfirst}) is presented in Table \ref{tab:modelcompcredit}. The 
common-weights model and the  general model yield comparable results, while both models outperform 
the parametric model in predicting the probability of the first four credit levels. As for credit 
level 5, the three models are comparable regarding the goodness-of-fit criterion, while the 
nonparametric models yield larger penalty terms. We notice that there are substantially fewer 
firms with credit level 5, which may be the reason the nonparametric models provide larger penalty.

\begin{table}[t!]
\centering 
\caption{Credit ratings data. Summary of the posterior predictive loss criteria for model comparison. 
Each pair of numbers corresponds to $(G_j(\mathcal{M}),P_j(\mathcal{M}))$, $j=1,\cdots,5$. 
``Parametric'' refers to the continuation-ratio logits model. The values for model with the smallest $G_j(\mathcal{M})+P_j(\mathcal{M})$ are highlighted in bold. } 
\label{tab:modelcompcredit}
\begin{tabular}{ccccc} 
\hline \hline 
 & Parametric & Common-weights & Common-atoms & General \\ 
\hline 
Credit level 1 & $(92.65,90.17)$ & \textbf{(88.07,92.38)} & $(92.64,104.97)$ & $(86.61,95.79)$ \\
Credit level 2 & $(158.71,158.72)$ & $(153.13,158.92)$ & $(156.04,163.96)$ & \textbf{(153.10,158.07)}\\
Credit level 3 & $(150.18,150.82)$ & \textbf{(145.40,150.38)} & $(149.00,152.03)$ & $(148.11,148.60)$\\
Credit level 4 & $(95.95,96.29)$ & $(95.08,97.23)$ & $(97.41,100.10)$ & \textbf{(94.20,94.24)}\\
Credit level 5 & \textbf{(17.80,17.46)} & $(17.85,20.57)$ & $(21.19,31.04)$ & $(17.74,20.47)$\\
\hline 
\hline 
\end{tabular} 
\end{table} 

\begin{figure}[t!]
\centering
\includegraphics[width=16cm,height=4cm]{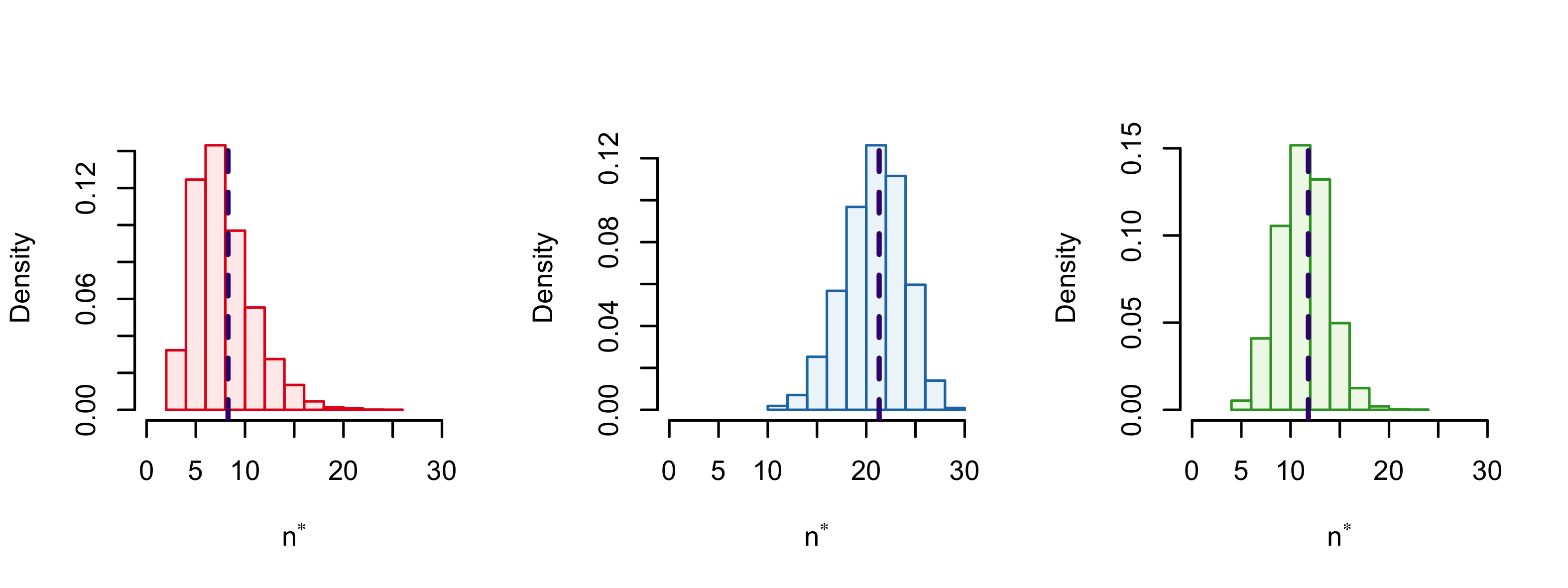}
\caption{\small Credit ratings data. Posterior distributions for the number of distinct 
components. The dashed line indicates the posterior mean. The panels correspond to, from 
left to right, the common-weights model, the common-atoms model, and the general model, 
respectively.}
\label{fig:numcompcredit}
\end{figure}

We also notice the significantly larger penalty terms for the common-atoms model. This is to be 
expected, since the shape of the regression curves are only allowed to be adjusted through the 
mixing weights under the common-atoms model. Therefore, to provide regression curve estimates 
as in Figure \ref{subfig:lsbpmargin}, it activates more mixing components. This is clear from 
Figure \ref{fig:numcompcredit} which shows the posterior distribution for the number of 
distinct components under the three nonparametric models. 

\begin{figure}[t!]
\centering
\includegraphics[width=16cm,height=18cm]{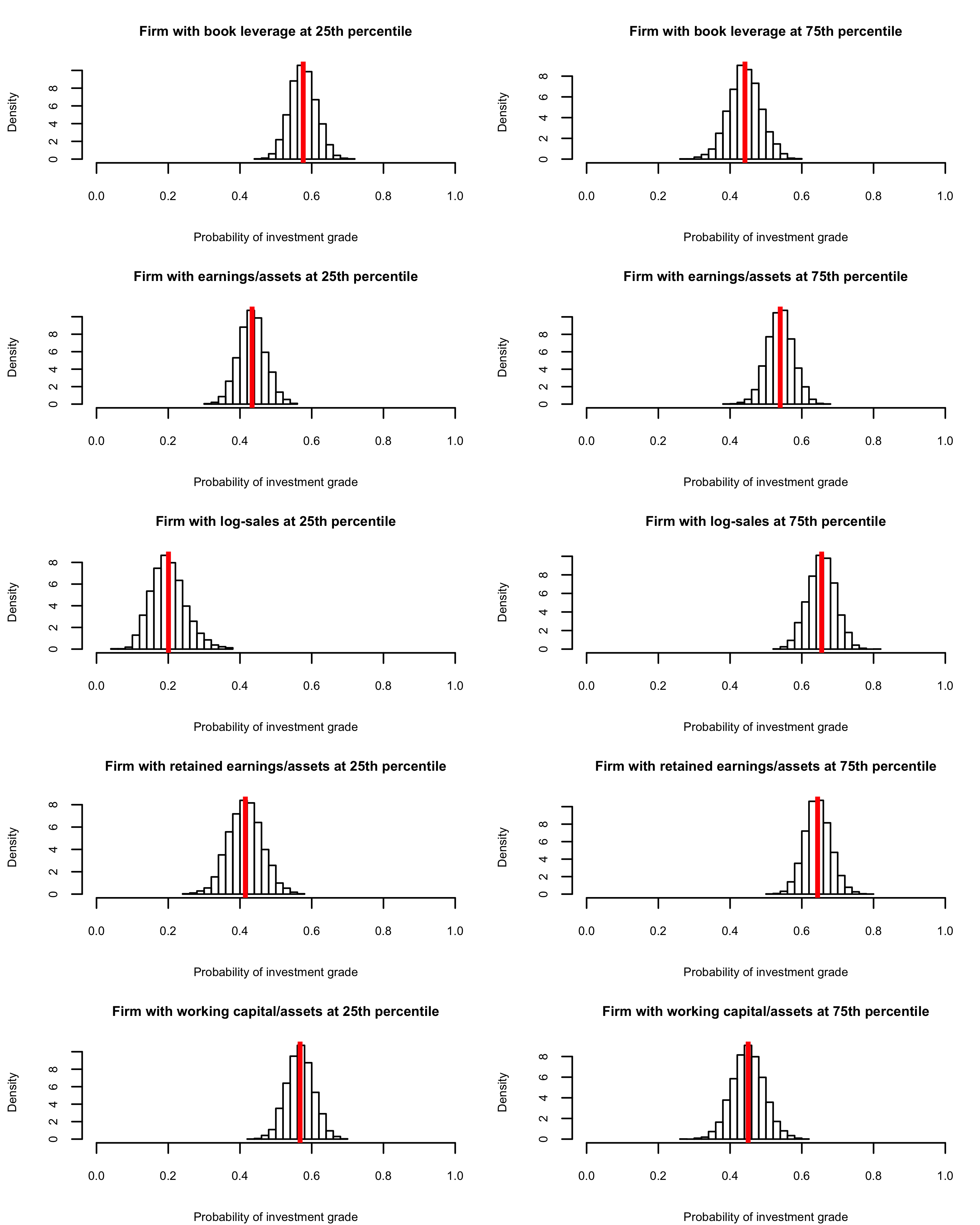}
\caption{\small Credit ratings data. Posterior distributions of the probability of obtaining 
investment grade rating under the common-weights model. The red solid lines indicate the posterior mean.}
\label{fig:credproblddp}
\end{figure}

Furthermore, it is also of interest to investigate the model performance on prediction. 
The credit rating of firms can be partitioned into two categories: investment grade (rating score 
is 3 or higher) and speculative grade. Because many bond portfolio managers are not allowed to 
invest in speculative grade bonds, firms with a speculative rating incur significant costs. 
It is helpful to check the models' implied posterior probability of obtaining an investment 
grade for a particular firm. We consider five prediction scenarios corresponding to the five 
covariates. In each scenario, we evaluate the change in the investment grade probability 
associated with one of the covariates changing from the 25th to the 75th percentile of the 
observed values, while holding all the other covariates at the average value of all observations. 
Figure \ref{fig:credproblddp} displays the posterior distribution of the probability of 
obtaining investment grade under the common-weights model.

Under the common-weights model, the probability moves along the expected direction concerning 
all covariates, except for the working capital, which coincides with the discovery in \citet{Verbeek2008}. 
The results indicate higher leverage, meaning that a firm is financed relatively more with debt, 
reduces the expected credit rating. This is due to the fact that firms with high leverage face 
substantially higher debt financing costs. In addition, the larger firms, indicated by larger log-sales, 
have significantly better credit ratings than smaller firms, ceteris paribus. Higher earnings before 
interest and taxes and higher retained earnings also improve credit ratings. Furthermore, one would 
expect that maintaining a high level of working capital would enhance a company's credit rating since 
it reduces risk. However, a high level of working capital reduces profits, raising concern about the 
company's ability to cover interest payments. This argument suggests a concave relationship between 
working capital and credit rating, postulating that firms could have an optimal working capital ratio. 
Our result indicates that the optimal ratio lies between the first and third quartiles.

\subsection{Retinopathy data}
\label{subsec:SMretdata}

\begin{figure}[t!]
\centering
\begin{subfigure}{\textwidth}
  \centering
  \includegraphics[width=16cm,height=4cm]{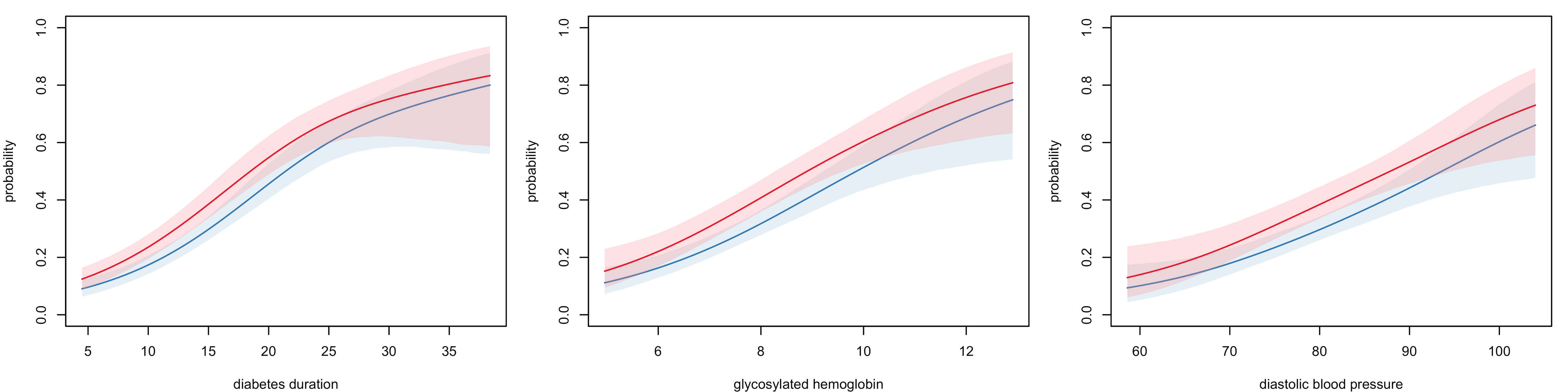}  
  \caption{\small Marginal probability response curve $\text{Pr}(\mathbf{Y}\geq 2\mid G_{\mathbf{x}})$.}
  \label{subfig:retprob23}
\end{subfigure}
\begin{subfigure}{\textwidth}
  \centering
  \includegraphics[width=16cm,height=4cm]{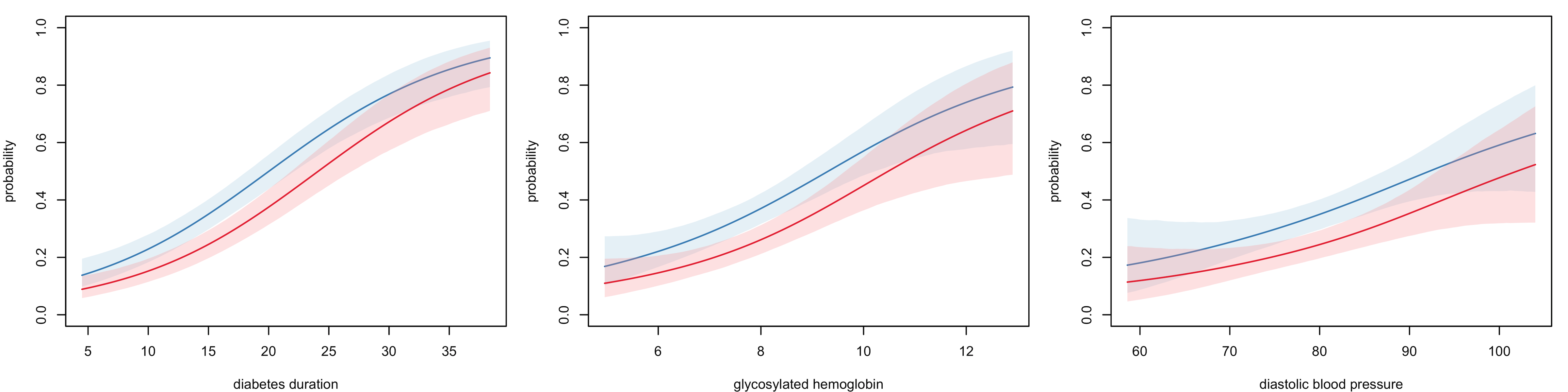}  
  \caption{\small Conditional probability response curve $\text{Pr}(\mathbf{Y}= 3\mid\mathbf{Y}\geq 2, G_{\mathbf{x}})$.}
  \label{subfig:retcondprob}
\end{subfigure}
\caption{\small Retinopathy data. Posterior mean (lines) and 95\% interval (shaded regions) 
estimates of the target probability response curves for smokers (in red) and non-smokers (in blue). 
Each panel corresponds to a risk factor.}
\label{fig:retprob}
\end{figure}

We examine the marginal effect of risk factors on developing retinopathy. We focus on the 
marginal probability of at least nonproliferative retinopathy 
$\text{Pr}(\mathbf{Y}\geq 2\mid G_{\mathbf{x}})$ and the conditional probability of advanced 
retinopathy or blindness, conditioning on at least nonproliferative retinopathy 
$\text{Pr}(\mathbf{Y}= 3\mid\mathbf{Y}\geq 2, G_{\mathbf{x}})$. Notice that the conditional 
probability response curves of advanced retinopathy or blindness are of particular interest, 
as they depict the effect of risk factors on development of more severe symptoms. 
Figure \ref{fig:retprob} shows the posterior mean and interval estimates of the marginal 
probability (Panel \ref{subfig:retprob23}) and conditional probability 
(Panel \ref{subfig:retcondprob}) response curves. They are obtained by computing, for smokers 
and non-smokers, posterior realizations for $\text{Pr}(\mathbf{Y}\geq 2\mid G_{\mathbf{x}})$ 
and $\text{Pr}(\mathbf{Y}= 3\mid\mathbf{Y}\geq 2, G_{\mathbf{x}})$ at a grid over the observed 
range for a risk factor, keeping the values of the other risk factors fixed at their 
observed average.

Figure \ref{fig:retprob} reveals the effect of risk factors on developing retinopathy. 
As expected, diabetes duration, glycosylated hemoglobin, and diastolic blood pressure all have 
a positive effect, not only on developing retinopathy, but also on developing further severe 
symptoms for patients with the disease. 
%

\begin{figure}[t!]
\centering
\includegraphics[width=16cm,height=4cm]{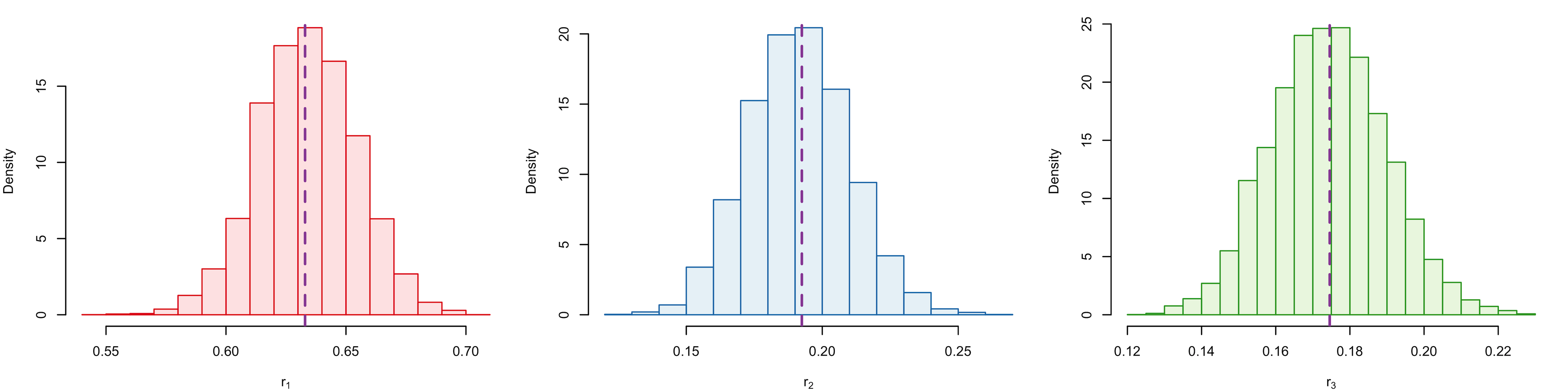}  
\caption{\small Retinopathy data. Posterior predictive distribution of the proportion for 
each ordinal response category. The dashed line indicates the observed proportion.}
\label{fig:retdatacheck}
\end{figure}

As one illustration of predictive model assessment of the general LSBP mixture model, we examine 
the posterior predictive distribution of the test statistics $r_j$, $j=1,2,3$, which represent 
the proportion of each ordinal response category among the $n$ responses. That is, for each 
posterior sample of model parameters, we obtain posterior predictive samples for all in-sample 
subjects, and compute the proportion of each category. We therefore obtain the posterior 
predictive distribution of each test statistic, which can be compared with the observed proportion. 
The results, displayed in Figure \ref{fig:retdatacheck}, suggest that the model generates predictions 
which are compatible with the observed responses. 

Moreover, we conduct model comparison using the posterior predictive loss criterion (see 
Table \ref{tab:retsummary}). The general model yields overall the best performance. The 
common-atoms model yields the smallest values for the goodness-of-fit component for 
category $1$ and $2$, but with larger penalty term values. All three nonparametric models 
show moderate to substantial improvement over the parametric continuation-ratio logits model.

\begin{table}[t!]
\centering 
\caption{Retinopathy data. Summary of model comparison using the posterior predictive loss criterion. 
Each pair of numbers corresponds to $(G_j(\mathcal{M}),P_j(\mathcal{M}))$, $j=1,\cdots,3$. 
``Parametric'' refers to the continuation-ratio logits model. The values for model with the smallest $G_j(\mathcal{M})+P_j(\mathcal{M})$ are highlighted in bold. } 
\label{tab:retsummary}
\begin{tabular}{ccccc} 
\hline \hline 
 & Parametric & Common-weights & Common-atoms & General \\ 
\hline 
Ordinal level 1 & $(110.07,119.01)$ & $(104.62,106.27)$ & $(101.96,112.20)$ & \textbf{(102.98,106.42)} \\
Ordinal level 2 & $(90.30,93.82)$ & $(90.00,91.16)$ & \textbf{(88.05,91.45)} & $(89.52,90.41)$\\
Ordinal level 3 & $(70.98,72.57)$ & $(63.66,64.42)$ & $(63.25,72.40)$ & \textbf{(63.17,64.15)}\\
\hline 
\hline 
\end{tabular} 
\end{table}

\end{document}